\documentclass[11pt]{article}
\usepackage{latexsym,amsmath,amsfonts}
\usepackage{mathrsfs}
\usepackage{caption}
\usepackage{amsmath,amsfonts,amssymb}
\usepackage{helvet}
\usepackage{booktabs}
\usepackage{url}
\usepackage{rotating}
\usepackage{multirow}
\usepackage{mathrsfs}
\usepackage{epsfig}
\usepackage{bm}
\usepackage{bbm}
\usepackage{color}

\def\text#1{\hbox{#1}}

\def\1{\mbox{1\hspace{-.20em}I}}

\newtheorem{proposition}{Proposition}[section]

\newtheorem{remark}{Remark}[section]
\newtheorem{corollary}{Corollary}[section]

\def\e{\varepsilon}
\def\r_\e{r_\epsilon}

\def\r{\right}

\numberwithin{equation}{section}
\def\AArm{\fam0 \mathrm}%
\def\AAk#1#2{\setbox\AAbo=\hbox{#2}\AAdi=\wd\AAbo\kern#1\AAdi{}}%
\def\AAr#1#2#3{\setbox\AAbo=\hbox{#2}\AAdi=\ht\AAbo\raise#1\AAdi\hbox{#3}}%
\def\BBn{{\AArm I\!N}}%
\def\BBp{{\AArm I\!P}}%
\def \I{\hbox{\rm 1\hskip -3pt I}}

\def\trasp{\mathsf{T}}
\newcommand\bbone{\ensuremath{\mathbbm{1}}}
\definecolor{darkross}{rgb}{0.008,0.412,0.471}
\definecolor{middleross}{rgb}{0.012,0.580,0.663}
\definecolor{lightross}{rgb}{0.016,0.749,0.855}
\definecolor{darkblue}{rgb}{0.067,0.008,0.471}
\definecolor{middleblue}{rgb}{0.094,0.012,0.663}
\definecolor{lightblue}{rgb}{0.122,0.016,0.855}
\definecolor{darkpurple}{rgb}{0.471,0.008,0.412}
\definecolor{middlepurple}{rgb}{0.663,0.012,0.580}
\definecolor{lightpurple}{rgb}{0.855,0.016,0.749}
\definecolor{darkbrown}{rgb}{0.471,0.067,0.008}
\definecolor{middlebrown}{rgb}{0.663,0.094,0.012}
\definecolor{lightbrown}{rgb}{0.855,0.122,0.016}
\definecolor{darkolive}{rgb}{0.412,0.471,0.008}
\definecolor{middleolive}{rgb}{0.580,0.663,0.012}
\definecolor{lightolive}{rgb}{0.749,0.855,0.016}
\definecolor{darkgreen}{rgb}{0.008,0.417,0.067}
\definecolor{middlegreen}{rgb}{0.012,0.663,0.094}
\definecolor{lightgreen}{rgb}{0.016,0.855,0.122}
\definecolor{darkocre}{rgb}{0.471,0.298,0.008}
\definecolor{middleocre}{rgb}{0.663,0.420,0.012}
\definecolor{lightocre}{rgb}{0.855,0.541,0.016}

\begin{document}

\newenvironment{proofdot}{{\bfseries Proof ---}
\parindent=0pt}{\parfillskip=0pt\hfil$\square\square$\par\bigbreak}

\title{\bf Bayesian inference  for CoVaR}

\author{M. Bernardi, \\ \textit{Sapienza University of Rome, Italy}\\
and \\
 G. Gayraud,  \\ \textit{Universit\'e de Technologie de Compi\`egne and CREST, France}\\ and \\  L. Petrella,\\ \textit{Sapienza University of Rome, Italy}}


\maketitle

\begin{abstract}
\vspace{0.5cm}
Recent financial disasters emphasised the need to investigate the consequence associated with the tail co-movements among institutions; episodes of contagion are frequently observed and increase the probability of large losses affecting market participants' risk capital. Commonly used risk management tools fail to account for potential spillover effects among institutions because they provide individual risk assessment. We contribute to analyse the interdependence effects of extreme events providing an estimation tool for evaluating the Conditional Value-at-Risk (CoVaR) defined as the Value-at-Risk of an institution conditioned on another institution being under distress. In particular, our approach relies on a Bayesian quantile regression framework. We propose a Markov chain Monte Carlo algorithm exploiting the representation of the Asymmetric Laplace distribution as a location-scale mixture of Normals. Moreover, since risk measures are usually evaluated on time series data and returns typically change over time, we extend the CoVaR model to account for the dynamics of the tail behaviour. An application to U.S. companies belonging to different sectors of the Standard and Poor's Composite Index is considered to evaluate the marginal contribution to the overall systemic risk of each individual institution.\newline\newline
%
%
\end{abstract}

\section{Introduction}
\label{sec:intro}
%
During the last years a particular attention has been devoted to measure and quantify the level of financial risk within a firm or investment portfolio. One of the most diffuse risk measurement has become the Value-at-Risk (VaR) which measures the maximum loss in value of a portfolio over a predetermined time period for a given confidence level. In fact, in the current banking regulation framework, the VaR becomes an important risk capital evaluation tool where different institutions are considered as independent entities. Unfortunately, such risk measure fails to consider the institution as part of a system which might itself experience instability and spread new sources of systemic risk. For a comprehensive and up to date overview of VaR and related risk measures see for example Jorion \cite{jorion.2007} and McNeil \textit{et al.} \cite{mcneil_etal.2005}. Recent financial disasters emphasised the need for a deep investigation of the co-movement among institutions in order to evaluate their tail interdependence relations. Especially during periods of financial distress, episodes of contagion among institutions are not rare and thus need to be taken into account in order to analyse the overall health level of a financial system: company specific risk can not be appropriately assessed in isolation, without accounting for potential spillover effects to and from other firms. For this reason different systemic risk measures have been proposed in literature to analyse the tail-risk interdependence (see Acharya \textit{et al}. \cite{acharya_etal.2012}, Acharya \textit{et al}. \cite{acharya_etal.2010}, Adams \textit{et al.} \cite{adams_etal.2010}, Brownlees and Engle \cite{brownlees_engle.2012} and Billio \textit{et al.} \cite{billio_etal.2012}). Recently, Adrian and Brunnermeier \cite{adrian_brunnermeier.2011} introduced the so called Conditional Value-at-Risk (CoVaR), which is defined as the overall VaR of an institution conditional on another institution being under distress. In this way the CoVaR not only capture the systematic risk embedded in each institution, but also reflects individual contributions to the systemic risk, capturing extreme tail risk interdependence.\newline
\indent There are many possible ways to infer on VaR and CoVaR. The most common approaches to estimate VaR are the variance--covariance methodology, historical and Monte Carlo simulations. For an overview of alternative parametric and nonparametric methodologies and processes to generate VaR estimates see Jorion \cite{jorion.2007} and Lee and Su \cite{lee_su.2012}.
Se also Chao \textit{et al.} \cite{chao_etal.2012} and Taylor \cite{taylor.2008} for recent developments.
%
%
Bernardi \cite{bernardi.2013} and Bernardi \textit{et al.} \cite{bernardi_etal.2012} propose to estimate VaR and related risk measures by fitting asymmetric mixture models to the unconditional distribution of returns. Moreover, Girardi and Erg\"{u}n \cite{girardi_ergun.2013} propose to estimate the CoVaR using multivariate Generalized ARCH models for the conditional returns, and Bernardi \textit{et al.} \cite{bernardi_etal.2013} consider the class of multivariate hidden Markov models.\newline
%
%
\indent In this paper, since both VaR and CoVaR are distribution quantiles we address the problem of their estimation using a quantile regression approach. 
Quantile regression has been popular as a simple, robust and distribution free modeling tool since the seminal work of Koenker and Basset \cite{koenker_basset.1978} and Koenker \cite{koenker.2005}. It provides a way to model the conditional quantiles of a response variable with respect to some covariates in order to have a more complete picture of the entire conditional distribution than traditional linear regression. In fact, sometimes problem specific features, like skewness, fat-tails, outliers, truncated and censored data, and heteroskedasticity, can shadow the nature of the dependence between the variable of interest and the covariates so that the conditional mean would not be enough to understand the nature of that dependence. In particular, the quantile regression approach is appropriate not only when the underlying model is nonlinear or the innovation terms are non-Gaussian, but also when modeling the tail behaviour of the underlying distribution is the primary interest. There is a number of papers on quantile regression both in frequentist and Bayesian framework dealing with parametric and nonparametric approaches. For a detailed review and references, see for example, Lum and Gelfand \cite{lum_gelfand.2012} and Koenker \cite{koenker.2005}.\newline  
\indent In quantile regression, the quantile of order $\tau$ of a dependent variable $Y$ is expressed as a function of covariates $\mathbf{X}$,  say $q^{\tau}\left(\mathbf{X}\right)$. In literature different representations have been proposed to specify the quantile function  $q^{\tau} (\mathbf{x})$; the most common specification is the linear one adopted hereafter: 
\begin{equation}
\label{eq:linear_quantile}
q^{\tau}\left(\mathbf{x}\right)=\mathbf{x}^{\trasp}\boldsymbol{\theta},
\end{equation} 
where $\mathbf{x}^{\trasp}$ denotes the transpose of $\mathbf{x}$.\newline 
\indent The problem of estimating $q^{\tau}\left(\mathbf{x}\right)$ through quantile regression has been considered both from the frequentist and Bayesian point of view. In the former case, Koenker and Basset \cite{koenker_basset.1978} show that the quantile estimation problem is solved by the following minimisation problem:
\begin{eqnarray}
\displaystyle{{\mbox{\rm argmin}}_{q^{\tau}}}\sum_{t=1}^T\rho_{\tau}\left(y_{t}-q^{\tau}\left(\mathbf{x}_t\right)\right), 
\label{eq:quantile_freq}
\end{eqnarray}
where $\left(y_t,\mathbf{x}_t\right)$ for $t=1, \ldots,T$ are observations from  $\left(Y,\mathbf{X}\right)$ and $\rho_{\tau}\left(y\right)=y\left(\tau-\bbone\left(y<0\right)\right)$ is the quantile loss function. The Bayesian quantile regression approach (see Yu and Moyeed \cite{yu_moyeed.2001}, Kottas and Gelfand \cite{kottas_gelfand.2001} and Kottas and Krnjajic \cite{kottas_krnjajic.2009}) instead considers the distribution of $Y\vert\mathbf{x}$ as belonging to the Asymmetric Laplace distribution family, denoted by $\text{ALD}\left(\tau,q^{\tau}\left(\mathbf{x}\right),\sigma\right)$, with positive $\sigma$, whose density function is given by: 
\begin{equation}
\label{eq:ald_distribution}
{\rm ald}\left(y\mid q^{\tau}(\mathbf{x}),\sigma\right)=\frac{\tau\left(1-\tau\right)}{\sigma}\exp\left\{-\frac{\rho_{\tau}\left(y-q^{\tau}\left(\mathbf{x}\right)\right)}{\sigma}\right\}\bbone_{\left(-\infty,\infty\right)}\left(y\right).
\end{equation} 
The nice feature of the $\text{ALD}\left(\tau,q^{\tau}\left(\mathbf{x}\right),\sigma\right)$ distribution is that the regression function $q^{\tau}\left(\mathbf{x}\right)$ corresponds exactly to the theoretical $\tau$-th quantile of $Y\mid\mathbf{x}$.\newline
%
%
\indent Quantile regression methods have been extensively considered in literature to evaluate the VaR (see, among others, Huang \cite{huang.2012}, Schaumburg \cite{schaumburg.2010}, Chernozhukov and Du \cite{chernozhukov_du.2008}, Kuester \text{et al.} \cite{kuester_etal.2006}, Taylor \cite{taylor.2008} and Gerlach \textit{et al.} \cite{gerlach_etal.2011}); recently Chao \textit{et al.} \cite{chao_etal.2012}, Fan \textit{et al.} \cite{fan_etal.2013}, Hautsch \textit{et al}. \cite{hautsch_etal.2011} and Chao \textit{et al.} \cite{chao_etal.2012} consider the same approach to calculate also the CoVaR. In this paper, we propose a Bayesian approach to cast the CoVaR within a quantile regression framework and we show how to model and evaluate it as a quantile of the conditional distribution of an institution $k$ given a particular quantile of another institution $j$. Bayesian methods are very useful and flexible tools of combing data with prior information in order to provide the entire posterior distribution of the parameters of interest. It also allows for parameter uncertainty to be taken into account when making predictions. In the context of the present paper, since the quantities of interest are risk measures, learning about the whole distribution becomes more relevant due to the interpretation of the VaR and CoVaR as financial losses. The use of Markov chain Monte Carlo methods (MCMC) has made Bayesian inference very attractive during the last decades, allowing for efficient inference for complex statistical models. In the Bayesian quantile regression framework, the inference on the unknown parameters is made analytically tractable because it relies on the exact likelihood function for the quantiles of interest, see equation (\ref{eq:ald_distribution}). Moreover, post--processing the MCMC output we are able to make inference on the VaR and CoVaR functions as well as to calculate their posterior credible sets useful to assess estimates' accuracy.
%
%
Up to our knowledge this is the first attempt to infer on CoVaR from a Bayesian point of view.\newline 
\indent As second step, since risk measures are usually evaluated on time series data and returns typically change over time, we extend the Adrian and Brunnermeier \cite{adrian_brunnermeier.2011} CoVaR approach to account for the dynamics of the tail behaviour. The idea is to consider time varying quantiles to link the future tail behaviour of a time series to its past movements which is important in  risk management contest.  In particular, starting from the idea stated in De Rossi and Harvey \cite{derossi_harvey.2009} we propose a dynamic model to capture the evolution of VaR and CoVaR. 
In order to provide a flexible solution for the quantile modelisation whilst retaining a parsimonious representation, the time evolution of the process should be carefully chosen. Hence, throughout the paper, we propose to model the dynamics of the quantile functions' parameters as local linear trends which represents a good compromise between degree of smoothness of the resulting quantiles and the ability of the model to capture changes over time. Time varying quantiles represent a valid alternative to conditional quantile autoregression proposed in different contexts by Engle and Manganelli \cite{engle_manganelli.2004}, Gerlach \textit{et al}. \cite{gerlach_etal.2011}, Gourieroux and Jasiak \cite{gourieroux_jasak.2008} and Koenker and Xiao \cite{koenker_xiao.2006}.\newline
\indent To implement the dynamic Bayesian inference, we cast VaR and CoVaR models in state space representation and we run a Gibbs sampler algorithm using the Exponential-Gaussian  mixture  representation of ALD distributions (Kotz \textit{et al.} \cite{kotz_etal.2001}). This approach allows us to obtain a conditionallly Gaussian state space representation which permits an efficient numerical solution to the inferential problem. 
%
%
In order to make posterior inference, we use the maximum a posteriori summarising criteria and we prove that it leads to estimated quantiles having good sample properties according to De Rossi and Harvey results \cite{derossi_harvey.2009}.\newline
\indent There are several applications of CoVaR which are interesting both in economics and finance. In this paper we analyse different U.S. companies belonging to several sectors of the Standard and Poor's Composite Index (S\&P500) in order to evaluate the marginal contribution to the overall systemic risk of a single institution belonging to it. The empirical results show that the proposed models provide realistic and informative characterisation of extreme tail co-movements. Moreover, our findings suggest that the dynamic model we propose is more appropriate when dealing with financial time series data.\newline
\indent The paper is organised as follows. In Section \ref{sec:var_covar_representations} we give a brief definition of Value-at-Risk and Conditional Value-at-Risk measures. In Section \ref{sec:bayesian_inference} we build the time invariant Bayesian model and we provide details on how to make inference using Markov chain Monte Carlo (MCMC) algorithms. In Section \ref{sec:DynTVQuantile} we extend the previous framework to the time varying case, allowing the representation of marginal and conditional quantiles as functions of latent processes. In Section \ref{sec:empirical_application} we apply the proposed models to real data, while Section \ref{sec:conclusion} concludes.
%
\section{VaR and CoVaR representations}
\label{sec:var_covar_representations}
%
Let $\left(Y_{1},\ldots,Y_{d}\right)$ be a $d$-dimensional $\left(d >1\right)$ random vector where each $Y_j$ is expressed through some covariates $\mathbf{X}=\left(X_{1},X_{2},\ldots, X_{M}\right)$, $\left(M \geq 1\right)$. We have in mind that, for any $j \in \{1,\ldots,d\}$, $Y_j$, the variable of interest of institution $j$, depends on some covariates $\mathbf{X}$ and that for any $k \in \{1,\ldots,d\}$, $k\neq j$, the behaviour of the variable $Y_k$, related to either institution $k$ or the whole system, depends on covariates $\mathbf{X}$ as well as on the  behaviour of the variable of institution $j$, $Y_j$. Without loss of generality, thereafter, we fix $\tau\in\left(0,1\right)$ and suppose that we are interested in institutions $j$ and $k$ for $j\neq k$ and $\left(j,k\right) \in \{1,\ldots,d\}\times \{1,\ldots,d\}$.\newline
%
%
\indent Let us recall that the Value-at-Risk, $\text{VaR}_j^{\mathbf{x},\tau}$ of institution $j$ is the $\tau$-th level conditional quantile of the random variable $Y_j\mid\mathbf{X}=\mathbf{x}$, i.e.
\begin{eqnarray*}
\BBp\left(Y_j \leq \text{VaR}_j^{\mathbf{x},\tau}\mid\mathbf{X}=\mathbf{x}\right)=\tau. 
\label{eq:def_VaR} 
\end{eqnarray*}
The Conditional Value-at-Risk $\left(\text{CoVaR}_{k \vert j}^{\mathbf{x},\tau}\right)$ is the Value-at-Risk of institution $k$ conditional on  $Y_j=\text{VaR}_j^{\mathbf{x},\tau}$ at the level 
$\tau$, i.e., $\text{CoVaR}_{k \vert j}^{\mathbf{x},\tau}$ satisfies the following equation
\begin{eqnarray}
\BBp\left(Y_k \leq \text{CoVaR}_{k \vert j}^{\mathbf{x},\tau} \mid \mathbf{X}=\mathbf{x},Y_j=\text{VaR}_j^{\mathbf{x},\tau}\right)=\tau. 
\label{eq:def_CoVaR}
\end{eqnarray}
Note that the CoVaR corresponds to the $\tau$-th quantile of the conditional distribution of $Y_k\mid\{\mathbf{X}=\mathbf{x},Y_j=\text{VaR}_j^{\mathbf{x},\tau}\}$.
Assuming the linear representation (\ref{eq:linear_quantile}) of the quantiles of interest, we can write: 
%
%
\begin{eqnarray}
\text{VaR}_j^{\mathbf{x},\tau}&=&\theta_{j,0}^{\tau}+\theta_{j,1}^{\tau} x_1+\theta_{j,2}^{\tau}x_2+\ldots +\theta_{j,M}^{\tau} x_M 
\label{eq:linear-reg-quant}\\
\text{CoVaR}_{k \vert j}^{\mathbf{x},\tau}&=&\theta_{k,0}^{\tau} +\theta_{k,1}^{\tau} x_1+\theta_{k,2}^{\tau}x_2+\ldots +\theta_{k,M}^{\tau} x_M+\beta^{\tau}\text{VaR}_j^{\mathbf{x},\tau}
\label{eq:linear-reg-covar-var},
\end{eqnarray}
where $\theta_{l,m}^ {\tau}$ and $\beta$ are unknown parameters with $l \in \{j,l\}$ and $m=0,\ldots,M$. For the sake of simplicity we consider the same $\tau$ for both VaR and CoVaR and for the ease of exposition we drop the $\tau$ index from all parameters. 
%
\section{Time invariant quantile model}
\label{sec:bayesian_inference}
%
The use of Bayesian inference in quantile regression contest is now quite standard although quite recent. In what follows, we adopt the approach used in Yu and Moyeed \cite{yu_moyeed.2001} where data come from an Asymmetric Laplace Distribution 
which is a convenient tool to deal with quantile regression problems in a Bayesian framework. Suppose that we  observe  $\left(\mathbf{y},\mathbf{x}\right)=\left(\mathbf{y}_t,\mathbf{x}_t\right)_{t=1}^{T}=\left(y_{j,t},y_{k,t},\mathbf{x}_t\right)_{t=1}^{T}$, $T$ independent realizations of $\left(Y_j,Y_k,\mathbf{X}\right)$.
To estimate $\text{VaR}_j^{\mathbf{x},\tau}$ and $\text{CoVaR}_{k \vert j}^{\mathbf{x},\tau}$ we consider the following  equations:
\begin{eqnarray}
y_{j,t}&=&\mathbf{x}_t^{\trasp}\boldsymbol{\theta}_j+\epsilon_{j,t}
\label{eq:model_static_var}\\
y_{k,t} &=& \mathbf{x}_t^{\trasp}\boldsymbol{\theta}_k+\beta y_{j,t}+\epsilon_{k,t}, 
\label{eq:model_static_covar}
\end{eqnarray}
for $t=1,2,\dots,T$, where $\beta$, $\boldsymbol{\theta}_j$ and $\boldsymbol{\theta}_k$ are unknown parameters of dimension $1$, $\left(M+1\right)$ and $\left(M+1\right)$ respectively, and the first component of  $\mathbf{x}_t$  is equal to 1 to include a constant term in the regression function. Here, for any $t\in\left\{1,\ldots,T\right\}$, $\epsilon_{j,t} $ and $\epsilon_{k,t}$ are independent random variables distributed according to $\text{ALD}(\tau,0,\sigma_j)$ and $\text{ALD}(\tau,0,\sigma_k)$ respectively, with positive $\sigma_j$ and $\sigma_k$. Due to the property of Asymmetric Laplace distributions, the functions $\mathbf{x}^{\trasp}\boldsymbol{\theta}_j$ and $\mathbf{x}^{\trasp}\boldsymbol{\theta}_k+\beta y_j$ correspond to the $\tau$-th quantiles of $Y_j\mid\mathbf{X}=\mathbf{x}$ and $Y_k\mid\left\{\mathbf{X}=\mathbf{x},Y_j=y_j\right\}$, respectively.\newline
\indent For a fully Bayesian modelling, we need to specify the prior distribution on  the unknown parameters vector $\boldsymbol{\gamma}=\left(\boldsymbol{\theta},\beta,\sigma_j,\sigma_k\right)$. We assume the following priors independent on the value of $\tau$:
\begin{equation} 
\label{eq:prior_Gibbs}
\pi\left(\boldsymbol{\gamma}\right)=\pi\left(\boldsymbol{\theta}\right)\pi\left(\beta\right)\pi\left(\sigma_j\right)\pi\left(\sigma_k\right),
\end{equation}
with $\boldsymbol{\theta}=\left(\boldsymbol{\theta}_j,\boldsymbol{\theta}_k\right)^{\trasp}\sim{\cal N}_{\left(2M+2\right)}\left(\boldsymbol{\theta}^0,\Sigma^0\right)$, $\beta\sim {\cal N}\left(\beta^0, \sigma_{\beta}^{2}\right)$, $\sigma_j\sim \mathcal{IG}\left(a_j^0,b_j^0\right)$ and $\sigma_k\sim \mathcal{IG}\left(a_k^0,b_k^0\right)$, and where $\Sigma^0=\text{diag}\left(\Sigma^0_j,\Sigma^0_k\right)$, $\boldsymbol{\theta}^0=\left(\boldsymbol{\theta}^0_j,\boldsymbol{\theta}^0_k\right)^{\trasp}$, $\beta^0$, $\sigma_{\beta}^{2}>0$, $a_j^0>0$, $b_j^0>0$, $a_k^0>0$ and $b_k^0>0$ are given hyperparameters. Notations ${\cal N}$  and $\mathcal{IG}$ refer to a Gaussian and an Inverse Gamma distributions, respectively. Typically vague priors are imposed on $\sigma_j$ and $\sigma_k$ because they are regarded as nuisance parameters, see e.g. Yu and Moyeed \cite{yu_moyeed.2001} and Tokdar and Kadane \cite{tokdar_kadane.2012}.\newline
\indent As discussed in Yu and Moyeed \cite{yu_moyeed.2001}, due to the complexity of the likelihood function, the resulting posterior density for the regression parameters $\boldsymbol{\theta}$ and $\beta$ does not admit a closed form representation for the full conditional distributions, and need to be sampled by using  MCMC-based  algorithms. According to Kozumi and Kobayashi \cite{kozumi_kobayashi.2011}, we instead adopt the following well-known representation (see e.g. Kotz {\it et al.} \cite{kotz_etal.2001} and Park and Casella \cite{park_casella.2008}) of $\epsilon\sim\text{ALD}\left(\tau,0,\sigma\right)$ as a location-scale mixture of Gaussian distributions:
\begin{equation}
\label{mixture-repres}
\epsilon=\lambda \omega+\delta\sqrt{\sigma \omega}z,
\end{equation}
%
%
where $\omega\sim \mathcal{E}\text{xp}\left(\sigma^{-1}\right)$ and $z \sim \mathcal{N}(0,1)$ are independent random variables and $\mathcal{E}\text{xp}(\cdot)$ denotes the Exponential distribution. Moreover, the parameters $\lambda$ and $\delta^2$ are fixed equal to
\begin{equation}
\label{eq:lambda_delta_def}
\lambda=\frac{1-2\tau}{\tau\left(1-\tau\right)},\qquad\delta^2=\frac{2}{\tau\left(1-\tau\right)},
\end{equation}
in order to ensure that the $\tau-$th quantile of $\epsilon$ be equal to zero. The previous representation (\ref{mixture-repres}) allows us to use a Gibbs sampler algorithm detailed in the next subsection.  
%
Exploiting this augmented data structure, the model defined by equations (\ref{eq:model_static_var}) and (\ref{eq:model_static_covar}) admits, conditionally on $w$, the following Gaussian representation: 
\begin{eqnarray}
y_{j,t}&=&\mathbf{x}_t^{\trasp}\boldsymbol{\theta}_j+\lambda \omega_{j,t}+\delta\sqrt{\sigma_j \omega_{j,t}}z_{j,t}  
\label{eq:model_var_GaussMixApp}\\
y_{k,t} &=& \mathbf{x}_t^{\trasp}\boldsymbol{\theta}_k + \beta y_{j,t}+\lambda \omega_{k,t}+\delta\sqrt{\sigma_k \omega_{k,t}}z_{k,t},
\label{eq:model_covar_GaussMixApp}
\end{eqnarray}
for $t=1,2,\dots,T$, where $z_{j,t}$,  $z_{k,t}$ are independent and  $\omega_{j,t}$,  $\omega_{k,t}$ are independently drawn from $\mathcal{E}\text{xp}\left(\sigma_j^{-1}\right)$ and $\mathcal{E}\text{xp}\left(\sigma_k^{-1}\right)$, respectively. From equations (\ref{eq:model_var_GaussMixApp}) and (\ref{eq:model_covar_GaussMixApp}), the distribution of $\mathbf{Y}$ conditional on the parameters vector $\boldsymbol{\gamma}$, the observed exogenous variables $\mathbf{x}$ and the  augmented variables $\boldsymbol{\omega}=(\omega_{j,t}, \omega_{k,t})_{t=1}^T$, becomes
\begin{eqnarray}
f\left(\mathbf{y}\mid\boldsymbol{\omega},\mathbf{x},\boldsymbol{\gamma}\right)
=\prod_{t=1}^T\mathcal{N}\left(y_{j,t}\mid\omega_{j,t},\mathbf{x}_t,\boldsymbol{\theta}_j,\sigma_j\right)
\prod_{t=1}^T\mathcal{N}\left(y_{k,t}\mid\omega_{k,t},y_{j,t},\mathbf{x}_t,\beta,\boldsymbol{\theta}_k,\sigma_k\right).\nonumber
\end{eqnarray}
%
%
\subsection{Computations}
\label{subsec:computations}
%
Due to the Gaussian representation shown above we are able to implement a partially collapsed Gibbs sampler algorithm based on data augmentation (see Liu \cite{liu.1994} and van Dyk and Park \cite{van_dyk_park.2008}). The key idea of the complete collapsed Gibbs sampler is to avoid simulations from the full conditional distributions of the all model parameters $\left(\boldsymbol{\theta}_j,\boldsymbol{\theta}_k,\beta,\sigma_j,\sigma_k\right)$  by analytically marginalizing them out. This approach has several advantages with respect to a systematic sampling because it reduces the computational time and increases the convergence rate of the sampler. In our model this complete collapsed approach is not possible since the predictive distribution of the augmented variables $\left(\omega_{j,t},\omega_{k,t}\right)$ has not a closed form expression. Instead, it is possible to integrate out the variables $\left(\omega_{j,t},\omega_{k,t}\right)$ given the observations, from the full conditionals of the scale parameters $\left(\sigma_{j},\sigma_{k}\right)$. The partially collapsed Gibbs sampler we implement is an iterative simulation procedure from the following full conditional distributions:\newline
\begin{enumerate}
\item the full conditional distribution of the scale parameters $\sigma_j$ and $\sigma_k$ are sampled by integrating out the augmented latent factors $(\omega_{j,t},\omega_{k,t})_{t=1}^T$, becoming:
\begin{equation} 
\pi\left(\sigma_l\mid\mathbf{y}_l,\boldsymbol{\theta}_l\right)\propto\mathcal{IG}\left(\widetilde{a}_l,\widetilde{b}_l\right),\quad\mathbf{y}_l=\left(y_{l,t}\right)_{t=1}^\trasp,\forall l \in \{j,k\}\nonumber
\end{equation}
where
\begin{eqnarray}
\begin{array}{llll}
 & \widetilde{a}_j=a_j^0+T,&& \widetilde{b}_j=b_j^0+\sum_{t=1}^T\rho_{\tau}\left(y_{j,t}-\mathbf{x}_t^\trasp\boldsymbol{\theta}_j\right),\\
& \widetilde{a}_k^0=a_k^0+T,&& \widetilde{b}_k=b_k^0+\sum_{t=1}^T\rho_{\tau}\left(y_{k,t}-\mathbf{x}_t^\trasp\boldsymbol{\theta}_k-\beta y_{j,t}\right)
\end{array}
\end{eqnarray}
\item $\pi\left(\omega_{j,t}^{-1}\mid y_{j,t},\mathbf{x}_t,\boldsymbol{\theta}_j,\sigma_j\right)\propto\mathcal{IN}\left(\psi_{j,t},\phi_{j}\right)$, $\forall t=1,\ldots,T$, i.e., an Inverse Gaussian with parameters
\begin{equation*}
\psi_{j,t}=\sqrt{\frac{\lambda^2+2\delta^2}{\left(y_{j,t} -\mathbf{x}_t^{\trasp}\boldsymbol{\theta}_j\right)^2}},\qquad
\phi_{j}=\frac{\lambda^2+2\delta^2}{\delta^2\sigma_j}
\end{equation*}
\item
$\pi\left(\omega_{k,t}^{-1}\mid\mathbf{y}_t,\mathbf{x}_t ,\boldsymbol{\theta}_k,\beta,\sigma_k\right)\propto\mathcal{IN}\left(\psi_{k,t},\phi_{k}\right)$, $\forall t=1,\ldots,T$,
with parameters
\begin{equation*}
\psi_{k,t}=\sqrt{\frac{\lambda^2+2\delta^2}{\left(y_{k,t} -\mathbf{x}_t^{\trasp}\boldsymbol{\theta}_k-\beta y_{j,t}\right)^2}},\qquad
\phi_{k}=\frac{\lambda^2+2\delta^2}{\delta^2\sigma_k}
\end{equation*}
\item
$\pi\left(\boldsymbol{\theta}_j\mid\mathbf{y}_j,\mathbf{x},\boldsymbol{\omega}_j,\sigma_j\right)\propto\mathcal{N}_{M+1}\left(\widetilde{\boldsymbol{\theta}}_j,\widetilde{\Sigma}_j\right)$, where $\boldsymbol{\omega}_j=(\omega_{j,t})_{t=1}^T$, with 
$$\begin{array}{llll}
& \widetilde{\boldsymbol{\theta}}_j&=&\boldsymbol{\theta}_{j}^{0}+{\rm K}_j\left(\mathbf{y}_j-\mathbf{x}^{\trasp}
\boldsymbol{\theta}_j^0-\lambda\boldsymbol{\omega}_j\right)\nonumber\\
&\widetilde{\Sigma}_j&=&\left(\mathbb{I}_{M+1}-{\rm K}_j\mathbf{x}\right)\Sigma_j^0\nonumber\\
&{\rm K}_j&=&\Sigma_j^0\mathbf{x}^{\trasp}\left({\rm W}_j+\mathbf{x}\Sigma_j^0\mathbf{x}^\trasp\right)^{-1}\\
& {\rm W}_j& =& {\rm diag}\left(\left(\omega_{j,t} \times \delta^2 \times \sigma_j\right)_{t=1}^T \right)
\end{array}$$
$\mathbb{I}_{M+1}$ denotes the identity matrix of size  $(M+1)$.
\item
$\pi\left(\left(\boldsymbol{\theta}_k, \beta\right)^{\trasp}\mid\mathbf{y},\mathbf{x},\boldsymbol{\omega}_k,\sigma_k\right)\propto\mathcal{N}_{M+2}\left(\left(\widetilde{\boldsymbol{\theta}}_k,\widetilde{\beta}\right),\widetilde{\Sigma}_k\right)$, where $\boldsymbol{\omega}_k=\left(\omega_{k,t}\right)_{t=1}^T$
with
\begin{eqnarray*}
\left(\widetilde{\boldsymbol{\theta}}_k,\widetilde{\beta}\right)^{\trasp}&=&
\left(\boldsymbol{\theta}_{k}^{0},\beta^{0}\right)^{\trasp}+{\rm K}_k\left(\mathbf{y}_k-\left(\mathbf{x}, \mathbf{y}_j\right)^{\trasp} \left(\boldsymbol{\theta}_k^0, \beta^0\right)^{\trasp}-\lambda\boldsymbol{\omega}_k\right)\nonumber\\
\widetilde{\Sigma}_k&=&\left(\mathbb{I}_{M+2}-{\rm K}_k\left(\mathbf{x}, \mathbf{y}_j \right)\left( 
\begin{array}{cc}\Sigma_k^0 & 0 \\
0 & \sigma_{\beta}^{2} \end{array}\right)\right)\nonumber\\
{\rm K_k}&=&\left( 
\begin{array}{cc}\Sigma_k^0 & 0 \\
0 & \sigma_{\beta}^{2} \end{array}\right)(\mathbf{x}, \mathbf{y}_j )^{\trasp}\left({\rm W}_k+(\mathbf{x},\mathbf{y}_j) \left( \begin{array}{cc}\Sigma_k^0 & 0 \\
0 & \sigma_{\beta}^{2} \end{array}\right) (\mathbf{x},\mathbf{y}_j)^\trasp\right)^{-1}\\
{\rm W}_k&=&{\rm diag}\left(\left(\omega_{k,t} \times \delta^2 \times \sigma_k\right)_{t=1}^T\right).
\end{eqnarray*}
%
%
\end{enumerate}
To initialise the Gibbs sampling algorithm we simulate a random draw from the joint prior distribution of the parameters defined in equation (\ref{eq:prior_Gibbs}), and conditionally on that, we simulate the initial values of the augmented variables  $\left(\omega_{j,t},\omega_{k,t}\right)_{t=1}^T$ from their exponential distributions. Updating the parameters in this order ensures that the posterior distribution is the stationary distribution of the generated Markov chain. This is because combining steps 1 and 2 essentially produces draws from the conditional posterior distribution $\pi\left(\sigma_j,\sigma_k,\boldsymbol{\omega}_{j},\boldsymbol{\omega}_{k}\mid\boldsymbol{\theta}_k, \boldsymbol{\theta}_j,\beta,\mathbf{y},\mathbf{x}\right)$.
%
\subsection{VaR and CoVaR posterior estimation} 
\label{subsec:Var-CovaR-est}
%
From a Bayesian point of view once we retrieve simulations from the posterior distribution we can choose several ways to summarise them. Lin and Chang \cite{lin_chang.2012} use the maximisation of the posterior density to make inference for the quantile regression parameters, showing that this is equivalent to the minimisation problem (\ref{eq:quantile_freq}) in the frequentist context. Those considerations bring us to consider the Maximum a Posteriori (MaP) criteria as an estimate of all the posterior parameters in equations 
(\ref{eq:model_static_var}) and (\ref{eq:model_static_covar}), assuming ALD distributions for the error terms and diffuse priors on the regressor parameters. From all the MaP parameters involved in the marginal and conditional quantiles, say $\left(\boldsymbol{\theta}_j^{\rm MaP},\boldsymbol{\theta}_k^{\rm MaP},\beta^{\rm MaP}\right)$, the estimators of $\text{VaR}_j^{\mathbf{x},\tau}$ and the $\text{CoVaR}_{k \vert j}^{\mathbf{x},\tau}$ are then derived from equations (\ref{eq:linear-reg-quant}) and (\ref{eq:linear-reg-covar-var}) as follows: 
\begin{eqnarray*} 
\left(\text{VaR}_j^{\mathbf{x},\tau}\right)^{\rm MaP}&=& \mathbf{x}^{\trasp} \boldsymbol{\theta}_{j}^{\rm MaP} 
\label{est-bay-var}\\
\left(\text{CoVaR}_{k \vert j}^{\mathbf{x},\tau}\right)^{\rm MaP}&=&\mathbf{x}^{\trasp} \boldsymbol{\theta}_{k}^{\rm MaP}
+\beta^{\rm MaP}\left(\text{VaR}_j^{\mathbf{x},\tau}\right)^{\rm MaP}. 
\label{est-bay-covar}
\end{eqnarray*}
%
%
%
Credible sets at a given confidence level for both $\text{VaR}_{k \vert j}^{\mathbf{x},\tau}$ and $\text{CoVaR}_{k \vert j}^{\mathbf{x},\tau}$ estimates can be calculated by marginalising out the scale parameters $\left(\sigma_j,\sigma_k\right)$ and the latent variables $\left(\boldsymbol{\omega}_j,\boldsymbol{\omega}_k\right)$, using the sample draws of the MCMC algorithm. Monte Carlo estimates of the marginal posterior densities of the quantile functions are given by
\begin{eqnarray*} 
\pi\left(\text{VaR}_j^{\mathbf{x},\tau}\mid\mathbf{y}_j\right)&=&\frac{1}{G}\sum_{g=1}^G\pi\left(\mathbf{x}^{\trasp} \boldsymbol{\theta}_{j}\mid\sigma_j^{(g)},\boldsymbol{\omega}_j^{(g)},\mathbf{y}_j\right),\\
\label{conf-bay-var}
\pi\left(\text{CoVaR}_{k \vert j}^{\mathbf{x},\tau}\mid\mathbf{y}_k\right)&=& \frac{1}{G}\sum_{g=1}^G\pi\left(\mathbf{x}^{\trasp}\boldsymbol{\theta}_{k}
+\beta\left(\text{VaR}_j^{\mathbf{x},\tau}\right)^{(g)}\mid\sigma_k^{(g)},\boldsymbol{\omega}_k^{(g)},\mathbf{y}_k\right),
\label{conf-bay-covar}
\end{eqnarray*}
where $G$ denotes the number of post burn-in iterations. The $95\%$ High Posterior Credible intervals $\text{HPD}_{95\%}$ for the $\tau$-th quantile can be obtained from the samples $\left\{\mathbf{x}^{\trasp} \boldsymbol{\theta}_{j}\mid\sigma_j^{(g)},\boldsymbol{\omega}_j^{(g)},\mathbf{y}_j\right\}_{g=1}^G$ and $\left\{\mathbf{x}^{\trasp}\boldsymbol{\theta}_{k}
+\beta\left(\text{VaR}_j^{\mathbf{x},\tau}\right)^{(g)}\mid\sigma_k^{(g)},\boldsymbol{\omega}_k^{(g)},\mathbf{y}_k\right\}_{g=1}^G$. 
\section{Time-varying quantile model}
\label{sec:DynTVQuantile}
%
As mentioned before VaR and CoVaR are respectively unconditional and conditional quantiles, given current information, of future portfolio values. It is typically the case that returns change over time and for this reason  it can be interesting to build suitable model for time varying VaR and CoVaR. In particular, when modeling time varying quantiles, it is important to link future tail behaviours of time series to its  past movements to take into account for risk management arguments. Recently, time varying quantile topic has been received increased attention and different econometric models have been proposed: the most known are the Conditional Autoregressive Valute-at-Risk (CAViaR) model of Engle and Manganelli \cite{engle_manganelli.2004}, the Quantile Autoregressive (QAR) model of Koenker and Xiao \cite{koenker_xiao.2006}, and the Dynamic Additive Quantile (DAQ) model of Gourieroux and Jasak \cite{gourieroux_jasak.2008}. Most of them introduce an autoregressive structure in their modeling, which is intuitively attractive, as series of financial returns tend to exhibit time varying conditional moments, fat tails and volatility clustering. More recently, Gerlach \textit{et al.} \cite{gerlach_etal.2011} deal with the problem of estimating the conditional dynamic VaR using a Bayesian approach. The resulting conditional quantile for the variable of interest is directly modeled as a smooth function of the observed past returns.\newline
\indent In this paper we propose a different approach to introduce dynamics in the quantiles, modelling both the VaR and CoVaR as a function of latent variables having their own time dependence. The introduction of latent states having a dynamic evolution allow for the future behaviour of the modeled quantiles to depend upon their past movements in a flexible way.
%
%
In particular we estimate, from a Bayesian point of view, the required quantiles simultaneously and we allow the quantiles depending on exogenous variables. In so doing we have in mind a generalisation of  De Rossi and Harvey \cite{derossi_harvey.2009} and Kurose and Omori \cite{kurose_omori.2012} results who proposed to model the unconditional quantile curve using smoothing spline interpolation. More precisely, we model the observed vector at each point in time $\left(y_{j,t},y_{k,t}\right)$, as a function of independent latent processes $\left(\mu_{j,t},\mu_{k,t}\right)$ and the regressor terms in the following way: $\forall t\in {1,\ldots, T}$, 
\begin{eqnarray}
\label{eq:VaR-varying}
y_{j,t}&=&\mu_{j,t}+\mathbf{x}_{t}^{\trasp}\boldsymbol{\theta}_{j}+\epsilon_{j,t}\\ 
y_{k,t}&=&\mu_{k,t}+\mathbf{x}_{t}^{\trasp}\boldsymbol{\theta}_{k}+\beta_t y_{j,t}+\epsilon_{k,t},
\label{eq:CoVaR-varying} 
\end{eqnarray}
where $\epsilon_{j,t}\sim \text{ALD}\left(\tau, 0,\sigma_j\right)$, $\epsilon_{k,t}\sim\text{ALD}\left(\tau,0,\sigma_k\right)$ are independent random variables. The intercept terms $\mu_{l,t}$ with $l\in\left\{j,k\right\}$ are introduced to account for time dependence in the quantile functions. In fact, for $l\in\left\{j,k\right\}$, we propose the following smooth time-varying dynamics for  $\mu_{l,t}$:
\begin{eqnarray}
\label{eq:quantile_dynamic_1}
\mu_{l,t+1}&=&\mu_{l,t}+\mu_{l,t}^{*}+\eta_{l,t} \\
%
%
\mu^{*}_{l,t+1}&=&\mu_{l,t}^{*}+\eta^{*}_{l,t},
\label{eq:quantile_dynamic_2}
\end{eqnarray}
where $\left(\mu_{l,1},\mu^{*}_{l,1}\right)^\trasp\sim {\cal N}_{2}\left(0,\kappa\mathbb{I}_2\right)$ with $\kappa>0$ sufficiently large, $(\eta_{l,t},\eta^{*}_{l,t})^\trasp\sim {\cal N}_{2}(0,S_l)$ and $S_l=s_l^2V=s_{l}^{2}\left( \begin{array}{lr}  \frac 1 3 & \frac 1 2 \\   \frac 1 2 &  1  \end{array} \right)$, with $s_l>0$ allows a certain degree of smoothness of the quantile process.
%
%
Since one of our main focuses is to analyze the dynamic co-movement of two institutions, we also allow the parameter $\beta_t$ to change over time. To reflect different impacts between institutions we consider the following evolution for $\beta_t$:
\begin{eqnarray}
\beta_{t+1}&=&\beta_{t}+\beta_{t}^{*}+\eta_{\beta,t}  \\
\label{eq:quantile_dynamic_4}
\beta^{*}_{t+1}&=&\beta_{t}^{*}+\eta^{*}_{\beta, t}  
\label{eq:quantile_dynamic_3},
\end{eqnarray}
$\left(\mu_{\beta,1},\mu^{*}_{\beta,1}\right)^\trasp\sim {\cal N}_{2}\left(0,\kappa\mathbb{I}_2\right)$ and $(\eta_{\beta,t},\eta^{*}_{\beta,t})^\trasp\sim {\cal N}_{2}(0,s^2_{\beta}V)$ and $\kappa$ is defined as before. Throughout the paper we assume that $\forall l\in\left\{j,k,\beta\right\}$, $(\eta_{l,t},\eta^{*}_{l,t})$ is independent of $(\epsilon_{j,t},\epsilon_{k,t})$ (here we use $\beta$ as an index since there is no ambiguity).\newline
\indent In order to estimate the model parameters we rewrite equations (\ref{eq:VaR-varying})-(\ref{eq:quantile_dynamic_3}) using a state space representation so that $\forall t\in {1,\ldots, T}$, 
\begin{eqnarray}
\mathbf{y}_t&=&Z_t\boldsymbol{\xi}_t + \mathbf{x}_t^\trasp\boldsymbol{\theta}+\boldsymbol{\epsilon}_t
\label{eq:non-GaussianSSM_measurement}\\
\boldsymbol{\xi}_{t+1}&=&A\boldsymbol{\xi}_t+\boldsymbol{\eta}_t\\
\label{eq:non-GaussianSSM_transition}
\boldsymbol{\xi}_1&\sim&\mathcal{N}_6\left(0,\kappa\mathbb{I}_6\right),
\label{eq:non-GaussianSSM_first_state}
\end{eqnarray}
where
\begin{itemize}
%
\item $\boldsymbol{\epsilon}_t=\left(\epsilon_{j,t},\epsilon_{k,t}\right)$ is the vector of independent ALDs as defined in equations (\ref{eq:model_static_var})--(\ref{eq:model_static_covar}),
\item $Z_t=\left(\begin{array}{cccccc}1 & 0 & 0 & 0 & 0 & 0\\0 & 0 & 1 & 0 & y_{j,t} & 0\end{array}\right)$ is the time varying matrix of loading factors,
\item $\boldsymbol{\xi}_t=\left(\mu_{j,t},\mu_{j,t}^*,\mu_{k,t},\mu_{k,t}^*,\beta_{t},\beta_{t}^*\right)^\trasp$ is the vector of latent states whose dynamic is given by the transition matrix $A$, with $A=\mathbb{I}_3\otimes B$, $B=\left(\begin{array}{cc} 1 & 1\\ 0& 1\end{array}\right)$; and $\otimes$ denotes the Kronecker product, 
\item $\boldsymbol{\theta}=\left(\boldsymbol{\theta}_j,\boldsymbol{\theta}_k\right)$ is a $(M \times 2)$ matrix of time invariant coefficients,
\item $\boldsymbol{\eta}_t=\left(\eta_{j,t},\eta_{j,t}^*,\eta_{k,t},\eta_{k,t}^*,\eta_{\beta,t},\eta_{\beta,t}^*\right)^\trasp$ is the time-varying error vector distributed according to $\mathcal{N}_6\left(0,\Omega\right)$, where $\Omega=diag\left(s_j^2,s_k^2,s_\beta^2\right)\otimes V$. 
\end{itemize}
To complete the Bayesian model specification we choose as prior distribution for the non-varying parameters $\boldsymbol{\theta}$ the same as in  Section \ref{sec:bayesian_inference}; the parameters in the variance-covariance matrices of all latent processes are distributed according to $\mathcal{IG}\left(r_l^0,v^0_l\right)$ with positive  $r_l^0$ and $v_l^0$ $\forall l\in \{j,k,\beta\}$. In addition we assume that the vector of first states $\boldsymbol{\xi}_1$ is distributed according to equation (\ref{eq:non-GaussianSSM_first_state}), with $\kappa>0$ sufficiently large.\newline 
\indent The linear state space model introduced in (\ref{eq:non-GaussianSSM_measurement})-(\ref{eq:non-GaussianSSM_transition}) for modeling time-varying conditional quantiles is non-Gaussian because of the assumption made on the innovation terms. So in those circumstances 
optimal filtering techniques used to analytically marginalize out the latent states based on the Kalman filter recursions can not be applied (see Durbin and Koopman, \cite{durbin_koopman.2012}). Considering the (\ref{mixture-repres}) representation of the innovation terms in (\ref{eq:non-GaussianSSM_measurement}) it easy to recognise that the non-Gaussian state space model admits a conditionally Gaussian representation.
%
%
More specifically equations  (\ref{eq:non-GaussianSSM_measurement}) and (\ref{eq:non-GaussianSSM_transition}) become: $\forall t\in {1,\ldots, T}$,  
%
\begin{eqnarray}
\mathbf{y}_t&=& \mathbf{c}_t+Z_t\boldsymbol{\xi}_t +\mathbf{x}_t^\trasp\boldsymbol{\theta}+G_t\boldsymbol{\nu}_t,\quad \boldsymbol{\nu}_t\sim\mathcal{N}_2\left(0,\mathbb{I}_2\right)
\label{eq:ApproxGaussianSSM_measurement}\\
\boldsymbol{\xi}_{t+1}&=&A\boldsymbol{\xi}_t+\boldsymbol{\eta}_t,\quad\boldsymbol{\eta}_t\sim\mathcal{N}\left(0,\Omega\right)\\
\label{eq:ApproxGaussianSSM_transition}
\boldsymbol{\xi}_1&\sim&\mathcal{N}_6\left(0,\kappa\mathbb{I}_6\right),
\label{eq:ApproxGaussianSSM_first_state}
\end{eqnarray}
where the time-varying vector $\boldsymbol{c}_t$, and matrix $G_t$ are respectively
$\boldsymbol{c}_t=\left(\lambda\omega_{j,t}, \lambda\omega_{k,t}\right)^\trasp$
and $G_t=\left(\begin{array}{cc}\delta\sqrt{\sigma_{j}\omega_{j,t}}& 0\\0 &\delta\sqrt{\sigma_{k}\omega_{k,t}}\end{array}\right)$; $\omega_{j,t}$ and $\omega_{k,t}$ are independent with $\omega_{l,t}\sim\mathcal{E}\text{xp}\left(\sigma^{-1}_l\right)$ for $l\in\left(j,k\right)$ and $\sigma_l>0$; $\lambda$ and $\delta$ are defined in equation (\ref{eq:lambda_delta_def}).\newline\newline
\noindent The complete-data likelihood of the unobservable components $\left(\boldsymbol{\xi}_t\right)_{t=1}^T$ and  $\boldsymbol{\omega}=(\omega_{j,t}, \omega_{k,t})_{t=1}^T$ and all parameters $\boldsymbol{\gamma}=\left(\boldsymbol{\theta},s_j^2,s_k^2,s_\beta^2,\sigma_j,\sigma_k\right)$ can be factorized as follows:
%
\begin{eqnarray}
&&\mathcal{L}\left(\left(\boldsymbol{\xi}_t\right)_{t=1}^T,\boldsymbol{\omega},\boldsymbol{\gamma}\mid\mathbf{y},\mathbf{x}\right)\nonumber\\
&&\propto\prod_{t=1}^Tf\left(y_{j,t}\mid\boldsymbol{\xi}_t,\omega_{j,t},\sigma_j,\mathbf{x}_t\right)\prod_{t=1}^Tf\left(y_{k,t}\mid y_{j,t},\boldsymbol{\xi}_t,\omega_{k,t},\sigma_k,\mathbf{x}_t\right)\nonumber\\
&&\times\prod_{t=1}^Tf\left(\omega_{j,t}\mid\sigma_j\right)\prod_{t=1}^Tf\left(\omega_{k,t}\mid\sigma_k\right)f\left(\boldsymbol{\xi_1}\right)\prod_{t=1}^{T-1}f\left(\boldsymbol{\xi_{t+1}}\mid\boldsymbol{\xi_t},s_j^2,s_k^2,s_\beta^2\right)\nonumber\\
&&\times\left(\sigma_j\sigma_k\right)^{-\frac{T}{2}}\exp\left\{-\frac{1}{2}\sum_{t=1}^T\left(\mathbf{y}_t-\mathbf{c}_t-Z_t\boldsymbol{\xi}_t-\mathbf{x}_t^\trasp\boldsymbol{\theta}\right)^\trasp\left(G_tG_t^{\trasp}\right)^{-1}\left(\mathbf{y}_t-\mathbf{c}_t-Z_t\boldsymbol{\xi}_t-\mathbf{x}_t^\trasp\boldsymbol{\theta}\right)\right\}\nonumber\\
&&\times\prod_{t=1}^T\left(\omega_{j,t}\omega_{k,t}\right)^{-\frac{1}{2}} \nonumber\left(\sigma_j\right)^{-T}\exp\left\{-\frac{\sum_{t=1}^T\omega_{j,t}}{\sigma_j}\right\}\left(\sigma_k\right)^{-T}\exp\left\{-\frac{\sum_{t=1}^T\omega_{k,t}}{\sigma_k}\right\}\nonumber\\
&&\times\exp\left\{-\frac{1}{2\kappa}\boldsymbol{\xi}_1^\trasp\boldsymbol{\xi}_1\right\}
\exp\left\{-\frac{1}{2}\sum_{t=1}^{T-1}\left(\boldsymbol{\xi}_{t+1}-A\boldsymbol{\xi}_t\right)^\trasp\Omega^{-1}\left(\boldsymbol{\xi}_{t+1}-A\boldsymbol{\xi}_t\right)\right\}. 
\label{compl-lik}
\end{eqnarray}
%
\subsection{Computations} 
\label{subsec:PostInf}
%
Using the complete-data likelihood in (\ref{compl-lik}) and the prior distributions stated in sections \ref{sec:bayesian_inference} and \ref{sec:DynTVQuantile}
 we are able to write the joint posterior distribution of the parameters and the unobservable components. The form of the posterior allows us to sample from the complete conditional distributions and to use the Gibbs sampler algorithm, as shown below.\newline
\indent After choosing a set of initial values for the parameter vector $\boldsymbol{\gamma}^{(0)}$, simulations from the posterior distribution at the $i$-th iteration of  $\boldsymbol{\gamma}^{(i)}$, $\left\{\boldsymbol{\xi}_t,t=1,2,\ldots,T\right\}^{(i)}$ and $\left\{\omega_{l,t}, l\in{j,k},t=1,2,\dots,T\right\}^{(i)}$ for $i=1,2,\dots$, are obtained by the following Gibbs sampling scheme:
%
\begin{enumerate}
\item For $ l \in \{j,k,\beta\}$, generate $s_l^2$ from $\mathcal{IG}\left(\widetilde{r}_l,\widetilde{v}_l\right)$ with parameters
%
\begin{equation} \nonumber
\widetilde{r}_l=r_l^0+\left(T-1\right),\qquad\widetilde{v}_l=v_l^0+\sum_{t=1}^{T-1}\left(\boldsymbol{\xi}_{t+1}^{l}-B\boldsymbol{\xi}_t^{l}\right)^\trasp V^{-1}\left(\boldsymbol{\xi}_{t+1}^{l}-B\boldsymbol{\xi}_t^{l}\right)
\end{equation}
where $\boldsymbol{\xi}_t^l$ denotes the vector consisting of the elements $\left(\mu_{j,t},\mu^{*}_{j,t}\right)$ for $l=j$, $\left(\mu_{k,t},\mu^{*}_{k,t}\right)$ for $l=k$, and $\left(\beta_{t},\beta^{*}_{t}\right)$ for $l=\beta$.
%
\item For $l=j,k$, generate $\sigma_l$  from $\mathcal{IG}\left(\widetilde{a}_l,\widetilde{b}_l\right)$ with parameters
\begin{equation} \nonumber
\widetilde{a}_l=a_l^0+T,\qquad\widetilde{b}_l=b_l^0+\sum_{t=1}^{T}\rho_\tau\left(y_{l,t}-z_{l,t}\boldsymbol{\xi}_t\right),
\end{equation}
where $z_{l,t}$ denotes the first row of the matrix $Z_t$ for $l=j$ or the second one for $l=k$.
%
\item For $l=j,k$, denote $\mathbf{y}_l=(y_{l,t})_{t=1}^T$, $\mathbf{z}_l\boldsymbol{\mu}_l=(z_{l,t}\times\mu_{j,t})_{t=1}^T$, $\boldsymbol{\beta}\mathbf {y}_j=(\beta_t\times y_{j,t})_{t=1}^T$. Then,
generate $\boldsymbol{\theta}_l\sim\mathcal{N}_M\left(\widetilde{\boldsymbol{\theta}_l},\widetilde{{\Sigma}}_l\right)$  with parameters
\begin{eqnarray} 
\widetilde{\boldsymbol{\theta}}_j&=&\boldsymbol{\theta}_{j}^{0}+{\rm K}_j\left(\mathbf{y}_j-z_j\boldsymbol{\mu}_j-\mathbf{x}^{\trasp}\boldsymbol{\theta}_j^0-\lambda\boldsymbol{\omega}_j\right)\nonumber\\
\widetilde{{\Sigma}}_j&=&\left(\mathbb{I}_{M}-{\rm K}_j\mathbf{x}\right)\Sigma_j^0\nonumber\\
K_j&=&\Sigma_j^0\mathbf{x}^{\trasp}\left({\rm W}_j+\mathbf{x}\Sigma_j^0\mathbf{x}^\trasp\right)^{-1} \nonumber\\
{\rm W}_j&=&{\rm diag}\left(\left(\omega_{j,t} \times \delta^2 \times \sigma_j\right)_{t=1}^T \right) \nonumber
\end{eqnarray}
and
\begin{eqnarray}
\widetilde{\boldsymbol{\theta}}_k&=&\boldsymbol{\theta}_{k}^{0}+{\rm K}_k\left(\mathbf{y}_k-z_k\boldsymbol{\mu}_k-\boldsymbol{\beta}\mathbf{y}_j-\mathbf{x}^{\trasp}\boldsymbol{\theta}_k^0-\lambda\boldsymbol{\omega}_k\right)\nonumber\\
\widetilde{\Sigma}_k&=&\left(\mathbb{I}_{M}-K_k\mathbf{x}\right)\Sigma_k^0\nonumber\\
{\rm K}_k&=&\Sigma_k^0\mathbf{x}^{\trasp}\left({\rm W}_k+\mathbf{x} \Sigma_k^0\mathbf{x}^\trasp \right)^{-1} \nonumber\\
{\rm W}_k&=&{\rm diag}\left(\left(\omega_{k,t} \times \delta^2 \times \sigma_k\right)_{t=1}^T\right) \nonumber
\end{eqnarray}
where $\Sigma_j^0$ and $\Sigma_k^0$ are square matrices of dimension $M$.
%
\item For $l=j,k$ and for all $t \in \{1,2,\dots,T\}$, generate $\omega_{l,t}^{-1}\sim\mathcal{IN}\left(\psi_{l,t},\phi_{l}\right)$,with parameters
\begin{equation} \nonumber
\psi_{j,t}=\sqrt{\frac{\lambda^2+2\delta^2}{\left(y_{j,t}-z_{j,t}\mu_{j,t} - \mathbf{x}_t^{\trasp}\boldsymbol{\theta}_j\right)^2}},\qquad
\phi_{j}=\frac{\lambda^2+2\delta^2}{\delta^2\sigma_j}
\end{equation}
and
\begin{equation} \nonumber
\psi_{k,t}=\sqrt{\frac{\lambda^2+2\delta^2}{\left(y_{k,t}-\beta_ty_{j,t}-z_{k,t}\mu_{k,t} - \mathbf{x}_t^{\trasp}\boldsymbol{\theta}_k\right)^2}},\qquad
\phi_{k}=\frac{\lambda^2+2\delta^2}{\delta^2\sigma_k}.
\end{equation}
%
\item Since conditionally on the augmented latent states $\left(\omega_{j,t},\omega_{k,t}\right)_{t=1}^\trasp$, the state space model defined in equations (\ref{eq:ApproxGaussianSSM_measurement})--(\ref{eq:ApproxGaussianSSM_transition}), is linear and Gaussian, the latent dynamics can be marginalized out by running the Kalman filter-smoothing algorithm. We draw $\left(\boldsymbol{\xi}_t,\beta_t\right)_{t=1}^{T}$ jointly using the multi-move simulation smoother of Durbin and Koopman \cite{durbin_koopman.2002}. This entails running a Kalman filter forward with the state equation defined as in (\ref{eq:ApproxGaussianSSM_transition}).  As in Johannes and Polson \cite{johannes_polson.2009}, equation (\ref{eq:CoVaR-varying}) for $y_{k,t}$ is a measurement equation with time-varying coefficients, because $y_{j,t}$ is known and represents a time varying factor loading. Once the Kalman filter is run forward, we run the Kalman smoother backward in order to get the moments of joint full conditional distribution of the latent states (\ref{compl-lik}). Finally, we simulate a sample path by drawing from this joint distribution. For a similar simulation algorithm based on forward-filtering backward-smoothing see also Carter and Kohn \cite{carter_kohn.1994}, \cite{carter_kohn.1996} and Fruhwirth-Schnatter \cite{fruhwirth-schnatter.1994}.
\end{enumerate}

\subsection{Maximum a Posteriori}
\label{subset:dynamic_map}
As in the time-invariant case, once we retrieve simulations from the posterior distribution, in order to make posterior inference, we use the maximum a posteriori summarising criteria.  
In what follows we prove that using this criteria, the estimated quantiles have good sample properties according to De Rossi and Harvey \cite{derossi_harvey.2009} Proposition 3, i.e. a generalization of the ``fundamental property'',  where the sample quantile has the appropriate number of observations above and below.
\begin{proposition} 
\label{Prop} 
For the state space model defined in equations (\ref{eq:VaR-varying})-(\ref{eq:quantile_dynamic_3}) with prior distributions specified in Sections \ref{sec:bayesian_inference} and \ref{sec:DynTVQuantile}, $\kappa$ large enough and a diffuse prior on $\boldsymbol{\theta}$, the MaP quantile estimates $\mu_{j,t}^{map}+\mathbf{x}_{t}^{\trasp}\boldsymbol{\theta}_{j}^{\rm MaP}$ and 
$\mu_{k,t}^{\rm MaP}+\mathbf{x}_{t}^{\trasp}\boldsymbol{\theta}_{k}^{\rm MaP}+y_{j,t} \beta_t^{\rm MaP}$
%
%
satisfy 
\begin{eqnarray*}
\sum_{t \notin C }\left(x_{m,t}+1\right)\chi_\tau\left(y_{j,t}-\left(\mu_{j,t}^{\rm MaP} + \mathbf{x}_{t}^{\trasp}\boldsymbol{\theta}_{j}^{\rm MaP}\right)\right)&=&0
\label{sample-marg}\\
\sum_{t \notin C}\left(y_{j,t}+x_{m,t}+1\right)\chi_\tau\left(y_{k,t}-\left(\mu_{k,t}^{\rm MaP}+\mathbf{x}_{t}^{\trasp}\boldsymbol{\theta}_{k}^{\rm MaP}+y_{j,t} \beta_t^{map}\right)\right) &=&0,  
\label{sample-cond}
\end{eqnarray*}
$\forall m \in \{1,\ldots,M\}$, where $C \subset \{1,\ldots,T\}$ is the set of all points such that the Map quantile estimate coincides with observations and 
\begin{equation}
\chi_\tau: \; z  \rightarrow \left\{ \begin{array}{lll} \tau - 1 & \mbox{{\rm if}} & z<0  \\
\tau & \mbox{{\rm if}} & z>0.\end{array}\right.
\label{eq:chi_function}
\end{equation} 
\end{proposition}
\begin{proofdot}
%
For  $t \in \{1,\ldots,T\}$  recall that $\boldsymbol{\xi}^j_{t}=\left(\mu_{j,t},\mu^{*}_{j,t}\right)^{\trasp}$ and define 
$\boldsymbol{\xi}^{k,\beta}_t=\left(\mu_{k,t},\mu^{*}_{k,t},\beta_t,\beta_t^{*}\right)^{\trasp}$,  $\boldsymbol{\xi}^{j}=\left(\boldsymbol{\xi}^{j}_t\right)_{t=1}^{T}$  and $\boldsymbol{\xi}^{k,\beta}=\left(\boldsymbol{\xi}^{k,\beta}_t\right)_{t=1}^{T}$.
From equations (\ref{eq:VaR-varying})-(\ref{eq:quantile_dynamic_3}), let us write the complete-posterior distribution $p\left(\boldsymbol{\theta},\boldsymbol{\xi}^{j},\boldsymbol{\xi}^{k,\beta}, s_j^2,s_k^2,s_\beta^2,\sigma_j,\sigma_k \mid \left(\mathbf{y}_t\right)_{t=1}^{T}\right)$  as proportional to the product of two parts $Post_{marg}$ and $Post_{cond}$ where 
 \begin{eqnarray*}
Post_{marg}
&=&\prod^{T}_{t=1}{\rm ald}\left(y_{j,t}\mid\mu_{j,t}+\mathbf{x}_{t}^{\trasp}\boldsymbol{\theta}_{j},\sigma_j\right)s_j^{-1}
\exp\left\{-\frac{1}{2\kappa}\left(\boldsymbol{\xi}^{j}_{1}\right)^\trasp\boldsymbol{\xi}^{j}_1\right\}\nonumber\\
&&\times\exp\left\{-\frac{12}{2s_j^{2}}\sum_{t=1}^{T-1}\left(\boldsymbol{\xi}^{j}_{t+1}-B\boldsymbol{\xi}^{j}_t\right)^T  V^{-1}\left(\boldsymbol{\xi}^{j}_{t+1}-B\boldsymbol{\xi}^{j}_t\right)\right\} \nonumber \\
&&\times\exp(-\frac 1 2 (\boldsymbol{\theta_j} -\boldsymbol{\theta}_j^{0})^{\trasp} (\Sigma^{0}_j)^{-1} (\boldsymbol{\theta_j} -\boldsymbol{\theta}_j^{0})) 
\times \mathcal{IG} (r^0_j,v^0_j) \times \mathcal{IG} (a^0_j,b^0_j)  \label{post-aug-marg}\\
Post_{cond}&=&
\prod^{T}_{t=1}{\rm ald}\left(y_{k,t}\mid\mu_{k,t}+ \mathbf{x}_{t}^{\trasp} \boldsymbol{\theta}_{k}+ \beta_t y_{j,t},\sigma_k\right)  
\exp\left\{-\frac{1}{2\kappa}
\left(\boldsymbol{\xi}^{k,\beta}_1\right)^\trasp\boldsymbol{\xi}^{k,\beta}_1\right\}\left(s_k s_{\beta}\right)^{-1}\nonumber\\
&&\times\exp\left\{-\frac{12}{2}\sum_{t=1}^{T-1}\left(\Delta\boldsymbol{\xi}^{k,\beta}_{t+1}\right)^\trasp\left(\left[\begin{array}{cc}s_k^{2} & 0\\ 0 & s_\beta^{2}\end{array}\right] \otimes V\right)^{-1}\Delta\boldsymbol{\xi}^{k,\beta}_{t+1}\right\}\nonumber\\
&&\times\exp\left(-\frac 12 \left(\boldsymbol{\theta}_k -\boldsymbol{\theta}_k^{0}\right)^{\trasp} \left(\Sigma^{0}_k\right)^{-1}\left(\boldsymbol{\theta}_k-\boldsymbol{\theta}_k^{0}\right)\right)\nonumber\\ 
&&\times\mathcal{IG}\left(r^0_k,v^0_k\right)\times\mathcal{IG}\left(a^0_k,b^0_k\right)\times  \mathcal{IG}\left(r^0_{\beta},v^0_{\beta}\right),
\label{post-aug-cond}
\end{eqnarray*} 
%
where $\Delta\boldsymbol{\xi}^{k,\beta}_{t+1}=\boldsymbol{\xi}^{k,\beta}_{t+1}-\left(\mathbb{I}_2\otimes B\right)\boldsymbol{\xi}^{k,\beta}_{t}$. Note that the MaP of $p\left(\boldsymbol{\theta},\boldsymbol{\xi}^{j},\boldsymbol{\xi}^{k,\beta},s_j^2,s_k^2,s_\beta^2,\sigma_j,\sigma_k \mid\left(\mathbf{y}_t\right)_{t=1}^{T}\right)$ in $\left(\boldsymbol{\theta},\left(\mu_{j,t}\right)_{t=1}^{T},\left(\mu_{k,t}\right)_{t=1}^{T},\left(\beta_t\right)_{t=1}^{T}\right)$ is obtained by maximizing separately $Post_{marg}$ and $Post_{cond}$ with respect to $\left(\boldsymbol{\theta}_j,\left(\mu_{j,t}\right)_{t=1}^{T}\right)$ and $\left(\boldsymbol{\theta}_k,\left(\mu_{k,t}\right)_{t=1}^{T},(\beta_t)_{t=1}^{T}\right)$, respectively.  
%
%
Note also that the check function $\rho_\tau\left(\cdot\right)$ of the Asymmetric Laplace distribution is derivable everywhere except in zero and its derivative corresponds to the  function  $\chi_\tau\left(\cdot\right)$ defined in equation (\ref{eq:chi_function}).\newline\newline
\noindent Differentiating $\log\left(Post_{marg}\right)$ $\forall t=\left\{1,2,\dots,T\right\}\backslash \left\{C\right\}$, we obtain:
\begin{eqnarray}
\frac{\partial\log\left(Post_{marg}\right)}{\partial \mu_{j,1}}&=&
\frac{-\mu_{j,1}-\mu_{j,1}^{*}}{\kappa}+\frac{1}{\sigma_j}\chi_\tau\left(y_{j,1}-\mu_{j,1}-\mathbf{x}_{1}^{\trasp}\boldsymbol{\theta}_{j}\right)\nonumber\\ 
&&\qquad\qquad+\frac{6}{s^{2}_j}\left\{2\left(\mu_{j,2} -\mu_{j,1}-\mu_{j,1}^{*}\right)-\left(\mu_{j,2}^{*}-\mu_{j,1}^{*}\right)\right\}\nonumber\\
\frac{\partial \log\left(Post_{marg}\right)}{\partial \mu_{j,t} }&=&  
\frac{1}{\sigma_j}\chi_\tau\left(y_{j,t}-\mu_{j,t}-\mathbf{x}_{t}^{\trasp}\boldsymbol{\theta}_{j}\right)
-\frac{12}{s^{2}_j}\left(\mu_{j,t} -\mu_{j,t-1}-\mu_{j,t-1}^{*}\right)\nonumber\\
&&\qquad\qquad+\frac{12}{s^{2}_j}\left(\mu_{j,t+1}-\mu_{j,t}-\mu_{j,t}^{*}\right) 
-\frac{6}{s^{2}_j}\left(\mu_{j,t+1}^{*}-2\mu_{j,t}^{*}+\mu_{j,t-1}^{*}\right)\nonumber
\end{eqnarray}
$\forall t\in\{2,\ldots,T-1\}$, and
\begin{eqnarray}
\frac{\partial \log\left(Post_{marg}\right)}{\partial\mu_{j,T}}&=&
\frac{1}{\sigma_j}\chi_\tau\left(y_{j,T}-\mu_{j,T}-\mathbf{x}_{T}^{\trasp}\boldsymbol{\theta}_{j}\right)\nonumber\\
&&\qquad\qquad+\frac{6}{s^{2}_j} \left\{-2\left(\mu_{j,T} -\mu_{j,T-1}-\mu_{j,T-1}^{*}\right)+\left(\mu_{j,T}^{*}-\mu_{j,T-1}^{*}\right)\right\}\nonumber\\ 
\frac{\partial\log\left(Post_{marg}\right)}{\partial \boldsymbol{\theta}_j }&=&
\frac{1}{\sigma_j}\sum_{t=1}^{T}\mathbf{x}_{t}\chi_\tau\left(y_{j,t} - \mu_{j,t}-\mathbf{x}_{t}^{\trasp}\boldsymbol{\theta}_{j}\right)+\left(\Sigma^{0}_j\right)^{-1}\left(\boldsymbol{\theta}_j - \boldsymbol{\theta}_j^{0}\right), 
\label{first-comp-used}
\end{eqnarray}
where we use $S_j^{-1}=\frac{12}{s_j^{2}}V^{-1}=\frac{12}{s_j^{2}} \left(\begin{array}{cc} 1 & -1/2 \\ -1/2 & 1/3 \end{array}\right)$.
It turns out that 
\begin{eqnarray}
\sum_{t }\frac{\partial \log\left(Post_{marg}\right)}{\partial\mu_{j,t}}=\frac{-\mu_{j,1}-\mu_{j,1}^{*}}{\kappa}+ \frac{1}{\sigma_j}\sum_{t}\chi_\tau\left(y_{j,t}-\mu_{j,t}-\mathbf{x}_{t}^{\trasp}\boldsymbol{\theta}_{j}\right),\nonumber
\end{eqnarray}
which combined with equation (\ref{first-comp-used}) and choosing $\kappa$ sufficiently large  and a diffuse prior on $\boldsymbol{\theta}_j$ implies that the maximizer of $Post_{marg}$ satisfies the following equation
\begin{eqnarray*}
 \sum_{t \notin C}\left(x_{m,t}+1\right)\chi_\tau (y_{j,t}- \mu_{j,t}-\mathbf{x}_{t}^{\trasp}\boldsymbol{\theta}_{j})=0,\quad\forall m\in\left\{1,2,\dots,M\right\}.
\end{eqnarray*}
%
%
\noindent The derivatives of $\log\left(Post_{cond}\right)$ with respect to $\left(\boldsymbol{\theta}_k,\left(\mu_{k,t}\right)_{t=1}^{T}\right)$ can be obtained as the ones for $\log\left(Post_{marg}\right)$. Deriving $\log\left(Post_{cond}\right)$ with respect to $\left(\beta_t\right)_{t=1}^{T}$, $\forall t=\left\{1,2,\dots,T\right\}\backslash \left\{C\right\}$, leads to: 
%
%
\begin{eqnarray}
\frac{\partial\log\left(Post_{cond}\right)}{\partial \beta_{1}}&=&  
\frac{-\beta_{1}-\beta_{1}^{*}}{\kappa}+\frac{y_{j,1}}{\sigma_k}\chi_\tau\left(y_{k,1}-\mu_{k,1}-\mathbf{x}_{1}^{\trasp}\boldsymbol{\theta}_{k}-y_{j,1} \beta_1\right)\nonumber \\ 
&&\qquad\qquad+\frac{6}{s^{2}_\beta}\left\{2\left(\beta_{2}-\beta_{1}-\beta_{1}^{*}\right)-
\left(\beta_{2}^{*}-\beta_{1}^{*}\right)\right\} \nonumber \\
\frac{\partial\log\left(Post_{cond}\right)}{\partial \beta_{t}}&=&  
\frac{y_{j,t}}{\sigma_k}\chi_\tau\left(y_{k,t}-\mu_{k,t}-\mathbf{x}_{t}^{\trasp}\boldsymbol{\theta}_{k}-y_{j,t}\beta_t\right)-
\frac{12}{s^{2}_\beta}\left(\beta_{t}-\beta_{t-1}-\beta_{t-1}^{*}\right)\nonumber\\ 
&&\qquad\qquad+\frac{6}{s^{2}_\beta}\left\{2\left(\beta_{t+1}-\beta_{t}-\beta_{t}^{*}\right)-\left(\beta_{t+1}^{*}-2\beta_{t}^{*}+\beta_{t-1}^{*}\right)\right\},\nonumber
\end{eqnarray}
$\forall t \in \{2,\ldots,T-1\}$, and
\begin{eqnarray}
\frac{\partial\log\left(Post_{cond}\right)}{\partial \beta_{T}}&=&  
\frac{y_{j,T}}{\sigma_k}\chi_\tau\left(y_{k,T}-\mu_{k,T}-\mathbf{x}_{T}^{\trasp}\boldsymbol{\theta}_{k}-y_{j,T}\beta_t\right)\nonumber\\
&&\qquad\qquad+\frac{6}{s^{2}_\beta}\left\{-2\left(\beta_{T}-\beta_{T-1}-\beta_{T-1}^{*}\right)+\left(\beta_{T}^{*}-\beta_{T-1}^{*}\right)\right\}.\nonumber
%
%
\end{eqnarray}
It turns out that 
\begin{eqnarray}
\sum_{t }\frac{\partial \log\left(Post_{cond}\right)}{\partial \beta_{t}}=  
\frac{-\beta_{1}-\beta_{1}^{*}}{\kappa}+ 
\frac{1}{\sigma_k}\sum_{t  }y_{j,t}\chi_\tau\left(y_{k,t}-\mu_{k,t}-\mathbf{x}_{t}^{\trasp}\boldsymbol{\theta}_{k}-y_{j,t}\beta_t\right).\nonumber
\end{eqnarray}
Choosing  a sufficiently large $\kappa$ and a diffuse prior on $\boldsymbol{\theta}_k$ implies 
that  the maximizer of $Post_{cond}$ satisfies the following equation
\begin{eqnarray*}
\sum_{t \notin C}\left(y_{j,t} +x_{m,t}+1\right)\chi_\tau\left(y_{k,t}-\mu_{k,t}-\mathbf{x}_{t}^{\trasp}\boldsymbol{\theta}_{k}-y_{j,t}\beta_t\right)=0,
\label{MaP-CoVar}
\end{eqnarray*}
$\forall m\in\left\{1,2,\dots,M\right\}$, which concludes the proof of Proposition \ref{Prop}.
\end{proofdot}
\begin{corollary} 
\label{Coro}
The MaP  quantiles estimates  
$\left(q_t^{\tau}\left(\mathbf{x}_t\right)\right)^{\rm MaP}=\mu_{j,t}^{\rm MaP}+\mathbf{x}_{t}^{\trasp}\boldsymbol{\theta}_{j}^{\rm MaP}$ 
and $\left(q_t^{\tau}\left(\mathbf{x}_t, y_{j,t}\right)\right)^{\rm MaP}=\mu_{k,t}^{\rm MaP}+\mathbf{x}_{t}^{\trasp}\boldsymbol{\theta}_{k}^{\rm MaP}  +y_{j,t} \beta_t^{\rm MaP}$ 
satisfies a generalization of the fundamental property of sample time-varying  quantiles, that is: 
\begin{eqnarray}
\displaystyle{ \sum_{t \in A} h_t \leq\left(1-\tau\right) \sum_{t =1}^Th_t  -  \sum_{t \in C^-}h_t} & \mbox{{\rm and }} & 
\displaystyle{\sum_{t \in B}h_t\leq \tau  \sum_{t =1}^Th_t  -  \sum_{t \in C^-}h_t }, 
\label{Fundamental-prop}
\end{eqnarray}
where 
\begin{equation}
h_t=\left\{ \begin{array}{ll} x_{m,t}+1 & \mbox{{\rm for the VaR}} \\ 
y_{j,t}+x_{m,t}+1 & \mbox{{\rm for the CoVaR}},\end{array}\right.
\end{equation}
and $A\cup B \cup C=\{1,\ldots,T\}$. Here, $A$ and $B$ denote the set of indices such that observations are respectively (strictly) above and (strictly) below the MaP quantile estimates, and $C=C^+ \cup C^-= \{t \in C: h_t \geq 0\}\cup \{t \in C: h_t < 0\}$ is the set of indices such that the observations coincide with the quantile estimates.
%
%
\end{corollary}
\begin{proofdot}
%
Following the proof of Proposition 3 in De Rossi and Harvey \cite{derossi_harvey.2009}, using Proposition \ref{Prop} and the following inequalities
\begin{equation} 
-\tau \sum_{t \in C^+}h_t+\left(1-\tau\right) \sum_{t \in C^-} h_t< \sum_{t \notin C} h_t\chi_{\tau} \left(y_{j,t}-\left(q_t^{\tau}\left(\mathbf{x}_t\right)\right)^{\rm MaP}\right) 
< \left(1-\tau\right) \sum_{t \in C^+} h_t-\tau \sum_{t \in C^-} h_t,
\label{eq:ghis}
\end{equation}
where $h_t=x_{m,t}+1$, the result in (\ref{Coro}) is obtained by rewriting the central term in (\ref{eq:ghis}) as follows:
%
\begin{equation}
\sum_{t \notin C} h_t\chi_{\tau}\left(y_{j,t}-\left(q_t^{\tau}\left(\mathbf{x}_t\right)\right)^{\rm MaP}\right)=
\tau \sum_{t \in A} h_t+\left(\tau-1\right) \sum_{t \in B} h_t.
\end{equation}
\noindent The same occurs for the CoVaR. 
\end{proofdot}
\begin{remark}
If $h_t >0 \quad \forall t=1,2,\dots,T$, or if the distribution of $\left(Y_{j,t},Y_{k,t}\right)$ is continuous, then inequalities (\ref{Fundamental-prop}) coincide 
with the ones stated Proposition 3 of De Rossi and Harvey \cite{derossi_harvey.2009}. 
When $h_t \equiv h \quad \forall t $, then (\ref{Fundamental-prop}) corresponds to the fundamental property of sample time-varying quantiles.
\end{remark} 
%
\section{Empirical application}
\label{sec:empirical_application}
Throughout this section we illustrate the methodology previously discussed to real data. In particular we analyse separately the time-invariant specification of CoVaR proposed in Section \ref{sec:bayesian_inference} and the time-varying version considered  in Section \ref{sec:DynTVQuantile}.  The idea is to study the tail co-movements between an individual institution $j$ and the whole system $k$ it belongs to. The financial data we rely on refer to the Standard and Poor's Composite Index ($k$) for the U.S market where different sectors ($j$) are included. For those institutions and for the whole system we consider both micro and macro variables in order to take into account for individual information and for global economic conditions respectively. The analysis is based on weekly observations; whereas the data are not observable at the same frequency we build a smoothing state space model to fill the missing values.
%
%
The focus of the empirical application is to show how CoVaR provides interesting insights about the tail risk interdependence. Moreover, we show the relevance of introducing dynamics in the extreme quantiles in order to effectively capture the contribution of individual institutions to the systemic risk evolution. Approaching VaR and CoVaR estimation in a Bayesian framework allows us to calculate their credible sets which are necessary to assess estimates' accuracy.
%
%
\subsection{The data}
\label{subsec:data}
%
%
The empirical analysis is based on the publicly traded U.S. companies listed in Table \ref{tab:SP500_company_list} belonging to different sectors of the Standard and Poor's Composite Index (S\&P500). The sectors considered are: financials, consumer goods, energy, industrials, technologies and utilities. Financials consist of banks, diversified financial services and consumer financial services. Consumer Goods consist of the food and beverage industry, primary food industry and producers of personal and household goods. The energy sector consists of companies producing or supplying energy and it includes companies involved in the exploration and development of oil or gas reserves, oil and gas drilling, or integrated power firms. Industrials consist of industries such as construction and heavy equipment, as well as industrial goods and services that include containers, packing and industrial transport, while technologies are related to the research, development and/or distribution of technologically based goods and services. Utilities consist of the provision of gas and electricity.\newline
\indent Daily equity price data are converted to weekly log-returns (in percentage points) for the sample period between January 2, 2004 to December 28, 2012, covering the recent global financial crisis.
Table \ref{tab:SP500_data_summary_stat} provides summary statistics for the weekly returns. Except for few companies, the mean return during the estimation period is positive. Consumer goods and Energy have the highest average return, while Banks and Financial services have the lowest one. Focusing on the sample correlation with the market index return, the correlation involving the financials and industrials is the largest on average. The correlation with the market index return varies substantially across sectors, ranging from 0.553 (PEG) to 0.772 (AXP). Bellwether sectors like consumer and utilities show a surprisingly high correlation level.
%
%
A possible explanation for this empirical evidence is that the correlation among financial stock increases dramatically during times of turbulence and in particular in late 2008 as the global financial
crisis intensified. Finally, the last column of Table \ref{tab:SP500_data_summary_stat} provides the sample 1\% stress level of each institution's return evaluated over the entire time period. By comparing these values with the number of standard deviations away from their mean, we can see that asset return distributions do not appear highly skewed. The characteristic of the data summarised in Table \ref{tab:SP500_data_summary_stat} induce to study the risk interdependence through the CoVaR tool.\newline
\indent To control for the general economic conditions we use observations of the following macroeconomic regressors as suggested by Adrian and Brunnermeier \cite{adrian_brunnermeier.2011} and Chao, H\"{a}rdle and Wang \cite{chao_etal.2012}:
\begin{enumerate}
\item[(i)] the VIX index (VIX), measuring the model-free implied stock market volatility as evaluated by the Chicago Board Options Exchange (CBOE).
\item[(ii)] a short term liquidity spread (LIQSPR), computed as the difference between the 3-months collateral repo rate and the 3-months Treasury Bill rate. 
\item[(iii)] the weekly change in the three-month Treasury Bill rate (3MTB).
\item[(iv)] the change in the slope of the yield curve (TERMSPR), measured by the difference of the 10-Years Treasury rate and the 3-months Treasury Bill rate.
\item[(v)] the change in the credit spread (CREDSPR) between 10-Years BAA rated bonds and the 10-Years Treasury rate. 
\item[(vi)] the weekly return of the Dow Jones US Real Estate Index (DJUSRE). 
\end{enumerate}
Historical data for the volatility index (VIX) can be downloaded from the Chicago Board Options Exchange's website, while the remaining variables are from the Federal Reserve Board H.15 database. Data are available on a daily frequency and subsequently converted to a weekly frequency.\newline
\indent To capture the individual firms' characteristics, we include observations from the following microeconomic regressors:
\begin{enumerate}
\item[(i)] leverage (LEV), calculated as the value of total assets divided by total equity (both measured in book values).
\item[(ii)] the market to book value (MK2BK), defined as the ratio of the market value to the book value of total equity.
\item[(iii)] the size (SIZE), defined by the logarithmic transformation of the market value of total assets.
\item[(iv)] the maturity mismatch (MM), calculated as short term debt net of cash divided by the total liabilities.
\end{enumerate}
%
%
\begin{table}[!t]
\captionsetup{font={small}, labelfont=sc}
%
\begin{small}
\centering
 \smallskip
  \begin{tabular}{lcc}\\
  \hline\hline
   Name & Ticker Symbol & Sector\\
    \hline
CITIGROUP INC.					&	C 			&	Financial		\\
BANK OF AMERICA CORP.			&	BAC  		&	Financial		\\
COMERICA INC.					&	CMA  		&	Financial		\\
JPMORGAN CHASE \& CO.			&	JPM  		&	Financial		\\
KEYCORP							&	KEY   		&	Financial		\\
GOLDMAN SACHS GROUP INC.		&	GS  		&	Financial		\\
MORGAN STANLEY					&	MS  		&	Financial		\\
MOODY'S CORP.					&	MCO  		&	Financial   	\\
AMERICAN EXPRESS CO.			&	AXP  		&	Financial		\\
MCDONALD'S CORP.				&	MCD  		&	Consumer		\\
NIKE INC. 						&	NKE  		&	Consumer		\\
CHEVRON CORP.					&	CVX  		&	Energy			\\
EXXON MOBIL CORP.				&	XOM  		&	Energy			\\
BOEING CO.						&	BA  		&	Industrial		\\
GENERAL ELECTRIC CO.			&	GE  		&	Industrial		\\
INTEL CORP.						&	INTC  		&	Technology		\\
ORACLE CORP.					&	ORCL  		&	Technology		\\
AMEREN CORPORATION.				&	AEE  		&	Utilities		\\
PUBLIC SERVICE ENTERPRISE INC.	&	PEG  		&	Utilities		\\
      \hline\hline
\end{tabular}
\caption{List of included companies. All the listed companies belong to the Standard and Poor's Composite Index (S\&P500) at the start of the trading day of February 15, 2013. The last column reports the sector each company belongs to.}
\label{tab:SP500_company_list}
\end{small}
\smallskip
\end{table}
%
%
Microeconomic variables are downloaded from the Bloomberg database and are available only on a quarterly basis. Since our analysis builds on weekly frequencies we choose to impute missing observations by smoothing spline interpolation. Details on the procedure are given in Appendix \ref{sec:appendix_A}. 
%
\begin{figure}[!ht]
\begin{center}
\captionsetup{font={small}, labelfont=sc}
\includegraphics[width=1.0\linewidth]{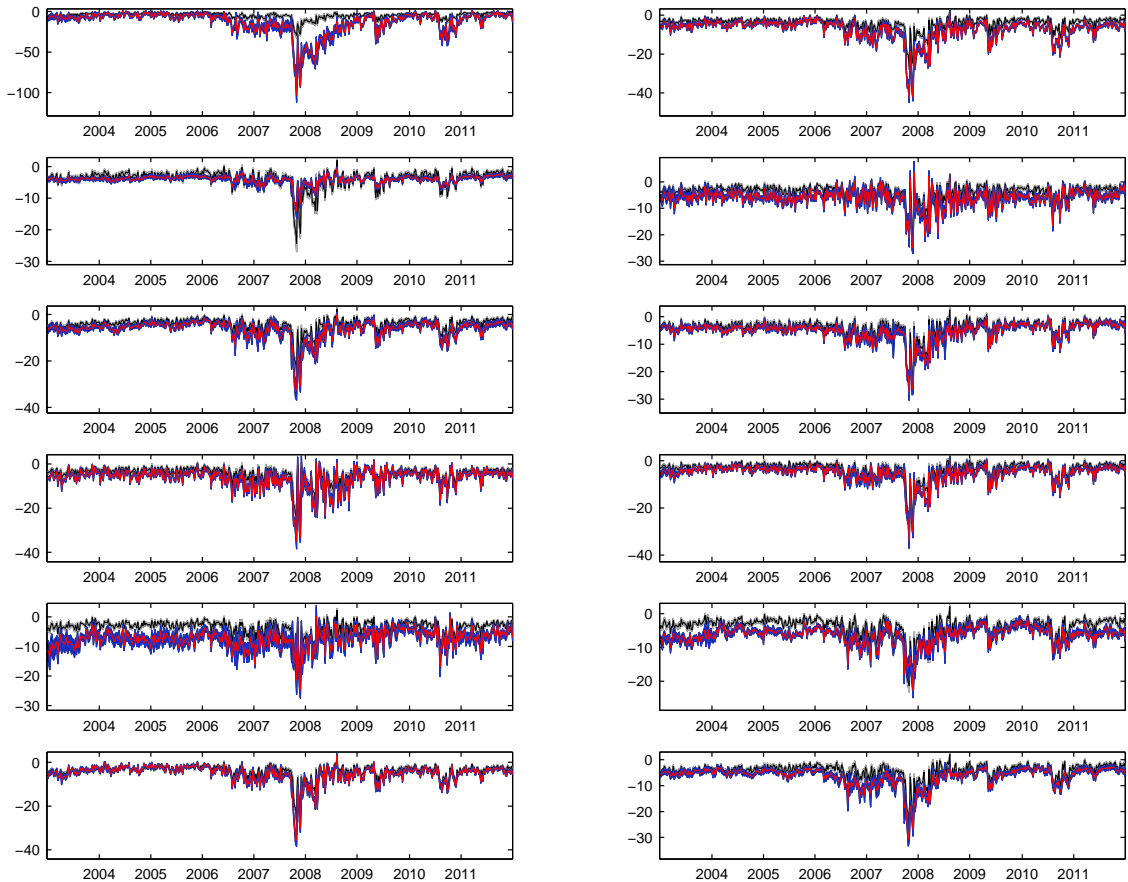}
\caption{\small{Time series plot of the $\text{VaR}^{\mathbf{x},\tau}_{j}$ (red line) and $\text{CoVaR}^{\mathbf{x},\tau}_{k\vert j}$ (gray line) at the confidence level $\tau= 0.025$ along with $\text{HPD}_{95\%}$ credible sets, obtained by fitting the model defined in equations (\ref{eq:model_static_var})--(\ref{eq:model_static_covar}) for the following assets: first panel (financial): C (left), GS (right); second panel (consumer): MCD (left) and NKE (right); third panel (energy): CVX (left), XOM (right); fourth panel (industrial): BA (left), GE (right); fifth panel (technology): INTC (left), ORCL (right); last panel (utilities): AEE (left), PEG (right).}}
\label{fig:Model_Static_Var_CoVaR}
\end{center}
\end{figure}
%
%
%
\begin{figure}[!ht]
\begin{center}
\captionsetup{font={small}, labelfont=sc}
\includegraphics[width=1.0\linewidth]{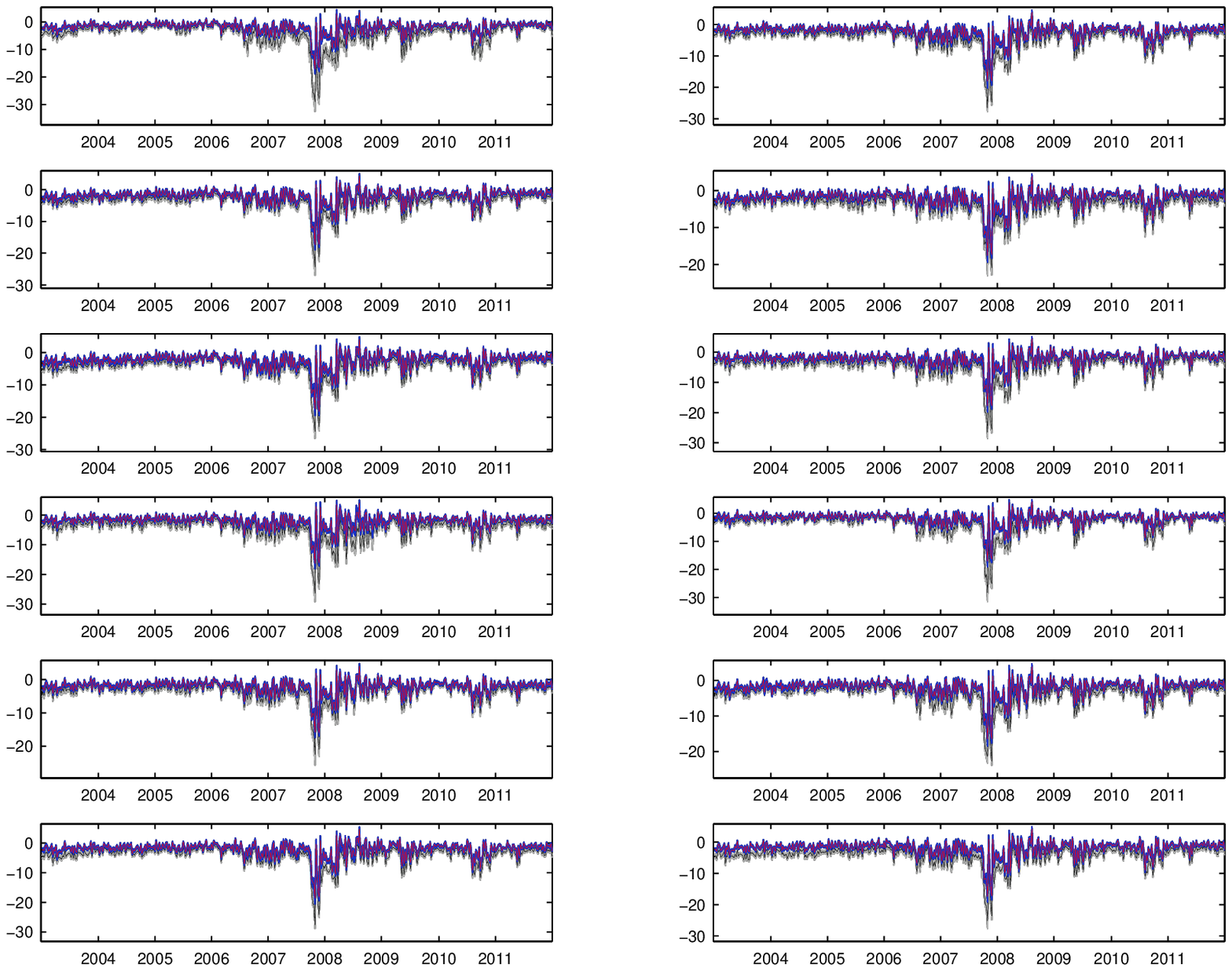}
\caption{\small{Time series plot of the $\text{CoVaR}^{\mathbf{x},\tau}_{k\vert j}$ at the confidence levels $\tau= 0.025$ (gray line) and $\tau= 0.1$ (red line) along with $\text{HPD}_{95\%}$ credible sets, obtained by fitting the model defined in equations (\ref{eq:model_static_var})--(\ref{eq:model_static_covar}) for the following assets: first panel (financial): C (left), GS (right); second panel (consumer): MCD (left) and NKE (right); third panel (energy): CVX (left), XOM (right); forth panel (industrial): BA (left), GE (right); fifth panel (technology): INTC (left), ORCL (right); last panel (utilities): AEE (left), PEG (right).}}
\label{fig:Model_Static_Double_CoVaR}
\end{center}
\end{figure}
%

%
\begin{figure}[!t]
\begin{center}
\captionsetup{font={small}, labelfont=sc}
\includegraphics[width=1.0\linewidth]{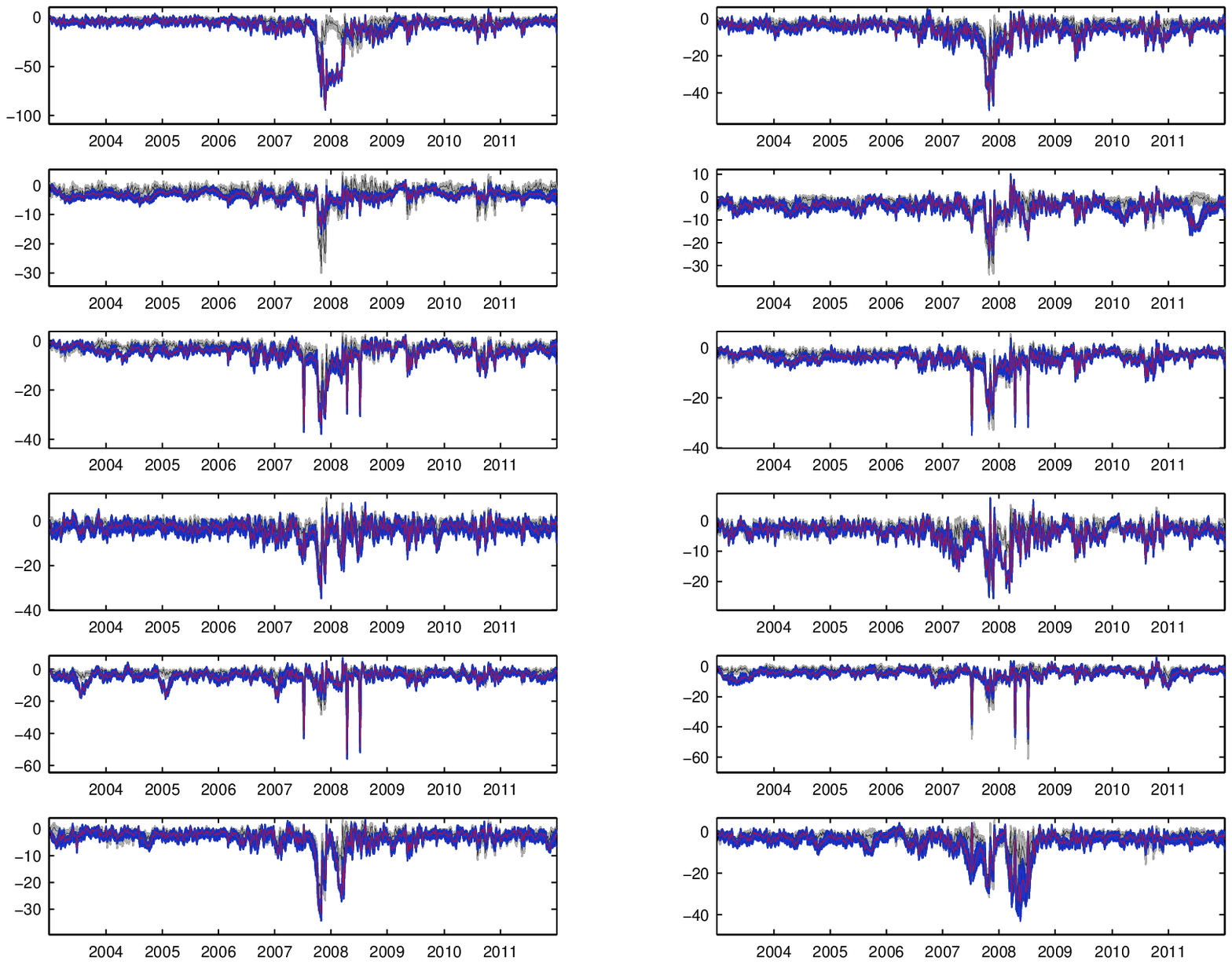}
\caption{\small{Time series plot of the $\text{VaR}^{\mathbf{x},\tau}_{j}$ (red line) and $\text{CoVaR}^{\mathbf{x},\tau}_{k\vert j}$ (gray line) at the confidence level $\tau=0.025$ along with $\text{HPD}_{95\%}$ credible sets, obtained by fitting the model defined in equations (\ref{eq:VaR-varying})--(\ref{eq:quantile_dynamic_3}) for the following assets: first panel (financial): C (left), GS (right); second panel (consumer): MCD (left) and NKE (right); third panel (energy): CVX (left), XOM (right); forth panel (industrial): BA (left), GE (right); fifth panel (technology): INTC (left), ORCL (right); last panel (utilities): AEE (left), PEG (right).}}
\label{fig:Model_Dynamic_Var_CoVaR}
\end{center}
\end{figure}
%
\subsection{Time-invariant risk beta}
\label{subsec:time_invariant_betas}
%
%
\begin{figure}[!t]
\begin{center}
\captionsetup{font={small}, labelfont=sc}
\includegraphics[width=1.0\linewidth]{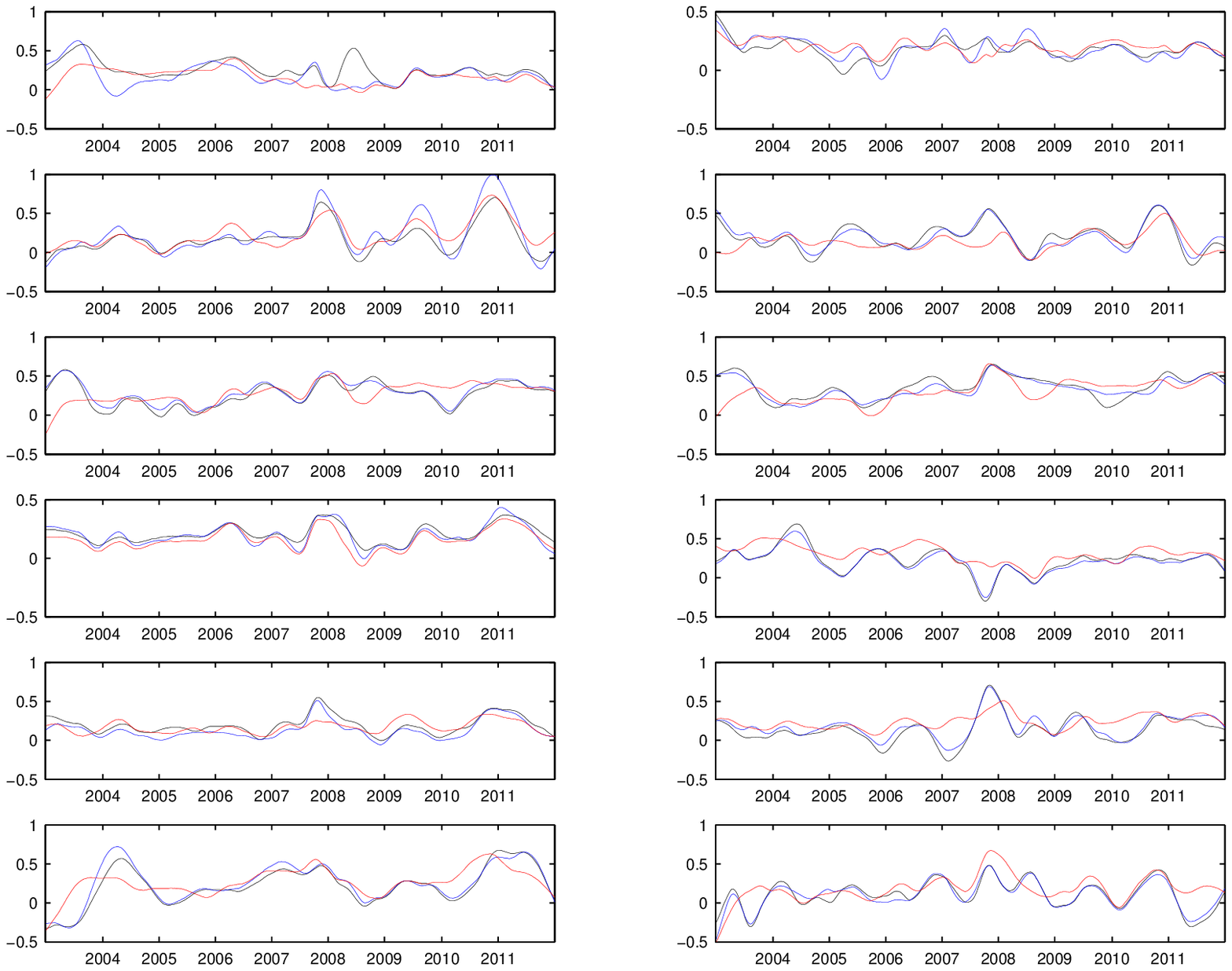}
\caption{\small{Time series plot of the dynamic $\beta$ at three confidence level $\tau=0.025$, (dark line), $\tau=0.05$, (blue line) and $\tau=0.5$ (red line), obtained by fitting the model defined in equations (\ref{eq:VaR-varying})--(\ref{eq:quantile_dynamic_3}) for the following assets, (denoted by $j$): first panel (financial): C (left), GS (right); second panel (consumer): MCD (left) and NKE (right); third panel (energy): CVX (left), XOM (right); forth panel (industrial): BA (left), GE (right); fifth panel (technology): INTC (left), ORCL (right); last panel (utilities): AEE (left), PEG (right).}}
\label{fig:Model_Dynamic_BETA_ALL}
\end{center}
\end{figure}
%
In what follows we provide the Bayesian empirical analysis for the time-invariant CoVaR model stated in Section \ref{sec:bayesian_inference}. In order to implement the inference we specify the hyper-parameters values for each prior distribution defined therein. We set the first moment of the exogenous regressor's parameters $\boldsymbol{\theta}$ equal to zero and let their variance-covariance matrices to be diagonal, i.e. $\Sigma_l^0=100\times\mathbb{I}_{11}$, for $l=j,k$. The $\beta$ parameter is assumed to have a Gaussian prior centered on $\beta^0=0$ with large variance, $\sigma_\beta^2=100$. Vague prior for the nuisance parameters $\sigma_l$, for $l=j,k$ are imposed by setting $a^0_l=b^0_l=0.0001$ which correspond to an Inverse Gamma distribution with infinite variance.\newline
%
\indent The MCMC algorithm illustrated in Section \ref{sec:bayesian_inference} run for 200,000 times with a burn-in phase of 100,000 iterations. 
%
%
%
For all the considered institutions, Tables \ref{tab:Model_Static_ParEst_0025}--\ref{tab:Model_Static_ParEst_005} report the estimated systemic risk $\beta$ and the exogenous parameters as well as the $\text{HPD}_{95\%}$, for $\tau=\left(0.025,0.05\right)$ credible sets. To check the MCMC convergence we also calculate the Geweke's convergence diagnostics (see Geweke \cite{geweke_1992}, \cite{geweke_2005}) which are not reported to save space but confirm the convergence of the chain.\newline 
\indent For all the reported institutions the $\beta$'s parameters are positive and significantly different from zero. Note that a positive $\beta$ indicates that a decrease in $\text{VaR}^{\mathbf{x},\tau}_j$ (expressed as a larger negative value) yields a greater negative $\text{CoVaR}^{\mathbf{x},\tau}_{k\vert j}$, i.e. a higher risk of system losses. Moreover, by comparing $\text{HPD}_{95\%}$ for the $\beta$ parameters, it is also evident that the extent of the systemic risk contribution is significantly different across institutions belonging to different sectors. Hence the result highlights the empirical evidence of a sector specific effect of individual losses to the overall systemic risk. We also observe that on average the systemic $\beta$ has lower value for institutions belonging to the financial sector and is higher for institutions belonging to consumer and energy sectors. This evidence gives the idea of existence of sectors having different sensitivity to the risk exposure.
In addition, it is worth noting that sometimes the $\beta$ parameter displays a huge variation also within the same sector. It is the case for example for the industrial one, where the estimated $\beta$ coefficient for GE is significantly different from the one of BA whose credible sets are not overlapping. Finally, comparing the $\beta$s for the two different values of $\tau$ considered we observe that on average higher values of the parameter tend to be associated with smaller values of the confidence level $\tau$, meaning that the co-movement between asset and market is stronger for extreme returns.\newline
\indent We now concentrate our attention to the influence of macroeconomic variables. From Tables \ref{tab:Model_Static_ParEst_0025}--\ref{tab:Model_Static_ParEst_005} we detect some remarkable differences among assets and in particular we observe that:
\begin{itemize}
%
%
\item[-] except for the VIX and the US Real Estate indices, the impact of the remaining variables change in magnitude and significance as we move from one asset to another. This heterogeneous behaviour seems to be transversal with respect to sectors at least in some cases such as liquidity spread (LIQSPRD). As expected, the VIX index estimated parameters are always significantly negative while the opposite happens for the US Real Estate Index. This is true for both the VaR and CoVaR regressions and for all the considered $\tau$.
\item[-] some macroeconomic variables, such as the changes in three-months Treasury Bill rate (3MTB) or the term spread (TERMSPR) display a different impact on the CoVaR and VaR being always positive or not significant in the first case and also negative for some sectors (financial and utilities) in the latter case. This means that an increase in the spread between ten years Treasury Bond rates and three-months Treasury Bill rates (CREDSPR) produces a decrease of the CoVaR while reducing individual risks in the case of financial and utilities sectors. This means that in principle large traded firms benefit from an increased interest rate because it increases the opportunity cost of different financing strategies. As expected, the change in credit spread has a negative impact on the firms' riskiness.
\item[-] for different values of the confidence level $\tau$ the macroeconomic variables show a different impact on the marginal and conditional quantiles becoming in general less significant as the $\tau$-level increases.
\end{itemize}
For the micro exogenous regressors, we note that:
\begin{itemize}
\item[-] the leverage regressor (LEV) has always a negative (when significant) coefficient for the VaR, indicating that individual risk is greatly enhanced in highly leveraged companies, while for the CoVaR regression it is positive for AXP (which belong to consumer finance sub-sector)  but negative for GS, MS and MCO (which belongs to the diversified financial services sub-sector);
%
%
\item[-] the market capitalisation (SIZE) has a significantly positive impact on the VaR while its sign varies also across institutions belonging to the same sector for the CoVaR regression. This evidence suggests that large institutions are more risky if considered in isolation. Moreover, the extent to which large companies contribute to the overall risk is not clear, depending on the ``degree of connection'' among institutions and on diversification of their portfolios;
%
%
\item[-] the maturity mismatch coefficient (MM) is always negative for the VaR regression while it is positive for the CoVaR regression. This denotes the existence of positive correlation between financial imbalances and individual riskiness (measured by the VaR), as we would expect;
\item[-] the market-to-book ratio (MK2BK) is significantly different from zero for the VaR only in two cases: CMA and KEY which belong to the bank sector, and it is always positive for the CoVaR regression. 
\end{itemize}
%
%
To have a complete picture of the individual and systemic risk contributions we plot in Figure \ref{fig:Model_Static_Var_CoVaR} together the estimated VaR and CoVaR along with their $\text{HPD}_{95\%}$ for some of the assets listed in Table \ref{tab:SP500_company_list}.
%
%
%
Starting with individual risk assessment, we clearly find that the VaR profiles are relatively similar across institutions, displaying strong negative downside effects upon the occurrence of the recent financial crisis of 2008, and 2010 and the sovereign debt crisis of 2012. However, the analysis of the time series evolution of the marginal contribution to the systemic risk, measured by CoVaR, reveals different behaviours for the considered assets. In particular, Citygroup (C), which belongs to the bank sector, seems to contribute more to the overall risk than other assets do. Inspecting Figure \ref{fig:Model_Static_Var_CoVaR} and Tables  \ref{tab:Model_Static_ParEst_0025}--\ref{tab:Model_Static_ParEst_005}, we note that institutions having low $\beta$ coefficients provide major contribution to the 2008 financial crisis. On the contrary, McDonalds Corp. (MCD), which belongs to the consumer sector, has a large estimated $\beta$, and inspecting the CoVaR plot in Figure \ref{fig:Model_Static_Var_CoVaR} its contribution to the overall systemic risk is much lower than that of financial institutions.\newline
%
%
%
\indent For the selected companies, Figure \ref{fig:Model_Static_Double_CoVaR} plots the CoVaR for two different confidence levels $\tau=0.025$ and $\tau=0.1$. From this figure we can highlight the different impact of the crisis among sectors.
%
%
For the financial one, for example, we note that the difference between $\text{CoVaR}_j^{\mathbf{x},0.025}$ and $\text{CoVaR}_j^{\mathbf{x},0.1}$ is much larger than for assets belonging to other sectors, meaning that the financial sector probably had a huge impact on the extreme systemic risk during the 2008 crisis, as withnessed by the extremely large losses.
\subsection{Time-varying risk beta}
\label{subsec:time_varying_betas}
%
%
\begin{figure}[!t]
\begin{center}
\captionsetup{font={small}, labelfont=sc}
\includegraphics[width=1.0\linewidth]{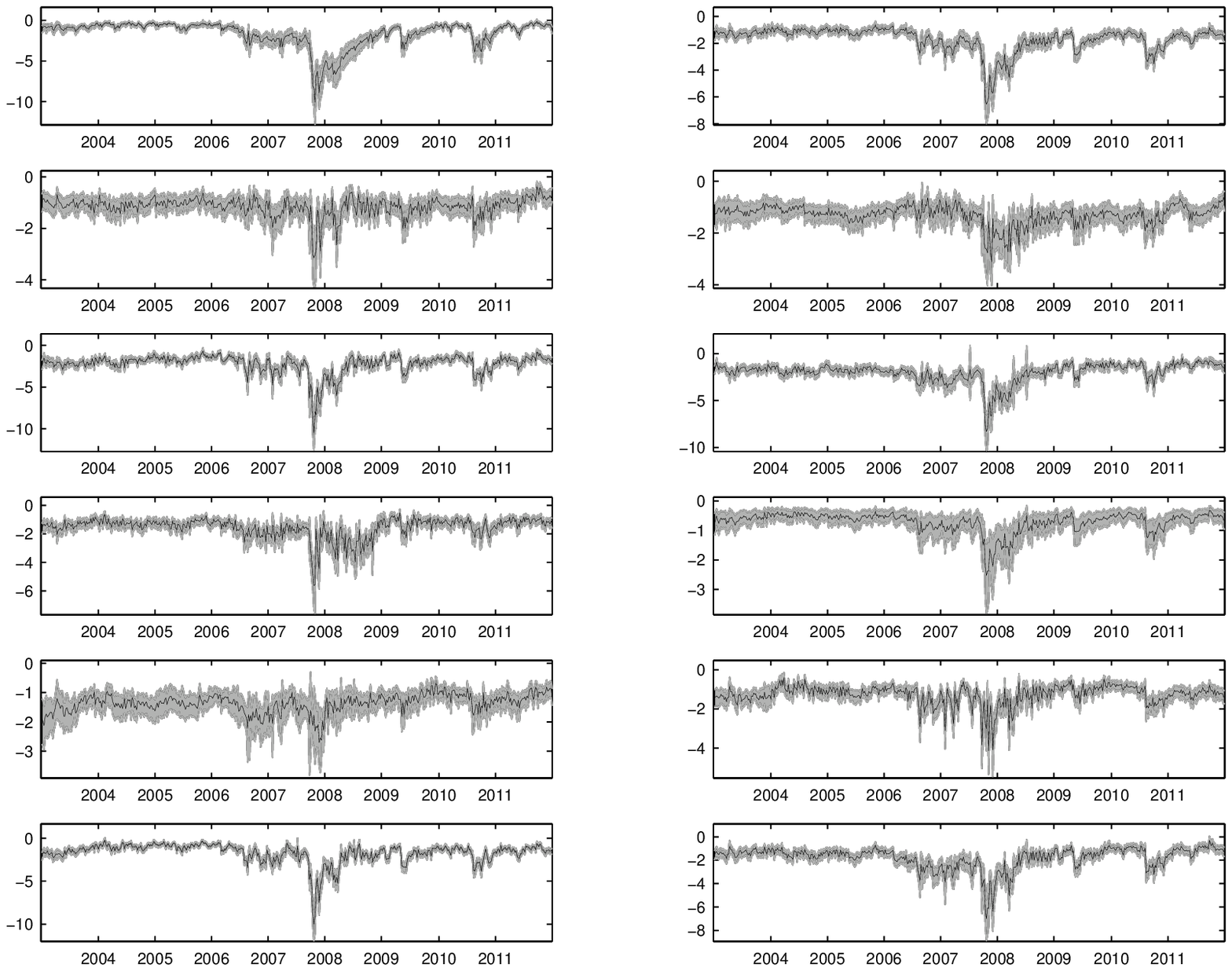}
%
\caption{\small{Time series plot of the static $\Delta\text{CoVaR}^{\mathbf{x},\tau}_{k\vert j}$ at the confidence level $\tau= 0.025$ along with $\text{HPD}_{95\%}$ credible sets, obtained by fitting the model defined in equations (\ref{eq:model_static_covar}) for the following assets, (denoted by $j$): first panel (financial): C (left), GS (right); second panel (consumer): MCD (left) and NKE (right); third panel (energy): CVX (left), XOM (right); forth panel (industrial): BA (left), GE (right); fifth panel (technology): INTC (left), ORCL (right); last panel (utilities): AEE (left), PEG (right).}}
\label{fig:Model_Static_DeltaCoVaR}
\end{center}
\end{figure}
%

%
\begin{figure}[!t]
\begin{center}
\captionsetup{font={small}, labelfont=sc}
\includegraphics[width=1.0\linewidth]{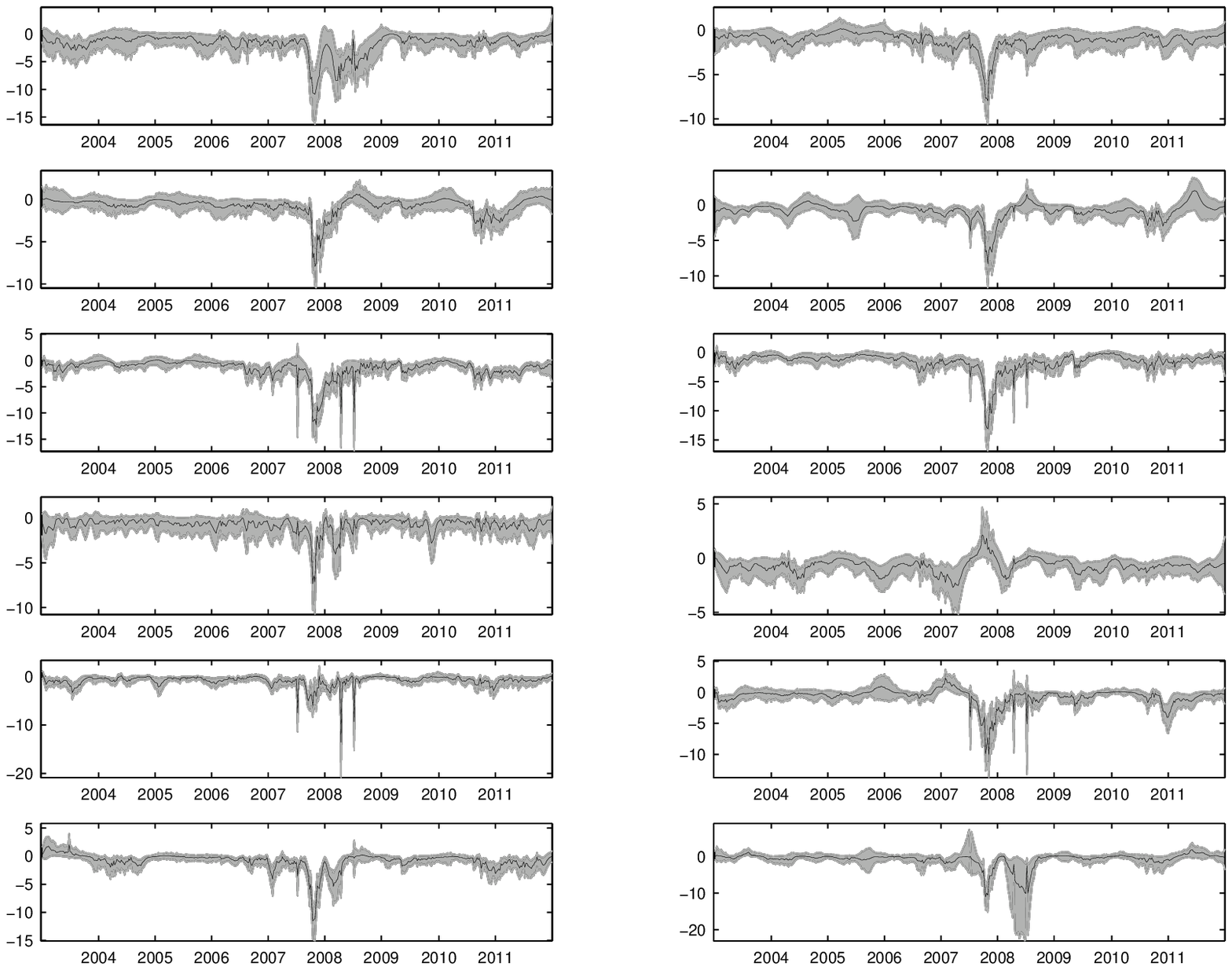}
\caption{\small{Time series plot of the dynamic $\Delta\text{CoVaR}^{\mathbf{x},\tau}_{k\vert j}$ at the confidence level $\tau= 0.025$ along with $\text{HPD}_{95\%}$ credible sets, obtained by fitting the model defined in equations (\ref{eq:VaR-varying})--(\ref{eq:quantile_dynamic_3}) for the following assets, (denoted by $j$): first panel (financial): C (left), GS (right); second panel (consumer): MCD (left) and NKE (right); third panel (energy): CVX (left), XOM (right); forth panel (industrial): BA (left), GE (right); fifth panel (technology): INTC (left), ORCL (right); last panel (utilities): AEE (left), PEG (right).}}
\label{fig:Model_Dynamic_DeltaCoVaR_0025}
\end{center}
\end{figure}
%
We estimate time-varying systemic risk betas according to the dynamic model defined in equations (\ref{eq:VaR-varying})--(\ref{eq:quantile_dynamic_4}) using the same exogenous variables described in Section \ref{subsec:data}. Tables \ref{tab:Model_Dynamic_ParEst_0025} and \ref{tab:Model_Dynamic_ParEst_005} list the posterior estimates of the exogenous regressor parameters $\boldsymbol{\theta}$ for both the $\text{VaR}_{j}^{\tau,\mathbf{x}}$ and $\text{CoVaR}_{k\vert j}^{\tau,\mathbf{x}}$ regressions, at the confidence level $\tau=0.025$ and $\tau=0.05$, respectively. We observe that all the macroeconomic variables, except for the volatility index (VIX) and the liquidity spread (LIQSPR), either have a positive impact on both VaR and CoVaR or are non significant. This means that an upward shift in the term structure of interest rate provides a marginal positive contribution to individual and systemic risks. Interestingly, American Express Co. (AXP) is the only financial institution displaying a negative coefficient for the variable credit spread (CREDSPR) in the individual VaR regression. This essentially means that an increase in the credit spread increases the individual riskyness of that institution. This result seems to be coherent with the American Express' institutional activity, since it is a global financial services institution whose main offerings are charge and credit cards. As expected, the volatility index (VIX) and the rate of change of the Dow Jones US real estate index (DJUSRE) have different effects on the individual and overall risks. Concerning the individual variables a clear positive effect for all the considered institutions is evident only for the SIZE and just for the VaR regression. The maturity mismatch (MM) is non significant almost for all the reported institutions for both VaR and CoVaR regression, while the market-to-book ratio (MK2BK) and the leverage (LEV) variables reveal an heterogeneous sector-specific impact on the risk measures. In this respect the analysis is not exhaustive and the identification of sector specific risk factors deserves further investigation using a larger number of institutions.\newline
\indent Figure \ref{fig:Model_Dynamic_Var_CoVaR}, which is the dynamic counterpart of Figure \ref{fig:Model_Static_Var_CoVaR}, shows that the dynamic CoVaR risk measure suddenly adapts to capture extreme negative losses especially during the 2008 financial crisis. Comparing this evidence with the one shown in Figure \ref{fig:Model_Static_Var_CoVaR} it is clear that the dynamic model provides a better characterisation of extreme tail co-movements when dealing with time series data.\newline
\indent Turning our attention to the time varying $\beta$'s, Figure \ref{fig:Model_Dynamic_BETA_ALL} plots the evolution of the MaP estimates for three different confidence levels $\tau=0.025$, $\tau=0.05$ and $\tau=0.5$. As expected the evolution of the $\beta$'s for $\tau=0.025$ (dark line) and $\tau=0.05$ (blue line), almost coincides for all the reported institutions. Moreover, the systemic risk betas behaviours display a huge cross-sectional heterogeneity. For example, the systemic risk beta of the energy institutions (third panel in Figure \ref{fig:Model_Dynamic_BETA_ALL}) increase over time, showing their highest values during the financial crisis at the end of 2008 and 2011. Conversely, during the same period, we observe a large drop down for the $\beta_t$ of General Electric (fourth panel in Figure \ref{fig:Model_Dynamic_BETA_ALL}). The time series behaviour of the systemic risk betas reveals a different impact of the crisis periods on the overall marginal risk contribution of each institution.
%
%
%
%
\subsection{Measuring marginal contribution to systemic risk}
%
In their paper Adrian and Brunnermeier \cite{adrian_brunnermeier.2011} introduced as a measure of the marginal contribution to system risk, the $\Delta\text{CoVaR}^{\mathbf{x},\tau}_{k\vert j}$ defined as
%
%
%
\begin{eqnarray*}
\Delta\text{CoVaR}^{\mathbf{x},\tau}_{k\vert j}
=\text{CoVaR}_{k \vert Y_j=\text{\footnotesize VaR}_{j}^{\mathbf{x},\tau}}^{\mathbf{x},\tau}-\text{CoVaR}_{k \vert Y_j=\text{\footnotesize VaR}_{j}^{\mathbf{x},0.5}}^{\mathbf{x},\tau}
\end{eqnarray*}
where $\text{CoVaR}_{k \vert Y_j=\text{\footnotesize VaR}_{j}^{\mathbf{x},\tau}}^{\mathbf{x},\tau}=\text{CoVaR}_{k  
\vert j}^{\mathbf{x},\tau}$ and $\text{CoVaR}_{k \vert Y_j=\text{\footnotesize VaR}_{k \vert j}^{\mathbf{x},0.5}}^{\mathbf{x},\tau}$ satisfies equation (\ref{eq:def_CoVaR}) with $\text{VaR}_{j}^{\mathbf{x},0.5}$ instead of $\text{VaR}_{j}^{\mathbf{x},\tau}$.
%
%
To illustrate the behaviour of $\Delta\text{CoVaR}^{\mathbf{x},\tau}_{k\vert j}$ in our application we plot in Figure \ref{fig:Model_Static_DeltaCoVaR} the time-invariant one, while in Figure \ref{fig:Model_Dynamic_DeltaCoVaR_0025} its dynamic version. As expected, for all the considered companies, the marginal contribution to systemic risk increases during market turbulences, showing their lowest values during the financial crisis of 2008. Moreover, some relevant differences among companies belonging to different sectors are evident: in particular, the financial sector, (first panel of Figures \ref{fig:Model_Static_DeltaCoVaR}--\ref{fig:Model_Dynamic_DeltaCoVaR_0025}), the energy sector (third panel of Figures \ref{fig:Model_Static_DeltaCoVaR}-\ref{fig:Model_Dynamic_DeltaCoVaR_0025}) and the utilities sector (bottom panel of Figures \ref{fig:Model_Static_DeltaCoVaR}-\ref{fig:Model_Dynamic_DeltaCoVaR_0025}), display the largest drop. Vice versa, the technology sector displays the lowest variations of the $\Delta\text{CoVaR}$ measure during the 2008 recession. The grey areas correspond to the $\text{HPD}_{95\%}$ associated to the $\Delta$CoVaR contributions providing information about the size of the risk contribution for the given confidence level. We note a huge cross-sectional heterogeneity on the credible sets behaviour with some sectors such as the consumer, industrial and technology being characterised by large uncertainty of the $\Delta$CoVaR estimates.\newline
\indent Comparing the two figures, it is evident that the dynamic $\Delta$CoVaR estimates are smoother than the corresponding time-invariant one, leading to a less noise-corrupted signal useful for policy maker purposes. Moreover, Figure \ref{fig:Model_Dynamic_DeltaCoVaR_0025} provides a clear indication of the high flexibility of the dynamic model implying a promptly reaction of the risk measure to the economic and financial downturns. These considerations argue in favour of the dynamic model when dealing with time series data.\newline
%
%
\indent Finally, we consider the time series relationship between VaR and CoVaR or $\Delta$CoVaR. In what is perhaps the key result of Adrian and Brunnermeier \cite{adrian_brunnermeier.2011}, they find that the CoVaR ($\Delta$CoVaR) of two institutions may be significantly different even if the VaR of the two institutions are similar. On this basis, they suggest the policy maker to employ the  CoVaR ($\Delta$CoVaR) risk measure, as a valid alternative to the VaR, while forming policy regarding institution's risk. Results obtained with our modelling support their thesis. In fact, building the following simple regression model for each institution $j$
\begin{eqnarray}
\text{CoVaR}_{k\vert j}^{\tau,\mathbf{x}}=\delta_0+\delta_1\text{VaR}_k^{\tau,\mathbf{x}}+\nu,
\label{eq:covar_versus_var_regression}
\end{eqnarray}
or
\begin{eqnarray}
\Delta\text{CoVaR}_{k\vert j}^{\tau,\mathbf{x}}=\delta_0+\delta_1\text{VaR}_k^{\tau,\mathbf{x}}+\nu,
\label{eq:deltacovar_versus_var_regression}
\end{eqnarray}
with $\mathbb{E}\left(\nu\vert\text{VaR}_k^{\tau,\mathbf{x}}\right)=0$, where $k$ is the Standard and Poor's Index we test $H_0:\delta_1=1$ to show that $\text{CoVaR}_{k\vert j}^{\tau,\mathbf{x}}$ ($\Delta\text{CoVaR}_{k\vert j}^{\tau,\mathbf{x}}$) is significantly different from the $\text{VaR}_{k}^{\tau,\mathbf{x}}$. Bold numbers in Table \ref{tab:CoVaR_summary} indicate cases where the corresponding coefficient $\delta_1$ is not significantly different from one. Except for few cases the null hypothesis is rejected and this is more evident for the dynamic model than for the time invariant one. In particular, for the $\Delta$CoVaR regression there is enough evidence that the estimated $\delta_1$ parameter is related to the sector the institutions belong to, being higher for financials, energies and utilities. Interestingly, the regression coefficients is almost zero for the technology sector. 
\section{Conclusion}
\label{sec:conclusion}
%
One of the major issue policy makers deal with during financial crisis is the evaluation of the extent to which riskness tail events spread across financial institutions. In fact, during financial turmoils, the correlations among asset returns tend to rise, a phenomenon known in the economic and financial literature as contagion. From a statistical point of view the risk of contagion essentially implies that the joint probability of observing large losses increases during recessions. The common risk measures recently imposed by the public regulators, (the Basel Committee, for the bank sector) such as the VaR, fail to account for such risk spillover among institutions. The CoVaR risk measure recently introduced by Adrian and Brunnermeier \cite{adrian_brunnermeier.2011} overcomes this problem being able to account for the dependence among institutions' extreme events.\newline
%
%
\indent In this paper we address the problem of estimating the CoVaR in a Bayesian framework using quantile regression. We first consider a time-invariant model allowing for interactions only among contemporaneous variables. The model is subsequently extended in a time-varying framework where the constant part and the CoVaR parameter $\beta$ are modelled as functions of unobserved processes having their own dynamics. 
In order to make posterior inference, we use the maximum a posteriori summarising criteria and we prove that it leads to estimated quantiles having good sample properties according to De Rossi and Harvey results \cite{derossi_harvey.2009} and we 
efficient Gibbs sampler algorithms based on data augmentation for the two models considered.\newline 
\indent The Bayesian approach used throughout the paper allows to infer on the whole posterior distribution of the quantities of interest and their credible sets which are important to assess the accuracy of the point estimates. Since the quantities of interest in this contest are risk measures, learning about the whole distribution becomes more relevant due to the interpretation of the VaR and CoVaR as financial losses. In addition, credible sets provide upper and lower limits for the capital requirements for banks and financial institutions.\newline
%
%
\indent To verify the reliability of the built models we analyse weekly time series for nineteen institutions belonging to six different sectors of the Standard and Poor 500 composite index spanning the period from 2nd January, 2004 to 31st December, 2012. We use micro and macro exogenous variable to characterise the quantile functions. From the empirical results it is clear that the model and the proposed approach are able to sharply estimate marginal and conditional quantiles providing a more realistic and informative characterisation of extreme tail co-movements. In particular, the dynamic version of the model we propose outperforms the time invariant specification when the analysis is based on time series data. Up to our knowledge this is the first attempt to implement a Bayesian inference for the CoVaR.\newline
%
%

\textbf{Acknowledgements.} This research is supported by the Italian Ministry of Research PRIN 2013--2015, ``Multivariate Statistical Methods for Risk Assessment'' (MISURA), and by the ``Carlo Giannini Research Fellowship'', the ``Centro Interuniversitario di Econometria'' (CIdE) and ``UniCredit Foundation''. The authors are grateful to the ANR-Blanc ``Bandhits'' for granting them the opportunity to work together in Paris on this paper. The second author is also grateful to the ``MEMOTEF'' Department for granting her a three-month visiting fellowship at Sapienza University in fall 2012. We are also very grateful to Roberto Casarin and Monica Billio for their helpful comments and suggestions. 

\newpage

\appendix
\section{Missing values treatment} 
\label{sec:appendix_A}
%
\begin{figure}[!ht]
\begin{center}
\captionsetup{font={small}, labelfont=sc}
\includegraphics[width=0.8\linewidth]{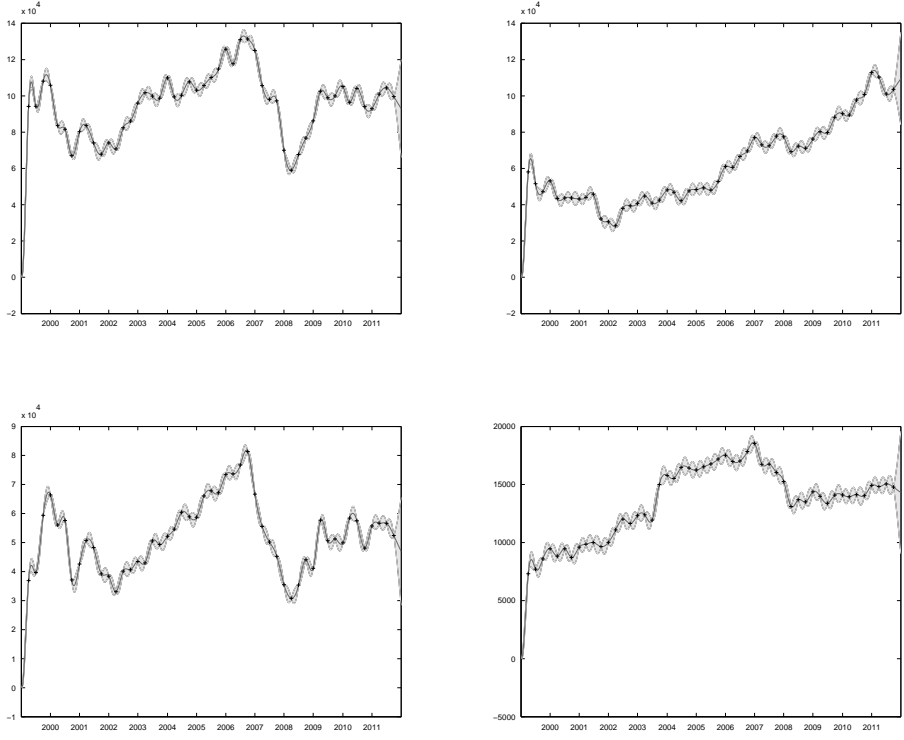}
\caption{\small{Missing values imputation using smoothing spline of the SIZE variable for AXP (top, left), MCD (top, right), BA (bottom, left) and AEE (bottom, right). In each plot predicted values (gray line) as well as 95\% HPD credible sets (gray area) are displayed. Dark cross denotes observed values.}}
\label{fig:missing_SIZE_appendix_A}
\end{center}
\end{figure}
%
%
In this appendix we give details on the procedure used to impute missing observations. In particular for each variable, starting from balance sheets data available only on a quarterly basis, we implement a nonparametric smoothing cubic spline, see for example Koopman \cite{koopman.1991} and Koopman \textit{et al}. \cite{koopman_etal.1998}. Formally, suppose we have an univariate time series $y_{\tau_1},y_{\tau_2},\dots,y_{\tau_T}$ not necessarily equispaced in time and define $\delta_t=\tau_t-\tau_{t-1}$, for $t=1,2,\dots,T$ as the difference between two consecutive observations, and assume the time series is not entirely observed, we approximate the series by a sufficiently smooth function $s\left(\tau_t\right)$. Standard approach chooses $s\left(\tau_t\right)$ by minimizing the following penalized-least squares criterion:
\begin{equation}
\mathcal{L}\left(\lambda\right)=\sum_{i=1}^T\left[y_{\tau_t}-s\left(\tau_{i}\right)\right]+\lambda \sum_{i=1}^T\left[\Delta^ms\left(\tau_{i}\right)\right]^2,
\label{eq:PLS_criterion}
\end{equation}
with respect to $s\left({\tau_t}\right)$, for a given penalization term $\lambda$. The function $s\left(\tau_t\right),\forall t=1,2,\dots,T$ is a polynomial spline of order $m+1$ and when $m=2$, we have a smoothing cubic spline model.\newline 
\indent To estimate the smoothing parameter $\lambda$ and to forecast missing observations model (\ref{eq:PLS_criterion}) can be cast in State Space form that is, for $m=2$:
\begin{eqnarray*}
y\left(\tau_t\right)&=&s\left(\tau_t\right)+\epsilon\left(\tau_t\right)\label{eq:cubic_smooth_sline}\\
s\left(\tau_{t+1}\right)&=&s\left(\tau_t\right)+\delta_t \ell\left(\tau_t\right)+\zeta_1\left(\tau_t\right)\\
\ell\left(\tau_{t+1}\right)&=&\ell\left(\tau_t\right)+\zeta_2\left(\tau_t\right)
\end{eqnarray*}
where $\left[s\left(\tau_1\right),\ell\left(\tau_1\right)\right]^\trasp\sim\mathcal{N}\left(\mathbf{0}_2,\kappa\mathbb{I}_2\right)$, with $\kappa$ sufficiently large to ensure a diffuse initialization of the latent states, the transition equation innovations vector is $\left[\zeta_1\left(\tau_t\right),\zeta_2\left(\tau_t\right)\right]^\trasp\sim\mathcal{N}\left(\mathbf{0}_2,\mathbf{S}_{\zeta}\right)$ with $\mathbf{S}_{\zeta}=\sigma^2_{\zeta}\left[\begin{array}{cc}\frac{1}{3}\delta_{t}^3 & \frac{1}{2}\delta_{t}^2 \\ \frac{1}{2}\delta_{t}^2  & \delta_{t}\end{array}\right]$, the measurement innovation $\epsilon\left(\tau_t\right)\sim\mathcal{N}\left(0,\sigma_{\epsilon}^2\right)$ and the penalty parameter $\lambda$ coincides with the signal-to-noise ratio $\lambda=\sigma_{\zeta}^2/\sigma_{\epsilon}^2$. We fix the parameter $\sigma_{\epsilon}$ to one and estimate $\lambda=\sigma_{\zeta}^2$ in a Bayesian framework imposing a diffuse Inverse Gamma prior distribution, i.e. $\lambda=\sigma_{\zeta}^2\sim\mathcal{IG}\left(\alpha_{\lambda}^0,\beta_{\lambda}^0\right)$. We fit the model using MCMC techniques, in particular, the Gibbs sampler with data augmentation (see Geman and Geman \cite{geman_geman.1984}, Tanner and Wong \cite{tanner_wong.1987} and Gelfand and Smith \cite{gelfand_smith.1990}). The Gibbs sampler consist in the following two steps: 
\begin{itemize}
\item[S1.] simulate the latent process $\boldsymbol{\xi}_t^{(i+1)}=\left[s\left(\tau_{t}\right), \ell\left(\tau_{t}\right)\right]^\trasp$, $\forall t=1,2,\dots,T$, using the disturbance simulation smoothing algorithm of de Jong and Shephard \cite{dejong_shephard.1995} appropriately adjusted to handle missing observations and the diffuse initialization of the state vector (see also the augmented Kalman filter and smoother of de Jong, \cite{dejong.1991}).
\item[S2.] simulate the $\lambda^{(i+1)}$ parameter from the complete full conditional distribution which is an Inverse Gamma distribution $\lambda^{(i+1)}\sim\mathcal{IG}\left(\widetilde{\alpha}^{(i+1)}_{\lambda},\widetilde{\beta}^{(i+1)}_{\lambda}\right)$ with parameters:
\begin{eqnarray*}
\widetilde{\alpha}^{(i+1)}_{\lambda}&=&\alpha_{\lambda}^0+\frac{T-1}{2}\nonumber\\
\widetilde{\beta}^{(i+1)}_{\lambda}&=&\beta_{\lambda}^0+\frac{1}{2}\sum_{t=1}^{T-1}\left(\boldsymbol{\xi}^{(i+1)}_{t+1}-A\boldsymbol{\xi}^{(i+1)}_t\right)^\trasp\boldsymbol{S}_{\zeta}^{-1}\left(\boldsymbol{\xi}^{(i+1)}_{t+1}-A\boldsymbol{\xi}^{(i+1)}_t\right),\nonumber
\end{eqnarray*}
\end{itemize}
for $i=1,2,\dots,G$, with $G$ equal to the number of draws. The algorithm is initializated at $i=0$ by simulating the $\lambda$ parameter from its prior distribution. To estimate the missing observations we average simulated points of $s\left(\tau_t\right)$ across Gibbs sampler draws. 

\newpage

\section{Tables}
\label{sec:appendix_B}
%
\begin{table}[!ht]
\captionsetup{font={small}, labelfont=sc}
%
\begin{small}
\centering
 \smallskip
  \begin{tabular}{lcccccc}\\
  \hline\hline
   Name & Mean & Min & Max & Std. Dev. & Corr. & 1\% Str. Lev.\\
    \hline
C 		&  -0.490  & -92.632  & 78.797  & 9.756  & 0.673  & -26.267  \\
BAC 	& - 0.212  & -59.287  & 60.672  & 7.990  & 0.705  & -27.231  \\
CMA 	&  -0.072  & -31.744  & 33.104  & 5.945  & 0.718  & -19.794  \\
JPM 	&   0.086  & -41.684  & 39.938  & 5.826  & 0.711  & -12.785  \\
KEY 	&  -0.210  & -61.427  & 40.976  & 7.428  & 0.694  & -24.231  \\
GS		& 	0.071  & -36.564  & 39.320  & 5.641  & 0.719  & -16.850  \\
MS   	&  -0.170  & -90.465  & 69.931  & 8.342  & 0.699  & -19.706  \\
MCO 	&   0.125  & -27.561  & 28.300  & 5.615  & 0.688  & -20.719  \\
AXP		&   0.091  & -28.779  & 24.360  & 5.104  & 0.772  & -16.108  \\
MCD		&   0.320  & -12.130  & 11.878  & 2.606  & 0.554  & -4.978  \\
NKE		&   0.260  & -18.462  & 18.723  & 3.746  & 0.655  & -11.596  \\
CVX		&   0.254  & -31.674  & 15.467  & 3.585  & 0.745  & -8.275  \\
XOM		&   0.197  & -22.301  & 8.717   & 3.081  & 0.696  & -7.123  \\
BA		&   0.164  & -25.294  & 16.034  & 4.258  & 0.727  & -12.468  \\
GE		&  -0.023  & -18.680  & 30.940  & 4.311  & 0.715  & -15.578  \\
INTC	&  -0.052  & -17.038  & 16.935  & 4.117  & 0.671  & -12.533  \\
ORCL	&   0.203  & -15.518  & 12.135  & 3.762  & 0.632  & -9.986  \\
AEE 	&   0.015  & -29.528  & 9.485   & 3.118  & 0.682  & -8.172  \\
PEG 	&   0.143  & -26.492  & 10.568  & 3.318  & 0.553  & -8.191  \\
S\&P500 &   0.052  & -20.083  & 11.355  & 2.637  & 1.000  & -7.258  \\
      \hline\hline
\end{tabular}
\caption{Summary statistics of the company's returns and market index (S\&P500) returns (in percentage). The sixth column, denoted by ``Corr'', is the correlation coefficient with the market returns while the last column, denoted by ``1\% Str. Lev.'' is the 1\% empirical quantile of the returns distribution.}
\label{tab:SP500_data_summary_stat}
\end{small}
%
%
\end{table}
%
%
\begin{table}[!h]
\captionsetup{font={small}, labelfont=sc}
%
\begin{small}
\centering
 \smallskip
  \begin{tabular}{lcccc}\\
  \hline\hline
   & \multicolumn{2}{c}{CoVaR} & \multicolumn{2}{c}{$\Delta$CoVaR}\\
   Name & Time Invariant  & Time Varying & Time Invariant & Time Varying\\
    \hline
C 		& 1.325  & 1.035  & 0.401  & 0.325  \\
BAC		& 1.084  & 0.983  & 0.230  & 0.298  \\
CMA		& 1.109  & 1.264  & 0.109  & 0.177  \\
JPM		& 1.259  & 0.573  & 0.136  & 0.111  \\
KEY		& 1.078  & -0.481  & 0.229  & 0.165  \\
GS		& 1.085  & 1.276  & 0.234  & 0.188  \\
MS		& {\bf 1.003}  & 0.604  & 0.275  & 0.132  \\
MCO		& 1.085  & -4.139  & 0.130  & 0.052  \\
AXP		& 1.110  & 18.352  & 0.171  & 0.205  \\
MCD		& 0.933  & 3.500  & 0.062  & 0.192  \\
NKE		& 0.842  & 1.251  & 0.095  & 0.222  \\
CVX		& {\bf 0.986}  & 2.376  & 0.291  & 0.354  \\
XOM		& 1.099  & 1.280  & 0.276  & 0.365  \\
BA		& {\bf 1.017}  & 2.666  & 0.169  & 0.156  \\
GE		& 1.179  & 21.245  & 0.078  & -0.039  \\
INTC	& 0.874  & 5.928  & 0.060  & 0.097  \\
ORCL	& {\bf 0.998}  & 3.985  & 0.066  & 0.229  \\
AEE		& 1.054  & 1.441  & 0.346  & 0.296  \\
PEG		& {\bf 1.001}  & 5.123  & 0.205  & 0.271  \\
      \hline\hline
\end{tabular}
\caption{$\delta_1$ estimates for the regressions in equations (\ref{eq:covar_versus_var_regression})--(\ref{eq:deltacovar_versus_var_regression}) for both the time invariant and time varying models. Bold numbers indicate that the corresponding coefficients are not significantly different from 1.}
\label{tab:CoVaR_summary}
\end{small}
%
%
\end{table}
%
%
\begin{sidewaystable}[!hp]
\captionsetup{font={small}, labelfont=sc}
\centering
 \smallskip
 \begin{small} 
 \resizebox{\columnwidth}{!}{%
 \setlength\tabcolsep{1pt}
 \begin{tabular}{cccccccccccccccccccc}\\
 \toprule
%
\multicolumn{1}{c}{\multirow{2}{*}{$\text{VaR}$}} & \multicolumn{9}{c}{Financial} & \multicolumn{2}{c}{Consumer} & \multicolumn{2}{c}{Energy} & \multicolumn{2}{c}{Industrial} & \multicolumn{2}{c}{Technology} & \multicolumn{2}{c}{Utilities}\\
\cmidrule(lr){2-10}\cmidrule(lr){11-12} \cmidrule(l){13-14} \cmidrule(l){15-16} \cmidrule(l){17-18} \cmidrule(l){19-20}
& C & BAC & CMA & JPM & KEY & GS & MS & MCO & AXP & MCD & MKE & CVX & XOM & BA & GE & INTC & ORCL & AEE & PEG\\
\cline{2-20}
\multicolumn{1}{c}{\multirow{2}{*}{CONST}} & -23.453  & -23.730  & -24.575  & -3.706  & -25.803  & -13.518  & -40.232  & 0.151  & -30.096  & 3.999  & -0.260  & -18.940  & -23.383  & -30.230  & -30.936  & -9.147  & -31.432  & 7.560  & -19.406  \\
& (-41.309,-5.136) & (-42.958,-11.088) & (-28.871,-15.193) & (-28.752,2.264) & (-40.714,-3.834) & (-22.909,-0.779) & (-58.293,-21.802) & (-13.902,17.419) & (-44.705,-13.426) & (-32.696,1.486) & (-24.156,12.067) & (-23.875,13.031) & (-22.646,8.844) & (-38.347,-14.434) & (-54.967,-19.101) & (-35.447,2.443) & (-46.134,-18.303) & (-5.195,17.662) & (-32.558,-0.159) \\
\multicolumn{1}{c}{\multirow{2}{*}{VIX}} & -0.881  & -0.766  & -0.300  & -0.371  & -0.905  & -0.398  & -0.603  & -0.329  & -0.332  & -0.092  & -0.104  & -0.314  & -0.237  & -0.164  & -0.321  & -0.168  & -0.166  & -0.267  & -0.301  \\
& (-1.295,-0.747) & (-0.855,-0.552) & (-0.350,-0.246) & (-0.459,-0.364) & (-0.975,-0.717) & (-0.436,-0.350) & (-0.778,-0.466) & (-0.399,-0.231) & (-0.346,-0.228) & (-0.104,-0.064) & (-0.106,-0.020) & (-0.381,-0.273) & (-0.321,-0.244) & (-0.189,-0.112) & (-0.326,-0.234) & (-0.178,-0.094) & (-0.220,-0.153) & (-0.318,-0.238) & (-0.337,-0.262) \\
\multicolumn{1}{c}{\multirow{2}{*}{LIQSPR}}& 0.014  & 0.097  & -0.034  & -0.000  & 0.022  & 0.042  & -0.018  & 0.025  & -0.048  & 0.017  & 0.008  & -0.010  & 0.037  & -0.030  & -0.002  & 0.006  & -0.041  & 0.016  & -0.054  \\
& (-0.076,0.041) & (0.003,0.088) & (-0.058,-0.012) & (-0.014,0.028) & (-0.029,0.078) & (0.000,0.057) & (-0.064,0.043) & (0.003,0.063) & (-0.080,-0.009) & (0.005,0.034) & (-0.007,0.049) & (-0.037,0.019) & (0.010,0.061) & (-0.039,0.007) & (-0.031,0.019) & (-0.081,0.028) & (-0.068,-0.026) & (-0.026,0.022) & (-0.094,-0.009) \\
\multicolumn{1}{c}{\multirow{2}{*}{3MTB}} & -0.007  & 0.092  & -0.084  & -0.025  & -0.159  & 0.116  & 0.090  & 0.025  & -0.110  & 0.015  & 0.067  & 0.017  & 0.029  & 0.014  & -0.010  & 0.208  & 0.167  & -0.046  & -0.036  \\
& (-0.207,0.071) & (-0.117,0.141) & (-0.107,-0.019) & (-0.049,0.054) & (-0.235,0.020) & (0.054,0.134) & (-0.081,0.101) & (0.006,0.159) & (-0.240,-0.079) & (-0.031,0.034) & (0.038,0.144) & (-0.041,0.094) & (-0.021,0.088) & (-0.034,0.056) & (-0.054,0.034) & (0.058,0.242) & (0.109,0.200) & (-0.091,-0.026) & (-0.062,0.064) \\
\multicolumn{1}{c}{\multirow{2}{*}{TERMSPR}} & 0.110  & 0.093  & 0.083  & 0.014  & -0.137  & 0.053  & -0.027  & -0.011  & -0.036  & -0.004  & 0.057  & 0.006  & 0.005  & 0.036  & 0.033  & 0.175  & 0.081  & -0.044  & -0.042  \\
& (0.005,0.179) & (-0.017,0.145) & (0.059,0.106) & (0.003,0.062) & (-0.233,-0.050) & (0.015,0.075) & (-0.117,0.008) & (-0.023,0.112) & (-0.083,-0.004) & (-0.025,0.009) & (-0.006,0.109) & (-0.050,0.017) & (-0.017,0.038) & (-0.035,0.051) & (0.002,0.050) & (0.105,0.208) & (0.054,0.124) & (-0.095,-0.029) & (-0.066,0.003) \\
\multicolumn{1}{c}{\multirow{2}{*}{CREDSPR}} & -0.003  & 0.072  & -0.130  & -0.038  & 0.033  & -0.018  & -0.213  & 0.203  & -0.097  & -0.075  & 0.056  & -0.109  & -0.024  & -0.101  & 0.011  & 0.123  & 0.022  & -0.167  & -0.092  \\
& (-0.130,0.110) & (-0.136,0.118) & (-0.153,-0.061) & (-0.071,0.002) & (-0.043,0.119) & (-0.133,-0.023) & (-0.411,-0.161) & (0.073,0.257) & (-0.159,-0.068) & (-0.088,-0.044) & (-0.010,0.057) & (-0.169,-0.047) & (-0.032,0.054) & (-0.150,-0.044) & (-0.045,0.040) & (0.036,0.152) & (-0.033,0.074) & (-0.198,-0.126) & (-0.116,-0.042) \\
\multicolumn{1}{c}{\multirow{2}{*}{DJUSRE}} & 0.776  & 0.566  & 0.368  & 0.670  & 0.576  & 0.492  & 0.788  & 0.785  & 0.348  & 0.112  & 0.592  & 0.217  & 0.287  & 0.757  & 0.488  & 0.423  & 0.095  & 0.410  & 0.106  \\
& (0.319,0.917) & (0.456,0.737) & (0.298,0.506) & (0.496,0.689) & (0.277,0.788) & (0.340,0.510) & (0.517,0.917) & (0.484,0.914) & (0.318,0.460) & (0.077,0.165) & (0.527,0.794) & (0.145,0.333) & (0.248,0.436) & (0.676,0.838) & (0.311,0.517) & (0.225,0.486) & (-0.031,0.200) & (0.327,0.455) & (0.022,0.193) \\
\multicolumn{1}{c}{\multirow{2}{*}{LEV}} & -1.046  & 0.093  & -0.147  & -0.081  & -0.463  & -0.281  & -0.025  & -0.008  & -0.546  & 0.288  & -4.157  & -1.443  & -0.671  & 0.027  & 0.690  & 2.208  & 4.951  & -7.226  & 0.121  \\
& (-1.268,-0.095) & (-1.027,1.295) & (-0.505,0.176) & (-0.186,0.336) & (-1.042,0.007) & (-0.312,-0.053) & (-0.193,0.126) & (-0.057,0.030) & (-1.165,-0.205) & (-0.057,6.935) & (-10.445,3.260) & (-8.269,-0.155) & (-0.363,14.448) & (0.021,0.036) & (-0.447,0.625) & (-7.393,7.797) & (4.051,10.956) & (-10.479,-3.950) & (-0.697,0.468) \\
\multicolumn{1}{c}{\multirow{2}{*}{MK2BK}} & -1.610  & 0.477  & 0.773  & 1.191  & -4.699  & 1.860  & -3.526  & -0.000  & 0.072  & 0.482  & 1.518  & -0.177  & -1.013  & -0.005  & -1.675  & -3.986  & -0.007  & -3.435  & -1.473  \\
& (-3.923,0.055) & (-0.730,2.861) & (0.524,2.384) & (-0.152,2.452) & (-6.073,-1.249) & (-0.144,2.376) & (-5.719,0.044) & (-0.000,0.001) & (-0.285,0.601) & (-0.671,0.389) & (1.177,3.623) & (-1.937,1.344) & (-2.299,-0.483) & (-0.011,0.005) & (-2.534,-0.481) & (-4.901,-0.982) & (-0.267,0.386) & (-4.128,-1.743) & (-2.744,-0.237) \\
\multicolumn{1}{c}{\multirow{2}{*}{SIZE}} & 4.276  & 3.033  & 2.689  & 0.215  & 5.003  & 1.701  & 4.328  & -0.493  & 3.392  & -0.859  & -0.540  & 2.127  & 2.284  & 2.705  & 2.632  & 1.016  & 1.745  & 2.030  & 2.348  \\
& (2.574,5.648) & (1.881,4.307) & (1.570,3.023) & (-0.199,2.292) & (2.287,6.438) & (0.518,2.311) & (2.398,6.036) & (-2.255,0.992) & (1.910,4.684) & (-0.885,1.978) & (-1.869,0.903) & (0.146,2.636) & (-1.739,1.377) & (1.232,3.415) & (1.774,4.884) & (-0.200,3.634) & (0.168,2.671) & (1.544,2.673) & (0.482,3.732) \\
\multicolumn{1}{c}{\multirow{2}{*}{MM}} & -16.018  & -40.628  & -65.857  & 8.852  & -5.523  & -5.775  & 4.814  & -9.419  & 5.392  & -2.905  & -5.080  & 22.866  & 13.566  & 16.945  & -6.272  & -6.329  & 0.686  & 3.878  & -1.645  \\
& (-33.629,1.883) & (-47.632,-24.672) & (-58.769,-23.794) & (-9.480,11.641) & (-14.424,16.149) & (-13.134,3.581) & (-10.206,19.449) & (-11.613,-5.467) & (-0.754,16.497) & (-7.443,0.710) & (-7.222,0.039) & (6.004,22.529) & (-1.730,12.212) & (14.240,36.859) & (-11.209,-2.179) & (-9.374,2.354) & (-1.218,2.378) & (-0.720,17.106) & (-12.453,19.996) \\
\multicolumn{1}{c}{\multirow{2}{*}{$\sigma_j$}} & 0.388  & 0.317  & 0.212  & 0.225  & 0.306  & 0.195  & 0.324  & 0.305  & 0.179  & 0.122  & 0.216  & 0.197  & 0.165  & 0.187  & 0.157  & 0.242  & 0.176  & 0.173  & 0.179  \\
& (0.342,0.410) & (0.290,0.348) & (0.191,0.229) & (0.197,0.235) & (0.301,0.362) & (0.193,0.232) & (0.300,0.362) & (0.279,0.334) & (0.167,0.200) & (0.117,0.140) & (0.186,0.223) & (0.172,0.206) & (0.145,0.174) & (0.175,0.210) & (0.148,0.178) & (0.220,0.263) & (0.180,0.216) & (0.147,0.175) & (0.172,0.205) \\
\hline
\multicolumn{1}{c}{\multirow{2}{*}{$\text{CoVaR}$}} & \multicolumn{9}{c}{Financial} & \multicolumn{2}{c}{Consumer} & \multicolumn{2}{c}{Energy} & \multicolumn{2}{c}{Industrial} & \multicolumn{2}{c}{Technology} & \multicolumn{2}{c}{Utilities}\\
\cmidrule(lr){2-10}\cmidrule(lr){11-12} \cmidrule(l){13-14} \cmidrule(l){15-16} \cmidrule(l){17-18} \cmidrule(l){19-20}
& C & BAC & CMA & JPM & KEY & GS & MS & MCO & AXP & MCD & MKE & CVX & XOM & BA & GE & INTC & ORCL & AEE & PEG\\
\cline{2-20}
\multicolumn{1}{c}{\multirow{2}{*}{CONST}} & 21.132& -0.745& 12.509& -8.071& 0.218& 0.027& -7.597& 16.835& 25.524& -9.039& -10.414& 9.677& -12.747& -2.660& 38.010& -8.476& -2.013& 8.342& -12.301\\
& (7.972,27.108) & (-0.198,14.951) & (-3.539,10.333) & (-15.303,0.607) & (-10.848,8.853) & (-6.376,2.947) & (-17.399,-5.490) & (13.627,22.211) & (6.678,26.090) & (-25.983,7.463) & (-15.093,8.644) & (3.317,23.778) & (-18.034,-7.015) & (-13.725,3.097) & (22.465,48.669) & (-24.689,-1.164) & (-7.721,2.699) & (-9.701,5.974) & (-14.436,-2.596) \\
\multicolumn{1}{c}{\multirow{2}{*}{VIX}} & -0.252& -0.188& -0.253& -0.260& -0.195& -0.118& -0.100& -0.240& -0.243& -0.149& -0.128& -0.099& -0.152& -0.122& -0.325& -0.129& -0.159& -0.153& -0.108\\
& (-0.276,-0.211) & (-0.204,-0.160) & (-0.245,-0.176) & (-0.303,-0.258) & (-0.213,-0.154) & (-0.146,-0.109) & (-0.105,-0.044) & (-0.269,-0.228) & (-0.248,-0.201) & (-0.195,-0.139) & (-0.141,-0.091) & (-0.107,-0.065) & (-0.165,-0.105) & (-0.145,-0.097) & (-0.333,-0.283) & (-0.159,-0.114) & (-0.173,-0.129) & (-0.185,-0.134) & (-0.139,-0.087) \\
\multicolumn{1}{c}{\multirow{2}{*}{LIQSPR}} & -0.008& 0.001& 0.010& 0.014& 0.012& 0.004& 0.013& 0.022& 0.005& 0.010& -0.010& 0.012& 0.009& 0.013& 0.038& -0.003& 0.012& 0.013& 0.018\\
& (-0.013,0.008) & (-0.010,0.018) & (0.008,0.036) & (0.000,0.022) & (0.004,0.018) & (-0.000,0.015) & (-0.009,0.011) & (0.008,0.027) & (0.000,0.016) & (-0.001,0.021) & (-0.023,-0.004) & (0.005,0.016) & (-0.002,0.014) & (-0.010,0.020) & (0.018,0.049) & (-0.014,0.007) & (-0.000,0.014) & (0.007,0.018) & (0.008,0.023) \\
\multicolumn{1}{c}{\multirow{2}{*}{3MTB}} & 0.069& 0.023& 0.044& 0.046& 0.036& 0.014& 0.063& 0.039& 0.089& 0.007& 0.031& 0.027& 0.026& -0.004& 0.026& -0.006& 0.027& 0.033& 0.048\\
& (0.021,0.080) & (-0.008,0.050) & (0.028,0.081) & (0.000,0.092) & (0.020,0.058) & (-0.017,0.024) & (0.043,0.103) & (0.017,0.056) & (0.050,0.106) & (-0.023,0.032) & (0.001,0.064) & (0.006,0.045) & (0.006,0.045) & (-0.052,0.019) & (-0.005,0.068) & (-0.019,0.031) & (0.023,0.062) & (0.022,0.049) & (0.015,0.076) \\
\multicolumn{1}{c}{\multirow{2}{*}{TERMSPR}}& 0.003& 0.015& 0.042& 0.031& 0.029& 0.014& 0.037& 0.024& 0.047& 0.007& 0.037& 0.023& 0.020& -0.007& 0.009& 0.015& -0.006& 0.037& 0.042\\
& (-0.012,0.012) & (-0.001,0.026) & (0.026,0.050) & (0.017,0.044) & (0.024,0.053) & (0.002,0.021) & (0.032,0.053) & (0.012,0.037) & (0.030,0.053) & (-0.008,0.017) & (0.016,0.044) & (0.010,0.038) & (0.004,0.029) & (-0.028,0.000) & (-0.005,0.027) & (0.003,0.030) & (-0.007,0.020) & (0.030,0.056) & (0.026,0.059) \\
\multicolumn{1}{c}{\multirow{2}{*}{CREDSPR}} & 0.002& -0.009& 0.022& -0.005& 0.002& -0.013& 0.033& -0.019& 0.054& -0.044& -0.002& -0.040& -0.006& -0.046& 0.008& -0.054& -0.071& 0.009& -0.057\\
& (-0.022,0.031) & (-0.044,0.004) & (-0.025,0.032) & (-0.025,0.038) & (-0.012,0.030) & (-0.030,0.011) & (0.038,0.063) & (-0.034,-0.001) & (0.024,0.063) & (-0.085,-0.039) & (-0.027,0.022) & (-0.052,-0.014) & (-0.034,0.003) & (-0.084,-0.015) & (-0.031,0.025) & (-0.067,-0.033) & (-0.096,-0.056) & (-0.012,0.018) & (-0.067,-0.030) \\
\multicolumn{1}{c}{\multirow{2}{*}{DJUSRE}} & 0.151& 0.234& 0.209& 0.236& 0.242& 0.284& 0.316& 0.219& 0.136& 0.301& 0.286& 0.237& 0.318& 0.287& 0.228& 0.235& 0.273& 0.180& 0.271\\
& (0.137,0.223) & (0.188,0.274) & (0.191,0.277) & (0.209,0.311) & (0.193,0.258) & (0.229,0.309) & (0.302,0.348) & (0.177,0.235) & (0.127,0.210) & (0.228,0.303) & (0.235,0.326) & (0.211,0.282) & (0.277,0.348) & (0.217,0.351) & (0.184,0.284) & (0.222,0.275) & (0.223,0.294) & (0.157,0.212) & (0.236,0.310) \\
\multicolumn{1}{c}{\multirow{2}{*}{LEV}} & -0.019& 0.086& -0.023& 0.327& -0.093& -0.072& -0.027& -0.015& 0.257& -2.304& 1.907& -3.260& 8.835& 0.009& 0.685& -1.409& -1.496& 0.861& -0.399\\
& (-0.065,0.019) & (-0.140,0.245) & (-0.383,0.028) & (0.212,0.428) & (-0.204,0.039) & (-0.121,-0.055) & (-0.091,-0.012) & (-0.016,-0.006) & (0.114,0.436) & (-4.171,0.565) & (-3.872,3.391) & (-5.988,-2.268) & (5.627,10.242) & (0.004,0.014) & (0.272,0.865) & (-4.305,-0.949) & (-2.520,-0.357) & (-0.137,3.041) & (-0.636,-0.180) \\
\multicolumn{1}{c}{\multirow{2}{*}{MK2BK}}& -0.971& -1.415& -1.176& -3.896& -1.429& -0.073& -0.881& 0.000& 0.070& 0.226& -0.067& 0.154& -0.001& 0.002& -0.590& -0.230& -0.258& -1.540& -0.104\\
& (-1.533,-0.521) & (-1.486,-0.471) & (-2.168,-1.074) & (-4.775,-3.528) & (-2.010,-0.568) & (-0.214,0.620) & (-1.207,-0.284) & (-0.000,0.000) & (-0.031,0.249) & (-0.280,0.705) & (-0.293,0.537) & (-0.324,0.737) & (-0.404,0.450) & (-0.002,0.004) & (-0.975,0.025) & (-0.762,-0.044) & (-0.315,0.004) & (-2.621,-1.617) & (-0.533,0.282) \\
\multicolumn{1}{c}{\multirow{2}{*}{SIZE}} & -1.853& 0.221& -0.946& 0.687& 0.394& 0.228& 0.779& -1.584& -2.402& 1.246& 0.722& -0.337& -0.342& 0.211& -3.012& 1.014& 0.558& -0.961& 1.439\\
& (-2.333,-0.685) & (-1.141,0.128) & (-0.618,1.069) & (0.042,1.281) & (-0.635,1.624) & (-0.033,0.844) & (0.559,1.649) & (-2.133,-1.225) & (-2.499,-0.800) & (-0.217,2.637) & (-0.602,1.033) & (-1.233,0.102) & (-0.400,-0.042) & (-0.289,1.266) & (-3.923,-1.705) & (0.428,2.745) & (0.077,1.005) & (-0.574,0.623) & (0.437,1.684) \\
\multicolumn{1}{c}{\multirow{2}{*}{MM}} & 18.739& 0.459& -6.417& 12.946& 9.352& -2.369& 1.805& 0.893& -8.130& 5.641& -1.648& 0.145& -0.555& -0.504& -0.796& 0.470& -0.492& 8.684& -11.421\\
& (11.500,21.613) & (-1.346,6.252) & (-17.247,2.584) & (9.701,19.381) & (-2.890,11.002) & (-6.649,1.025) & (-0.543,5.387) & (0.716,1.954) & (-12.256,-6.441) & (1.580,6.941) & (-4.143,-1.265) & (-4.200,1.809) & (-3.485,2.361) & (-3.486,8.583) & (-5.450,2.629) & (0.202,1.696) & (-1.003,0.082) & (4.172,9.289) & (-14.582,-3.803) \\
\multicolumn{1}{c}{\multirow{2}{*}{$\beta$}} & 0.138& 0.116& 0.147& 0.140& 0.107& 0.243& 0.112& 0.153& 0.262& 0.327& 0.252& 0.297& 0.340& 0.219& 0.153& 0.194& 0.179& 0.353& 0.272\\
& (0.103,0.158) & (0.103,0.138) & (0.113,0.168) & (0.086,0.169) & (0.097,0.142) & (0.207,0.256) & (0.114,0.148) & (0.120,0.181) & (0.218,0.287) & (0.187,0.324) & (0.162,0.244) & (0.294,0.354) & (0.310,0.421) & (0.193,0.290) & (0.086,0.193) & (0.164,0.216) & (0.157,0.233) & (0.277,0.370) & (0.220,0.315) \\
\multicolumn{1}{c}{\multirow{2}{*}{$\sigma_k$}} & 0.076  & 0.076  & 0.072  & 0.081  & 0.076  & 0.076  & 0.065  & 0.073  & 0.071  & 0.080  & 0.079  & 0.066  & 0.059  & 0.085  & 0.075  & 0.073  & 0.074  & 0.065  & 0.081  \\
& (0.071,0.085) & (0.068,0.082) & (0.071,0.085) & (0.070,0.084) & (0.067,0.080) & (0.063,0.076) & (0.059,0.071) & (0.067,0.081) & (0.065,0.078) & (0.074,0.088) & (0.073,0.087) & (0.058,0.070) & (0.059,0.071) & (0.074,0.088) & (0.072,0.086) & (0.066,0.078) & (0.070,0.084) & (0.062,0.074) & (0.070,0.084) \\
      \bottomrule
\end{tabular}}
\caption{Parameter estimates obtained by fitting the CoVaR model to each of the 24 assets vs SP500 and all the exogenous variables, for the confidence levels $\tau=0.025$. For each regressor the first row reports parameter estimates by Maximum a Posteriori, while the second row reports the 95\% High Posterior Density (HPD) credible sets.}
\label{tab:Model_Static_ParEst_0025}
\end{small}
\smallskip
\end{sidewaystable}
%

%
\begin{sidewaystable}[!h]
\captionsetup{font={small}, labelfont=sc}
\centering
 \smallskip
 \begin{small}
 \resizebox{\columnwidth}{!}{%
 \setlength\tabcolsep{1pt}
  \begin{tabular}{cccccccccccccccccccc}\\
  \toprule
\multicolumn{1}{c}{\multirow{2}{*}{$\text{VaR}$}} & \multicolumn{9}{c}{Financial} & \multicolumn{2}{c}{Consumer} & \multicolumn{2}{c}{Energy} & \multicolumn{2}{c}{Industrial} & \multicolumn{2}{c}{Technology} & \multicolumn{2}{c}{Utilities}\\
\cmidrule(lr){2-10}\cmidrule(lr){11-12} \cmidrule(l){13-14} \cmidrule(l){15-16} \cmidrule(l){17-18} \cmidrule(l){19-20}
& C & BAC & CMA & JPM & KEY & GS & MS & MCO & AXP & MCD & MKE & CVX & XOM & BA & GE & INTC & ORCL & AEE & PEG\\
\cline{2-20}
\multicolumn{1}{c}{\multirow{2}{*}{CONST}} & -30.567  & -22.619  & -15.032  & -29.325  & -22.208  & -3.429  & -42.321  & 20.671  & -21.941  & -19.800  & -7.908  & 2.736  & -5.371  & -24.596  & -40.958  & -17.548  & 8.218  & -5.352  & -10.077  \\
& (-40.095,-5.584) & (-43.424,-9.687) & (-23.739,-8.861) & (-35.705,-2.935) & (-42.447,-5.758) & (-15.338,7.775) & (-51.416,-15.868) & (-1.387,21.765) & (-43.386,-12.236) & (-28.864,5.137) & (-19.865,15.642) & (-24.191,13.474) & (-11.947,15.798) & (-36.554,-12.861) & (-49.954,-15.272) & (-27.708,6.410) & (-34.212,-3.006) & (-12.458,14.775) & (-20.598,1.331) \\
\multicolumn{1}{c}{\multirow{2}{*}{VIX}} & -0.571  & -0.416  & -0.397  & -0.339  & -0.434  & -0.313  & -0.275  & -0.359  & -0.334  & -0.084  & -0.128  & -0.311  & -0.203  & -0.090  & -0.284  & -0.128  & -0.191  & -0.244  & -0.144  \\
& (-0.853,-0.424) & (-0.776,-0.446) & (-0.409,-0.222) & (-0.464,-0.313) & (-0.646,-0.366) & (-0.297,-0.169) & (-0.443,-0.233) & (-0.396,-0.264) & (-0.352,-0.233) & (-0.109,-0.070) & (-0.158,-0.089) & (-0.304,-0.186) & (-0.256,-0.120) & (-0.157,-0.077) & (-0.331,-0.219) & (-0.169,-0.087) & (-0.238,-0.172) & (-0.292,-0.161) & (-0.229,-0.096) \\
\multicolumn{1}{c}{\multirow{2}{*}{LIQSPR}} & 0.038  & 0.005  & -0.048  & 0.008  & 0.006  & 0.035  & -0.032  & 0.038  & 0.008  & 0.014  & 0.000  & -0.006  & 0.046  & -0.023  & 0.002  & -0.005  & -0.048  & 0.004  & 0.010  \\
& (-0.020,0.058) & (0.015,0.090) & (-0.056,-0.013) & (-0.017,0.027) & (-0.042,0.025) & (-0.002,0.050) & (-0.079,0.007) & (0.001,0.040) & (-0.011,0.038) & (0.007,0.031) & (-0.024,0.040) & (-0.041,0.006) & (0.020,0.068) & (-0.053,-0.002) & (-0.027,0.024) & (-0.005,0.035) & (-0.045,0.005) & (-0.016,0.033) & (-0.038,0.032) \\
\multicolumn{1}{c}{\multirow{2}{*}{3MTB}} & -0.051  & -0.051  & -0.065  & -0.011  & -0.104  & 0.105  & 0.132  & 0.102  & -0.050  & -0.002  & 0.084  & 0.013  & 0.013  & 0.004  & 0.041  & 0.126  & 0.159  & -0.073  & 0.021  \\
& (-0.137,0.058) & (-0.161,0.048) & (-0.110,-0.011) & (-0.079,0.030) & (-0.134,0.045) & (0.062,0.191) & (0.036,0.191) & (0.052,0.139) & (-0.107,0.019) & (-0.027,0.023) & (0.017,0.123) & (0.018,0.135) & (-0.024,0.108) & (-0.044,0.047) & (-0.023,0.059) & (0.080,0.157) & (0.070,0.165) & (-0.094,-0.014) & (-0.014,0.077) \\
\multicolumn{1}{c}{\multirow{2}{*}{TERMSPR}} & 0.071  & 0.058  & 0.081  & 0.008  & -0.075  & 0.042  & 0.028  & 0.087  & 0.007  & 0.001  & 0.056  & -0.010  & -0.014  & -0.012  & 0.078  & 0.121  & 0.108  & -0.084  & -0.023  \\
& (0.022,0.141) & (-0.078,0.062) & (0.017,0.098) & (-0.027,0.054) & (-0.081,0.016) & (0.021,0.112) & (-0.028,0.090) & (0.047,0.120) & (-0.040,0.024) & (-0.023,0.012) & (0.015,0.097) & (-0.010,0.078) & (-0.021,0.056) & (-0.033,0.034) & (0.017,0.080) & (0.067,0.133) & (0.039,0.108) & (-0.121,-0.053) & (-0.062,0.029) \\
\multicolumn{1}{c}{\multirow{2}{*}{CREDSPR}} & -0.088  & -0.073  & -0.022  & -0.011  & -0.090  & -0.082  & -0.158  & 0.087  & -0.062  & -0.057  & 0.014  & -0.051  & -0.055  & -0.037  & 0.044  & 0.068  & 0.012  & -0.151  & -0.024  \\
& (-0.189,0.004) & (-0.134,0.070) & (-0.202,-0.056) & (-0.083,0.038) & (-0.066,0.113) & (-0.162,-0.003) & (-0.254,-0.084) & (0.003,0.114) & (-0.118,-0.024) & (-0.080,-0.035) & (-0.043,0.059) & (-0.090,0.033) & (-0.084,-0.008) & (-0.163,0.002) & (-0.001,0.080) & (0.001,0.102) & (-0.007,0.095) & (-0.185,-0.101) & (-0.087,0.020) \\
\multicolumn{1}{c}{\multirow{2}{*}{DJUSRE}} & 0.531  & 0.450  & 0.376  & 0.819  & 0.520  & 0.419  & 0.533  & 0.449  & 0.493  & 0.168  & 0.384  & 0.246  & 0.249  & 0.601  & 0.436  & 0.330  & 0.072  & 0.295  & 0.223  \\
& (0.350,0.733) & (0.346,0.651) & (0.334,0.655) & (0.570,0.839) & (0.331,0.702) & (0.342,0.570) & (0.449,0.820) & (0.348,0.638) & (0.487,0.655) & (0.104,0.202) & (0.330,0.570) & (0.146,0.326) & (0.150,0.337) & (0.562,0.722) & (0.334,0.523) & (0.231,0.419) & (0.056,0.266) & (0.295,0.450) & (0.123,0.324) \\
\multicolumn{1}{c}{\multirow{2}{*}{LEV}} & -0.513  & -0.426  & 0.565  & 0.025  & -0.972  & -0.220  & -0.122  & 0.009  & -0.020  & 3.318  & -2.905  & -4.431  & 4.910  & 0.029  & 0.691  & -5.424  & 4.029  & -3.167  & -0.068  \\
& (-0.812,-0.108) & (-0.239,1.553) & (-0.275,0.730) & (-0.197,0.474) & (-1.278,-0.258) & (-0.285,-0.069) & (-0.224,0.039) & (-0.021,0.031) & (-0.401,0.347) & (-0.643,5.836) & (-10.835,2.790) & (-7.671,1.642) & (-2.016,9.679) & (0.006,0.038) & (-0.807,0.915) & (-12.964,-1.582) & (2.517,9.632) & (-8.613,-0.274) & (-0.672,0.450) \\
\multicolumn{1}{c}{\multirow{2}{*}{MK2BK}} & -1.548  & 2.418  & -0.225  & 1.114  & -0.654  & 2.424  & -4.885  & -0.000  & -0.224  & -0.375  & 0.124  & -1.514  & -1.044  & -0.001  & -2.181  & -0.742  & -0.564  & -2.288  & -2.250  \\
& (-2.669,0.452) & (-1.796,1.546) & (-0.570,1.529) & (-1.574,2.181) & (-3.448,0.637) & (0.778,3.299) & (-5.854,-0.994) & (-0.000,0.000) & (-0.543,0.171) & (-0.536,0.491) & (-0.816,1.518) & (-3.069,0.710) & (-2.365,-0.464) & (-0.007,0.008) & (-2.650,-0.889) & (-2.200,-0.673) & (-0.616,0.019) & (-4.353,-0.753) & (-3.168,-0.808) \\
\multicolumn{1}{c}{\multirow{2}{*}{SIZE}} & 3.796  & 2.604  & 1.262  & 2.546  & 3.932  & 0.811  & 3.450  & -2.326  & 2.254  & 1.105  & 0.925  & 0.984  & -0.049  & 2.138  & 3.495  & 2.136  & -1.190  & 2.053  & 1.365  \\
& (1.785,4.808) & (0.987,3.809) & (0.633,2.376) & (0.322,2.952) & (2.099,6.253) & (-0.492,1.409) & (1.129,4.616) & (-2.402,-0.003) & (1.369,4.078) & (-1.053,1.796) & (-0.812,1.816) & (-0.030,2.663) & (-1.449,0.795) & (1.066,3.274) & (1.610,4.523) & (0.181,3.605) & (-0.554,1.967) & (0.995,2.441) & (0.197,2.503) \\
\multicolumn{1}{c}{\multirow{2}{*}{MM}} & -7.724  & -19.824  & -36.894  & -6.028  & 12.850  & -12.917  & 29.372  & -5.023  & -5.437  & -2.498  & -1.230  & 12.417  & 9.032  & 19.721  & -4.243  & 1.605  & -0.012  & -0.605  & 3.511  \\
& (-30.441,1.276) & (-33.085,-10.879) & (-47.880,-13.141) & (-12.368,12.512) & (-7.523,22.752) & (-15.137,3.840) & (9.365,32.212) & (-7.120,-3.242) & (-12.994,2.934) & (-7.179,-0.608) & (-3.759,3.553) & (6.140,22.127) & (1.043,11.944) & (4.277,27.544) & (-10.316,2.410) & (-1.659,4.307) & (-1.844,1.166) & (-9.528,7.868) & (-9.541,17.568) \\
\multicolumn{1}{c}{\multirow{2}{*}{$\sigma_j$}} & 0.653  & 0.521  & 0.415  & 0.349  & 0.525  & 0.382  & 0.528  & 0.515  & 0.337  & 0.232  & 0.350  & 0.357  & 0.305  & 0.353  & 0.279  & 0.409  & 0.348  & 0.257  & 0.320  \\
& (0.591,0.708) & (0.489,0.588) & (0.359,0.431) & (0.343,0.411) & (0.484,0.580) & (0.350,0.420) & (0.489,0.588) & (0.440,0.528) & (0.290,0.347) & (0.212,0.254) & (0.311,0.373) & (0.306,0.367) & (0.258,0.310) & (0.313,0.375) & (0.259,0.312) & (0.351,0.421) & (0.320,0.384) & (0.254,0.304) & (0.304,0.365) \\
\hline
\multicolumn{1}{c}{\multirow{2}{*}{$\text{CoVaR}$}} & \multicolumn{9}{c}{Financial} & \multicolumn{2}{c}{Consumer} & \multicolumn{2}{c}{Energy} & \multicolumn{2}{c}{Industrial} & \multicolumn{2}{c}{Technology} & \multicolumn{2}{c}{Utilities}\\
\cmidrule(lr){2-10}\cmidrule(lr){11-12} \cmidrule(l){13-14} \cmidrule(l){15-16} \cmidrule(l){17-18} \cmidrule(l){19-20}
& C & BAC & CMA & JPM & KEY & GS & MS & MCO & AXP & MCD & MKE & CVX & XOM & BA & GE & INTC & ORCL & AEE & PEG\\
\cline{2-20}
\multicolumn{1}{c}{\multirow{2}{*}{CONST}} & 19.181& 5.401& 3.905& -0.906& 3.785& -5.954& 1.791& 16.287& 24.691& -2.188& -18.412& 12.451& -11.071& -0.134& 18.708& -4.144& -7.104& -2.633& -10.080\\
& (6.034,20.256) & (-1.032,9.738) & (-2.861,10.222) & (-11.724,2.908) & (-6.847,12.108) & (-8.639,1.038) & (-10.801,4.484) & (13.990,24.617) & (7.882,30.063) & (-19.552,9.145) & (-20.668,2.366) & (6.583,21.842) & (-17.948,-7.645) & (-12.840,1.177) & (4.842,36.457) & (-15.879,5.770) & (-11.422,0.227) & (-9.876,7.335) & (-11.788,3.139) \\
\multicolumn{1}{c}{\multirow{2}{*}{VIX}} & -0.167& -0.168& -0.126& -0.164& -0.167& -0.115& -0.096& -0.212& -0.229& -0.105& -0.106& -0.075& -0.103& -0.078& -0.193& -0.083& -0.092& -0.130& -0.120\\
& (-0.181,-0.117) & (-0.179,-0.103) & (-0.161,-0.093) & (-0.196,-0.148) & (-0.189,-0.130) & (-0.149,-0.106) & (-0.099,-0.051) & (-0.249,-0.164) & (-0.248,-0.177) & (-0.128,-0.081) & (-0.116,-0.080) & (-0.087,-0.055) & (-0.133,-0.093) & (-0.104,-0.057) & (-0.226,-0.146) & (-0.108,-0.069) & (-0.107,-0.072) & (-0.171,-0.129) & (-0.127,-0.090) \\
\multicolumn{1}{c}{\multirow{2}{*}{LIQSPR}} & 0.006& 0.007& 0.025& 0.012& 0.003& 0.002& 0.009& 0.017& 0.012& 0.002& -0.005& 0.007& 0.013& 0.018& 0.025& 0.001& 0.004& 0.017& 0.012\\
& (-0.002,0.013) & (-0.006,0.013) & (0.006,0.027) & (-0.000,0.014) & (0.001,0.016) & (-0.000,0.017) & (0.001,0.016) & (0.008,0.024) & (0.002,0.020) & (-0.006,0.013) & (-0.017,-0.001) & (0.003,0.012) & (-0.002,0.012) & (0.000,0.016) & (0.014,0.034) & (-0.005,0.008) & (-0.005,0.010) & (0.009,0.022) & (0.004,0.017) \\
\multicolumn{1}{c}{\multirow{2}{*}{3MTB}} & 0.050& 0.030& 0.058& 0.035& 0.037& 0.005& 0.044& 0.035& 0.081& 0.023& 0.028& 0.022& 0.019& 0.002& 0.022& 0.049& 0.035& 0.036& 0.038\\
& (0.015,0.059) & (0.011,0.059) & (0.032,0.081) & (0.022,0.064) & (0.022,0.060) & (-0.021,0.016) & (0.021,0.063) & (0.029,0.073) & (0.030,0.083) & (0.014,0.056) & (0.012,0.056) & (0.015,0.043) & (-0.002,0.033) & (-0.011,0.037) & (0.002,0.052) & (0.011,0.040) & (0.017,0.060) & (0.009,0.042) & (0.022,0.064) \\
\multicolumn{1}{c}{\multirow{2}{*}{TERMSPR}} & 0.026& 0.026& 0.041& 0.025& 0.038& 0.003& 0.021& 0.026& 0.029& 0.021& 0.030& 0.026& 0.012& 0.005& 0.019& 0.027& 0.018& 0.035& 0.047\\
& (0.010,0.034) & (-0.000,0.035) & (0.020,0.054) & (0.013,0.038) & (0.020,0.051) & (-0.006,0.018) & (0.011,0.033) & (0.028,0.058) & (0.016,0.043) & (0.011,0.043) & (0.020,0.046) & (0.014,0.035) & (0.002,0.028) & (0.001,0.033) & (0.004,0.036) & (0.014,0.036) & (0.004,0.029) & (0.012,0.038) & (0.032,0.059) \\
\multicolumn{1}{c}{\multirow{2}{*}{CREDSPR}} & 0.029& -0.015& 0.006& 0.006& 0.013& -0.004& 0.009& -0.033& 0.045& -0.026& -0.000& -0.018& -0.033& -0.013& -0.019& -0.029& -0.048& -0.017& -0.030\\
& (-0.008,0.034) & (-0.030,0.013) & (-0.030,0.019) & (-0.027,0.030) & (-0.033,0.036) & (-0.033,0.004) & (-0.010,0.039) & (-0.030,0.020) & (0.016,0.061) & (-0.040,0.010) & (-0.025,0.025) & (-0.036,-0.004) & (-0.048,-0.004) & (-0.045,0.019) & (-0.034,0.015) & (-0.047,-0.005) & (-0.060,-0.009) & (-0.037,0.019) & (-0.046,-0.007) \\
\multicolumn{1}{c}{\multirow{2}{*}{DJUSRE}} & 0.255& 0.227& 0.284& 0.335& 0.220& 0.256& 0.316& 0.208& 0.219& 0.338& 0.330& 0.301& 0.324& 0.273& 0.270& 0.309& 0.312& 0.184& 0.268\\
& (0.222,0.311) & (0.223,0.318) & (0.243,0.332) & (0.258,0.337) & (0.188,0.268) & (0.233,0.298) & (0.304,0.364) & (0.190,0.280) & (0.138,0.246) & (0.292,0.385) & (0.271,0.347) & (0.242,0.324) & (0.271,0.346) & (0.259,0.354) & (0.221,0.310) & (0.254,0.330) & (0.277,0.353) & (0.167,0.244) & (0.249,0.318) \\
\multicolumn{1}{c}{\multirow{2}{*}{LEV}} & 0.006& -0.027& -0.077& 0.064& -0.045& -0.051& -0.017& -0.011& 0.218& -1.148& 3.143& -3.680& 6.780& 0.006& 0.258& -2.954& 0.064& 0.535& -0.263\\
& (-0.046,0.037) & (-0.121,0.249) & (-0.240,0.028) & (0.032,0.237) & (-0.222,0.021) & (-0.114,-0.036) & (-0.054,0.003) & (-0.011,0.003) & (0.083,0.441) & (-3.442,1.094) & (-3.116,3.717) & (-5.408,-2.476) & (5.551,9.462) & (-0.000,0.009) & (-0.006,0.540) & (-4.081,0.188) & (-2.276,0.479) & (-0.317,3.537) & (-0.671,-0.104) \\
\multicolumn{1}{c}{\multirow{2}{*}{MK2BK}} & -0.283& -0.843& -1.236& -3.108& -0.820& 0.135& -0.361& -0.000& 0.069& 0.377& -0.055& -0.064& 0.169& 0.002& -0.689& -0.239& -0.069& -2.004& -0.249\\
& (-0.922,-0.118) & (-1.172,-0.318) & (-1.708,-0.632) & (-3.712,-2.428) & (-1.543,-0.184) & (-0.200,0.541) & (-0.913,-0.075) & (-0.000,0.000) & (-0.003,0.297) & (-0.111,0.742) & (-0.355,0.407) & (-0.449,0.510) & (-0.385,0.216) & (-0.003,0.003) & (-1.158,0.062) & (-0.592,0.002) & (-0.223,0.057) & (-2.553,-1.282) & (-0.507,0.270) \\
\multicolumn{1}{c}{\multirow{2}{*}{SIZE}} & -1.700& -0.245& -0.150& 0.185& -0.125& 0.783& -0.075& -1.540& -2.301& 0.281& 1.294& -0.479& -0.216& 0.004& -1.381& 0.776& 0.621& 0.410& 1.264\\
& (-1.734,-0.546) & (-0.753,0.146) & (-0.809,0.708) & (-0.151,0.998) & (-0.998,1.093) & (0.190,1.054) & (-0.381,1.071) & (-2.384,-1.355) & (-2.841,-0.889) & (-0.616,1.799) & (-0.013,1.608) & (-1.094,-0.101) & (-0.355,0.126) & (-0.109,1.182) & (-2.895,-0.288) & (-0.267,1.791) & (0.196,1.090) & (-0.751,0.546) & (-0.100,1.419) \\
\multicolumn{1}{c}{\multirow{2}{*}{MM}} & 10.060& -0.078& -10.758& 12.859& 3.783& -3.935& -0.132& 1.050& -8.901& -0.244& -2.492& -0.534& 1.287& 7.242& 0.405& 0.888& -0.532& 5.599& -10.513\\
& (3.613,13.537) & (-2.005,3.748) & (-16.063,3.034) & (7.380,17.038) & (-3.388,8.257) & (-7.353,0.088) & (-1.542,3.406) & (0.174,1.486) & (-12.656,-6.095) & (-2.352,2.512) & (-3.112,-0.722) & (-2.232,2.380) & (-3.180,0.446) & (1.734,11.166) & (-3.810,3.196) & (-0.032,1.554) & (-0.726,0.290) & (1.983,8.818) & (-15.061,-0.298) \\
\multicolumn{1}{c}{\multirow{2}{*}{$\beta$}} & 0.102& 0.124& 0.101& 0.123& 0.136& 0.222& 0.143& 0.153& 0.245& 0.282& 0.169& 0.347& 0.386& 0.230& 0.097& 0.174& 0.154& 0.328& 0.208\\
& (0.082,0.124) & (0.093,0.131) & (0.098,0.166) & (0.102,0.161) & (0.108,0.159) & (0.184,0.239) & (0.118,0.151) & (0.105,0.181) & (0.199,0.292) & (0.189,0.304) & (0.147,0.217) & (0.316,0.388) & (0.308,0.398) & (0.171,0.244) & (0.084,0.170) & (0.170,0.225) & (0.147,0.211) & (0.275,0.356) & (0.156,0.268) \\
\multicolumn{1}{c}{\multirow{2}{*}{$\sigma_k$}} & 0.130  & 0.124  & 0.136  & 0.133  & 0.128  & 0.120  & 0.116  & 0.144  & 0.133  & 0.132  & 0.140  & 0.110  & 0.110  & 0.133  & 0.142  & 0.119  & 0.135  & 0.119  & 0.140  \\
& (0.122,0.146) & (0.121,0.145) & (0.124,0.148) & (0.120,0.144) & (0.120,0.144) & (0.113,0.135) & (0.106,0.127) & (0.123,0.147) & (0.118,0.141) & (0.125,0.150) & (0.120,0.144) & (0.100,0.120) & (0.103,0.124) & (0.126,0.151) & (0.126,0.152) & (0.111,0.132) & (0.120,0.143) & (0.113,0.135) & (0.124,0.148) \\
      \bottomrule
\end{tabular}}
\caption{Parameter estimates obtained by fitting the CoVaR model to each of the 24 assets vs SP500 and all the exogenous variables, for the confidence levels $\tau=0.05$. For each regressor the first row reports parameter estimates by Maximum a Posteriori, while the second row reports the 95\% High Posterior Density (HPD) credible sets.}
\label{tab:Model_Static_ParEst_005}
\end{small}
\smallskip
\end{sidewaystable}
%

%
\begin{sidewaystable}[!h]
\captionsetup{font={small}, labelfont=sc}
\centering
 \smallskip
 \begin{small}
 \resizebox{\columnwidth}{!}{%
 \setlength\tabcolsep{1pt}
  \begin{tabular}{cccccccccccccccccccc}\\
  \toprule
\multicolumn{1}{c}{\multirow{2}{*}{$\text{VaR}$}} & \multicolumn{9}{c}{Financial} & \multicolumn{2}{c}{Consumer} & \multicolumn{2}{c}{Energy} & \multicolumn{2}{c}{Industrial} & \multicolumn{2}{c}{Technology} & \multicolumn{2}{c}{Utilities}\\
\cmidrule(lr){2-10}\cmidrule(lr){11-12} \cmidrule(l){13-14} \cmidrule(l){15-16} \cmidrule(l){17-18} \cmidrule(l){19-20}
& C & BAC & CMA & JPM & KEY & GS & MS & MCO & AXP & MCD & MKE & CVX & XOM & BA & GE & INTC & ORCL & AEE & PEG\\
\cline{2-20}
\multicolumn{1}{c}{\multirow{2}{*}{VIX}} & -0.372  & -0.402  & -0.366  & -0.143  & -0.606  & -0.275  & -0.715  & -0.034  & -0.211  & -0.178  & -0.257  & -0.401  & -0.308  & -0.411  & -0.082  & -0.074  & -0.226  & -0.131  & -0.164  \\
& (-0.602,-0.120) & (-0.464,-0.084) & (-0.469,-0.314) & (-0.256,-0.025) & (-0.824,-0.460) & (-0.400,-0.220) & (-1.019,-0.668) & (-0.018,0.900) & (-0.283,-0.130) & (-0.206,-0.114) & (-0.298,-0.156) & (-0.514,-0.350) & (-0.418,-0.282) & (-0.511,-0.262) & (-0.164,-0.010) & (-0.086,0.069) & (-0.317,-0.167) & (-0.331,-0.081) & (-0.179,0.016) \\
\multicolumn{1}{c}{\multirow{2}{*}{LIQSPR}} & -0.010  & 0.051  & -0.008  & 0.027  & 0.094  & 0.004  & 0.065  & -0.012  & -0.013  & 0.045  & 0.027  & 0.038  & 0.017  & 0.005  & -0.052  & 0.047  & 0.047  & 0.024  & 0.015  \\
& (-0.046,0.051) & (-0.021,0.062) & (-0.031,0.018) & (0.006,0.069) & (0.045,0.136) & (-0.048,0.039) & (0.037,0.140) & (-0.026,0.124) & (-0.036,0.024) & (0.030,0.060) & (-0.017,0.049) & (-0.009,0.063) & (-0.019,0.067) & (-0.048,0.026) & (-0.068,-0.005) & (0.002,0.082) & (-0.024,0.059) & (0.008,0.078) & (-0.043,0.077) \\
\multicolumn{1}{c}{\multirow{2}{*}{3MTB}} & 0.126  & 0.017  & -0.038  & 0.088  & -0.054  & 0.038  & -0.002  & 0.104  & -0.039  & -0.020  & 0.090  & 0.062  & 0.021  & 0.072  & 0.028  & 0.076  & 0.100  & -0.030  & 0.043  \\
& (0.036,0.186) & (-0.054,0.074) & (-0.097,0.006) & (-0.015,0.088) & (-0.118,-0.012) & (0.031,0.136) & (-0.087,0.060) & (0.024,0.215) & (-0.070,0.017) & (-0.030,0.023) & (0.010,0.095) & (-0.007,0.108) & (-0.019,0.109) & (0.016,0.102) & (0.008,0.081) & (0.015,0.105) & (0.028,0.119) & (-0.043,0.035) & (-0.006,0.091) \\
\multicolumn{1}{c}{\multirow{2}{*}{TERMSPR}} & 0.143  & 0.066  & 0.113  & 0.107  & 0.004  & 0.050  & 0.018  & 0.084  & 0.029  & -0.015  & 0.063  & 0.026  & 0.028  & 0.064  & 0.062  & 0.065  & 0.068  & -0.026  & 0.021  \\
& (0.060,0.194) & (-0.011,0.099) & (0.062,0.124) & (0.044,0.116) & (-0.048,0.027) & (0.039,0.134) & (-0.046,0.056) & (0.054,0.179) & (-0.001,0.059) & (-0.029,0.018) & (0.009,0.084) & (-0.004,0.060) & (-0.009,0.070) & (0.028,0.090) & (0.034,0.079) & (0.032,0.090) & (0.013,0.082) & (-0.043,0.015) & (-0.021,0.043) \\
\multicolumn{1}{c}{\multirow{2}{*}{CREDSPR}} & 0.042  & 0.135  & -0.006  & 0.121  & 0.067  & 0.014  & -0.068  & 0.073  & -0.107  & -0.009  & 0.079  & 0.047  & 0.051  & 0.020  & -0.011  & 0.129  & 0.107  & -0.164  & -0.067  \\
& (-0.029,0.169) & (-0.078,0.126) & (-0.074,0.058) & (0.024,0.150) & (-0.036,0.114) & (-0.087,0.063) & (-0.130,0.050) & (0.045,0.230) & (-0.168,-0.079) & (-0.027,0.068) & (0.019,0.122) & (0.000,0.108) & (0.006,0.144) & (-0.073,0.068) & (-0.050,0.035) & (0.063,0.163) & (0.024,0.146) & (-0.154,-0.033) & (-0.117,-0.002) \\
\multicolumn{1}{c}{\multirow{2}{*}{DJUSRE}} & 0.821  & 0.443  & 0.326  & 0.621  & 0.574  & 0.543  & 0.632  & 0.581  & 0.585  & 0.130  & 0.468  & 0.221  & 0.326  & 0.431  & 0.512  & 0.317  & 0.216  & 0.295  & 0.337  \\
& (0.588,0.980) & (0.387,0.690) & (0.309,0.539) & (0.541,0.830) & (0.396,0.671) & (0.380,0.637) & (0.524,0.797) & (0.497,0.863) & (0.485,0.672) & (0.061,0.205) & (0.337,0.518) & (0.122,0.276) & (0.171,0.344) & (0.354,0.527) & (0.406,0.571) & (0.210,0.374) & (0.136,0.329) & (0.229,0.399) & (0.278,0.476) \\
\multicolumn{1}{c}{\multirow{2}{*}{LEV}} & 4.010  & -2.095  & -2.126  & -5.057  & -9.054  & 3.715  & 0.784  & 13.564  & 4.993  & -3.784  & -3.996  & -8.426  & 1.008  & 0.872  & 1.752  & 10.882  & 3.213  & 29.995  & 7.278  \\
& (3.060,5.498) & (-2.776,0.234) & (-2.110,-0.036) & (-8.468,-5.457) & (-11.707,-5.887) & (3.426,4.314) & (0.496,1.740) & (13.316,13.575) & (3.357,5.100) & (-5.398,6.451) & (-9.360,1.539) & (-8.127,3.678) & (-4.034,8.182) & (0.875,1.003) & (-2.053,2.712) & (-0.518,13.613) & (-2.475,9.482) & (27.499,33.212) & (0.963,8.038) \\
\multicolumn{1}{c}{\multirow{2}{*}{MK2BK}} & -10.536  & -3.206  & -2.164  & -6.315  & -2.946  & -9.493  & -0.836  & 0.000  & 0.190  & 4.503  & 2.413  & -5.572  & 1.460  & 0.008  & -1.991  & -35.730  & -11.746  & 7.257  & 21.233  \\
& (-20.510,-6.433) & (-10.438,-1.068) & (-5.478,0.780) & (-11.379,-1.873) & (-7.522,-0.918) & (-12.312,-2.865) & (-6.877,3.179) & (0.000,0.001) & (0.914,3.565) & (1.988,5.062) & (-1.859,3.453) & (-8.929,-1.863) & (-3.095,2.299) & (0.003,0.011) & (-4.704,-0.329) & (-45.045,-35.591) & (-11.470,-5.786) & (6.441,15.177) & (15.872,23.850) \\
\multicolumn{1}{c}{\multirow{2}{*}{SIZE}} & 6.666  & 7.570  & 2.617  & 25.876  & 20.569  & 5.981  & 4.095  & 30.641  & 11.211  & 8.860  & 3.074  & 14.158  & 5.035  & 7.428  & 6.276  & 34.560  & 15.592  & 2.974  & 13.221  \\
& (6.514,9.782) & (6.104,8.709) & (0.795,2.780) & (26.485,30.494) & (17.537,24.030) & (4.867,6.426) & (2.716,4.767) & (29.507,30.644) & (10.611,12.669) & (7.376,9.753) & (2.399,4.800) & (12.581,14.745) & (3.936,6.351) & (7.432,7.850) & (5.967,8.552) & (34.792,38.321) & (12.874,16.147) & (1.877,3.235) & (13.295,17.800) \\
\multicolumn{1}{c}{\multirow{2}{*}{MM}} & 2.854  & 3.491  & 2.974  & -2.903  & -3.318  & -3.859  & 1.419  & 2.211  & -2.196  & -0.940  & 0.266  & -4.235  & 3.713  & -1.985  & 1.846  & 8.831  & -5.146  & -1.110  & 2.157  \\
& (-6.838,5.443) & (-5.925,6.244) & (-7.669,4.549) & (-6.271,5.793) & (-6.506,5.477) & (-7.390,4.397) & (-6.704,5.186) & (-2.740,12.412) & (-8.834,2.961) & (-9.186,0.908) & (-7.130,4.815) & (-7.558,3.962) & (-7.566,4.130) & (-5.881,6.317) & (-4.292,6.978) & (2.176,12.974) & (-5.219,3.953) & (-6.208,5.558) & (-5.796,6.345) \\
\multicolumn{1}{c}{\multirow{2}{*}{$\sigma_j$}} & 0.241  & 0.214  & 0.186  & 0.145  & 0.162  & 0.176  & 0.188  & 0.095  & 0.126  & 0.092  & 0.128  & 0.135  & 0.115  & 0.121  & 0.115  & 0.114  & 0.121  & 0.094  & 0.117  \\
& (0.203,0.249) & (0.197,0.239) & (0.171,0.211) & (0.133,0.165) & (0.144,0.190) & (0.143,0.179) & (0.181,0.222) & (0.085,0.115) & (0.114,0.139) & (0.090,0.112) & (0.108,0.135) & (0.122,0.149) & (0.104,0.133) & (0.093,0.115) & (0.093,0.115) & (0.107,0.133) & (0.114,0.141) & (0.086,0.107) & (0.121,0.153) \\
\hline
\multicolumn{1}{c}{\multirow{2}{*}{$\text{CoVaR}$}} & \multicolumn{9}{c}{Financial} & \multicolumn{2}{c}{Consumer} & \multicolumn{2}{c}{Energy} & \multicolumn{2}{c}{Industrial} & \multicolumn{2}{c}{Technology} & \multicolumn{2}{c}{Utilities}\\
\cmidrule(lr){2-10}\cmidrule(lr){11-12} \cmidrule(l){13-14} \cmidrule(l){15-16} \cmidrule(l){17-18} \cmidrule(l){19-20}
& C & BAC & CMA & JPM & KEY & GS & MS & MCO & AXP & MCD & MKE & CVX & XOM & BA & GE & INTC & ORCL & AEE & PEG\\
\cline{2-20}
\multicolumn{1}{c}{\multirow{2}{*}{VIX}} & -0.157& -0.066& -0.137& -0.186& 0.023& -0.141& -0.010& -0.604& -0.402& -0.217& -0.143& -0.073& -0.104& -0.321& -0.297& -0.185& -0.067& -0.165& -0.150\\
& (-0.332,-0.163) & (-0.115,-0.045) & (-0.194,-0.106) & (-0.258,-0.180) & (-0.080,0.030) & (-0.193,-0.118) & (0.033,0.179) & (-0.969,-0.556) & (-0.447,-0.330) & (-0.285,-0.209) & (-0.234,-0.113) & (-0.122,-0.059) & (-0.104,-0.064) & (-0.428,-0.297) & (-0.298,-0.216) & (-0.269,-0.186) & (-0.087,-0.014) & (-0.188,-0.111) & (-0.156,-0.025) \\
\multicolumn{1}{c}{\multirow{2}{*}{LIQSPR}} & 0.000& -0.002& -0.006& 0.004& 0.005& -0.021& 0.005& 0.005& -0.012& -0.004& -0.013& -0.010& -0.009& 0.015& 0.012& 0.002& -0.014& -0.003& 0.017\\
& (-0.032,0.013) & (-0.008,0.016) & (-0.018,0.010) & (0.004,0.028) & (-0.020,0.015) & (-0.039,-0.009) & (-0.032,0.008) & (0.004,0.041) & (-0.018,0.013) & (-0.010,0.017) & (-0.026,-0.002) & (-0.016,0.002) & (-0.015,-0.003) & (-0.027,0.020) & (-0.008,0.021) & (-0.008,0.010) & (-0.024,-0.001) & (-0.014,0.007) & (-0.012,0.025) \\
\multicolumn{1}{c}{\multirow{2}{*}{3MTB}} & 0.031& 0.063& 0.040& 0.040& 0.076& 0.025& 0.013& 0.022& 0.036& 0.051& 0.034& 0.040& 0.040& 0.042& 0.015& 0.037& -0.006& 0.058& 0.044\\
& (-0.003,0.052) & (0.027,0.068) & (0.035,0.073) & (0.020,0.056) & (0.044,0.088) & (0.002,0.042) & (0.004,0.061) & (-0.007,0.045) & (0.011,0.052) & (0.030,0.071) & (0.005,0.039) & (0.020,0.052) & (0.028,0.052) & (0.016,0.063) & (0.006,0.044) & (0.015,0.048) & (0.000,0.045) & (0.050,0.078) & (0.021,0.068) \\
\multicolumn{1}{c}{\multirow{2}{*}{TERMSPR}} & 0.017& 0.024& 0.025& 0.023& 0.065& 0.026& 0.023& 0.003& 0.027& 0.044& 0.035& 0.034& 0.037& 0.040& 0.025& 0.018& 0.016& 0.053& 0.036\\
& (0.009,0.049) & (0.009,0.037) & (0.017,0.045) & (0.010,0.036) & (0.040,0.071) & (0.012,0.040) & (0.017,0.058) & (0.002,0.045) & (0.008,0.037) & (0.033,0.065) & (0.020,0.047) & (0.016,0.044) & (0.029,0.049) & (0.020,0.063) & (0.009,0.036) & (0.010,0.033) & (0.014,0.042) & (0.042,0.065) & (0.027,0.062) \\
\multicolumn{1}{c}{\multirow{2}{*}{CREDSPR}} & 0.027& 0.005& 0.022& -0.007& 0.034& 0.026& 0.025& 0.023& 0.048& 0.038& 0.034& 0.028& 0.030& 0.042& 0.056& 0.008& 0.008& 0.043& 0.033\\
& (-0.026,0.054) & (-0.030,0.025) & (-0.003,0.045) & (-0.021,0.024) & (0.016,0.077) & (0.002,0.054) & (-0.042,0.028) & (0.019,0.100) & (0.015,0.065) & (0.017,0.061) & (0.010,0.056) & (0.008,0.052) & (0.005,0.041) & (0.042,0.108) & (0.007,0.058) & (-0.005,0.036) & (-0.001,0.046) & (0.022,0.064) & (-0.025,0.047) \\
\multicolumn{1}{c}{\multirow{2}{*}{DJUSRE}} & 0.221& 0.276& 0.301& 0.302& 0.361& 0.285& 0.365& 0.251& 0.264& 0.279& 0.332& 0.316& 0.330& 0.251& 0.329& 0.274& 0.230& 0.300& 0.339\\
& (0.145,0.279) & (0.263,0.336) & (0.234,0.332) & (0.252,0.329) & (0.259,0.371) & (0.238,0.313) & (0.306,0.423) & (0.163,0.282) & (0.181,0.272) & (0.275,0.347) & (0.255,0.332) & (0.271,0.336) & (0.288,0.343) & (0.186,0.303) & (0.270,0.358) & (0.254,0.309) & (0.230,0.316) & (0.256,0.333) & (0.298,0.426) \\
\multicolumn{1}{c}{\multirow{2}{*}{LEV}} & 7.634& 0.618& 4.470& 0.063& 10.988& 2.324& 4.355& 5.113& -1.802& 5.172& 7.650& 15.472& -13.124& -3.985& 14.072& 2.518& -0.079& 8.039& -6.957\\
& (7.664,10.884) & (0.383,1.344) & (4.459,5.943) & (-0.680,0.487) & (11.007,16.783) & (2.167,4.758) & (3.473,4.345) & (5.065,5.148) & (-4.170,-1.854) & (-2.759,5.557) & (1.598,14.057) & (11.412,17.700) & (-15.960,-12.372) & (-3.990,-3.962) & (13.437,17.983) & (-7.174,3.374) & (0.497,5.535) & (7.772,14.210) & (-12.426,-6.557) \\
\multicolumn{1}{c}{\multirow{2}{*}{MK2BK}} & 14.821& -0.346& -31.865& -3.014& -25.140& 0.628& -16.423& -0.000& 19.025& 8.285& -1.378& -0.858& -1.336& -0.003& -17.047& 0.554& -0.257& -6.849& 0.253\\
& (14.334,26.435) & (-2.572,0.101) & (-42.140,-31.434) & (-6.063,-0.162) & (-43.667,-25.572) & (-22.032,2.998) & (-28.846,-15.475) & (-0.000,0.000) & (18.810,23.131) & (8.536,12.628) & (-6.127,-0.223) & (-2.193,0.426) & (-1.808,-0.253) & (-0.006,-0.001) & (-22.498,-16.637) & (0.544,6.286) & (-2.604,0.236) & (-21.980,-7.119) & (-6.298,0.724) \\
\multicolumn{1}{c}{\multirow{2}{*}{SIZE}} & -1.482& 0.537& -0.628& -0.360& 8.400& 0.132& 7.960& -13.493& -15.499& -0.241& 3.798& 3.458& -0.169& -4.621& 0.181& 0.776& 16.263& -0.116& 0.330\\
& (-6.462,-1.366) & (0.080,1.116) & (-1.121,-0.179) & (-0.777,-0.085) & (3.858,8.642) & (-1.006,0.259) & (7.821,11.076) & (-13.792,-13.138) & (-15.773,-14.937) & (-1.371,0.580) & (2.302,6.045) & (3.002,4.214) & (-0.347,-0.037) & (-4.685,-4.427) & (-1.315,0.517) & (0.063,1.322) & (15.511,16.903) & (-0.570,0.313) & (0.058,3.843) \\
\multicolumn{1}{c}{\multirow{2}{*}{MM}} & 6.625& 1.549& 7.691& -2.843& -1.273& 1.333& 0.624& -13.168& 3.921& -0.965& 0.029& 1.628& 1.932& -1.984& 0.522& -6.696& 0.007& -1.442& -2.947\\
& (-3.881,8.264) & (-4.259,5.645) & (-4.822,7.239) & (-5.071,5.153) & (-7.136,5.921) & (-8.886,3.295) & (-5.515,7.515) & (-24.580,-12.211) & (-1.947,8.920) & (-5.739,5.604) & (-7.708,4.510) & (-6.276,4.218) & (-3.530,6.667) & (-4.162,7.792) & (-7.930,2.710) & (-14.254,-3.877) & (-2.214,0.916) & (-7.059,4.074) & (-6.749,4.022) \\
\multicolumn{1}{c}{\multirow{2}{*}{$\sigma_k$}} & 0.068  & 0.054  & 0.042  & 0.057  & 0.055  & 0.051  & 0.073  & 0.035  & 0.042  & 0.049  & 0.056  & 0.050  & 0.044  & 0.033  & 0.050  & 0.045  & 0.060  & 0.051  & 0.057  \\
& (0.061,0.079) & (0.049,0.060) & (0.039,0.048) & (0.050,0.062) & (0.049,0.061) & (0.047,0.059) & (0.069,0.087) & (0.033,0.044) & (0.037,0.047) & (0.043,0.053) & (0.048,0.062) & (0.045,0.055) & (0.043,0.052) & (0.029,0.038) & (0.043,0.054) & (0.042,0.052) & (0.058,0.071) & (0.044,0.055) & (0.051,0.064) \\
      \bottomrule
\end{tabular}}
\caption{Parameter estimates obtained by fitting the Dynamic CoVaR model to each of the 19 assets vs SP500 and all the exogenous variables, for the confidence levels $\tau=0.025$. For each regressor the first row reports parameter estimates by Maximum a Posteriori, while the second row reports the 95\% High Posterior Density (HPD) credible sets.}
\label{tab:Model_Dynamic_ParEst_0025}
\end{small}
\smallskip
\end{sidewaystable}
%

%
\begin{sidewaystable}[!h]
\captionsetup{font={small}, labelfont=sc}
\centering
 \smallskip
 \begin{small}
 \resizebox{\columnwidth}{!}{%
 \setlength\tabcolsep{1pt}
  \begin{tabular}{cccccccccccccccccccc}\\
  \toprule
\multicolumn{1}{c}{\multirow{2}{*}{$\text{VaR}$}} & \multicolumn{9}{c}{Financial} & \multicolumn{2}{c}{Consumer} & \multicolumn{2}{c}{Energy} & \multicolumn{2}{c}{Industrial} & \multicolumn{2}{c}{Technology} & \multicolumn{2}{c}{Utilities}\\
\cmidrule(lr){2-10}\cmidrule(lr){11-12} \cmidrule(l){13-14} \cmidrule(l){15-16} \cmidrule(l){17-18} \cmidrule(l){19-20}
& C & BAC & CMA & JPM & KEY & GS & MS & MCO & AXP & MCD & MKE & CVX & XOM & BA & GE & INTC & ORCL & AEE & PEG\\
\cline{2-20}
\multicolumn{1}{c}{\multirow{2}{*}{VIX}} & -0.289  & -0.449  & -0.328  & -0.255  & -0.575  & -0.195  & -0.415  & -0.214  & -0.267  & -0.157  & -0.036  & -0.510  & -0.230  & -0.450  & 0.020  & -0.090  & -0.261  & -0.144  & -0.195  \\
& (-0.582,-0.172) & (-0.608,-0.328) & (-0.461,-0.280) & (-0.311,-0.104) & (-0.810,-0.459) & (-0.249,-0.114) & (-0.579,-0.292) & (-0.390,-0.124) & (-0.312,-0.177) & (-0.228,-0.114) & (-0.351,-0.095) & (-0.575,-0.413) & (-0.335,-0.197) & (-0.506,-0.315) & (-0.140,0.016) & (-0.246,-0.063) & (-0.340,-0.168) & (-0.248,-0.097) & (-0.276,-0.093) \\
\multicolumn{1}{c}{\multirow{2}{*}{LIQSPR}} & -0.005  & 0.033  & 0.023  & 0.027  & 0.091  & 0.012  & -0.002  & 0.039  & -0.026  & 0.045  & -0.015  & 0.014  & 0.012  & -0.031  & -0.051  & 0.052  & 0.019  & 0.002  & -0.018  \\
& (-0.053,0.053) & (-0.026,0.052) & (-0.021,0.032) & (-0.008,0.053) & (0.038,0.151) & (-0.007,0.047) & (-0.075,0.009) & (-0.017,0.049) & (-0.053,0.024) & (0.026,0.056) & (-0.022,0.051) & (-0.020,0.058) & (-0.022,0.028) & (-0.071,0.020) & (-0.084,-0.014) & (0.033,0.098) & (-0.022,0.064) & (0.005,0.065) & (-0.036,0.033) \\
\multicolumn{1}{c}{\multirow{2}{*}{3MTB}} & 0.098  & -0.011  & -0.093  & 0.011  & -0.065  & 0.098  & 0.055  & 0.078  & 0.002  & -0.026  & 0.086  & 0.009  & -0.021  & 0.082  & 0.051  & 0.118  & 0.083  & -0.015  & -0.004  \\
& (0.042,0.189) & (-0.082,0.058) & (-0.102,0.019) & (-0.037,0.069) & (-0.121,0.002) & (0.076,0.184) & (0.050,0.190) & (0.009,0.133) & (-0.103,0.023) & (-0.032,0.032) & (0.011,0.097) & (-0.000,0.120) & (-0.044,0.047) & (0.027,0.131) & (0.015,0.089) & (0.065,0.152) & (0.033,0.123) & (-0.066,0.015) & (-0.034,0.065) \\
\multicolumn{1}{c}{\multirow{2}{*}{TERMSPR}} & 0.107  & 0.030  & 0.053  & 0.092  & -0.001  & 0.051  & -0.011  & 0.026  & 0.025  & -0.011  & 0.063  & 0.023  & -0.026  & 0.094  & 0.058  & 0.118  & 0.069  & -0.056  & -0.043  \\
& (0.066,0.186) & (-0.027,0.078) & (0.034,0.114) & (0.015,0.104) & (-0.048,0.028) & (0.024,0.103) & (-0.025,0.083) & (-0.023,0.084) & (-0.014,0.050) & (-0.026,0.026) & (0.005,0.084) & (-0.006,0.071) & (-0.030,0.028) & (0.040,0.114) & (0.039,0.089) & (0.061,0.131) & (0.020,0.085) & (-0.091,-0.026) & (-0.059,0.022) \\
\multicolumn{1}{c}{\multirow{2}{*}{CREDSPR}} & 0.066  & 0.038  & 0.002  & -0.006  & 0.045  & -0.116  & -0.167  & 0.037  & -0.070  & -0.026  & 0.023  & 0.041  & -0.009  & 0.107  & -0.024  & 0.141  & 0.079  & -0.203  & -0.037  \\
& (-0.047,0.169) & (-0.106,0.077) & (-0.065,0.100) & (-0.048,0.081) & (-0.039,0.121) & (-0.117,0.013) & (-0.153,0.003) & (-0.065,0.147) & (-0.164,-0.048) & (-0.054,0.061) & (-0.003,0.127) & (-0.015,0.097) & (-0.040,0.049) & (-0.031,0.120) & (-0.040,0.038) & (0.055,0.156) & (0.023,0.142) & (-0.231,-0.135) & (-0.083,0.029) \\
\multicolumn{1}{c}{\multirow{2}{*}{DJUSRE}} & 0.621  & 0.642  & 0.461  & 0.602  & 0.385  & 0.569  & 0.875  & 0.624  & 0.493  & 0.092  & 0.478  & 0.183  & 0.276  & 0.346  & 0.591  & 0.244  & 0.176  & 0.358  & 0.329  \\
& (0.548,0.922) & (0.432,0.698) & (0.346,0.590) & (0.562,0.835) & (0.384,0.690) & (0.403,0.609) & (0.549,0.829) & (0.409,0.676) & (0.423,0.614) & (0.065,0.208) & (0.324,0.533) & (0.110,0.277) & (0.174,0.335) & (0.324,0.520) & (0.426,0.590) & (0.144,0.350) & (0.130,0.329) & (0.323,0.482) & (0.223,0.398) \\
\multicolumn{1}{c}{\multirow{2}{*}{LEV}} & -4.748  & -2.016  & 7.434  & -2.195  & -6.049  & -0.180  & -2.172  & 0.120  & -4.647  & -3.147  & -1.379  & -2.262  & -1.543  & -1.657  & -1.076  & -4.507  & 0.570  & 22.493  & 9.380  \\
& (-4.609,-1.123) & (-2.480,0.960) & (5.754,8.236) & (-4.715,-1.883) & (-6.803,1.339) & (-1.131,0.000) & (-4.018,-1.410) & (-0.112,1.184) & (-8.058,-4.422) & (-7.432,4.933) & (-10.645,1.433) & (-6.651,4.404) & (-3.639,7.919) & (-1.753,-1.567) & (-2.138,4.223) & (-7.159,4.734) & (-2.984,8.990) & (21.227,33.193) & (5.406,9.456) \\
\multicolumn{1}{c}{\multirow{2}{*}{MK2BK}} & -1.813  & 3.965  & -5.967  & -5.682  & -3.652  & 2.577  & -2.421  & 0.000  & -0.919  & -1.065  & -8.302  & -5.970  & -2.912  & -0.005  & -5.311  & -3.947  & -6.556  & 6.416  & 7.269  \\
& (-6.966,-1.801) & (-5.516,2.773) & (-7.438,-0.040) & (-7.437,1.429) & (-9.469,0.754) & (0.075,4.741) & (-5.909,2.442) & (-0.000,0.000) & (-3.836,0.050) & (-2.061,1.416) & (-9.582,0.143) & (-7.866,-0.416) & (-4.142,-0.508) & (-0.009,0.007) & (-9.451,-5.360) & (-8.723,-2.496) & (-13.354,-5.472) & (1.652,10.367) & (5.484,14.125) \\
\multicolumn{1}{c}{\multirow{2}{*}{SIZE}} & 10.337  & 4.014  & 1.037  & 11.342  & 21.974  & 1.990  & 3.975  & 9.592  & 1.528  & 0.113  & -0.279  & 4.189  & 7.332  & 1.060  & 6.710  & 2.545  & 12.098  & 3.623  & 4.409  \\
& (6.938,10.738) & (1.450,4.901) & (0.589,2.404) & (10.561,14.765) & (14.293,23.179) & (1.376,5.079) & (2.931,6.892) & (9.239,11.702) & (1.399,3.866) & (-2.178,0.983) & (-3.704,1.133) & (2.326,4.373) & (5.671,7.854) & (0.427,1.090) & (5.197,7.616) & (1.133,3.660) & (11.082,15.720) & (2.771,4.066) & (3.580,6.515) \\
\multicolumn{1}{c}{\multirow{2}{*}{MM}} & 2.416  & -2.600  & 2.491  & -2.095  & -5.669  & -4.732  & -1.288  & -1.058  & -3.205  & -5.392  & 1.594  & -0.998  & -1.434  & -4.732  & -3.612  & 4.854  & -0.169  & 0.299  & -1.372  \\
& (-6.762,5.264) & (-5.379,6.361) & (-7.012,5.229) & (-5.415,6.715) & (-6.873,5.029) & (-7.044,4.453) & (-7.058,5.047) & (-4.568,5.630) & (-8.965,3.998) & (-9.433,-0.051) & (-4.613,8.208) & (-6.219,5.500) & (-6.526,4.561) & (-6.274,6.346) & (-6.257,5.845) & (-5.056,6.705) & (-4.141,3.884) & (-7.265,4.380) & (-5.711,6.326) \\
\multicolumn{1}{c}{\multirow{2}{*}{$\sigma_j$}} & 0.398  & 0.423  & 0.330  & 0.349  & 0.353  & 0.344  & 0.490  & 0.294  & 0.326  & 0.193  & 0.250  & 0.242  & 0.264  & 0.202  & 0.191  & 0.280  & 0.257  & 0.263  & 0.263  \\
& (0.397,0.486) & (0.424,0.512) & (0.320,0.393) & (0.294,0.362) & (0.279,0.389) & (0.335,0.403) & (0.432,0.527) & (0.253,0.345) & (0.255,0.320) & (0.192,0.234) & (0.205,0.271) & (0.239,0.298) & (0.230,0.277) & (0.176,0.220) & (0.182,0.226) & (0.250,0.315) & (0.220,0.273) & (0.242,0.291) & (0.219,0.274) \\
\hline
\multicolumn{1}{c}{\multirow{2}{*}{$\text{CoVaR}$}} & \multicolumn{9}{c}{Financial} & \multicolumn{2}{c}{Consumer} & \multicolumn{2}{c}{Energy} & \multicolumn{2}{c}{Industrial} & \multicolumn{2}{c}{Technology} & \multicolumn{2}{c}{Utilities}\\
\cmidrule(lr){2-10}\cmidrule(lr){11-12} \cmidrule(l){13-14} \cmidrule(l){15-16} \cmidrule(l){17-18} \cmidrule(l){19-20}
& C & BAC & CMA & JPM & KEY & GS & MS & MCO & AXP & MCD & MKE & CVX & XOM & BA & GE & INTC & ORCL & AEE & PEG\\
\cline{2-20}
\multicolumn{1}{c}{\multirow{2}{*}{VIX}} & -0.094& -0.121& -0.129& -0.279& -0.157& -0.124& -0.333& -0.172& -0.188& -0.161& -0.110& -0.080& -0.106& -0.314& -0.296& -0.143& -0.132& -0.169& -0.089\\
& (-0.166,-0.055) & (-0.243,-0.099) & (-0.165,-0.086) & (-0.238,-0.128) & (-0.228,-0.127) & (-0.162,-0.100) & (-0.419,-0.310) & (-0.309,-0.122) & (-0.292,-0.186) & (-0.193,-0.119) & (-0.182,-0.107) & (-0.120,-0.067) & (-0.147,-0.081) & (-0.393,-0.300) & (-0.319,-0.217) & (-0.207,-0.107) & (-0.140,-0.063) & (-0.206,-0.132) & (-0.136,-0.063) \\
\multicolumn{1}{c}{\multirow{2}{*}{LIQSPR}} & -0.005& 0.017& 0.011& -0.006& -0.009& -0.000& -0.006& 0.023& -0.004& 0.006& -0.008& 0.002& -0.009& 0.003& 0.008& 0.007& -0.007& 0.002& 0.001\\
& (-0.029,0.003) & (-0.014,0.020) & (-0.013,0.017) & (-0.029,0.007) & (-0.014,0.012) & (-0.016,0.001) & (-0.022,0.006) & (0.007,0.050) & (-0.009,0.018) & (-0.005,0.015) & (-0.015,0.002) & (-0.011,0.002) & (-0.020,-0.004) & (-0.019,0.018) & (-0.006,0.025) & (-0.008,0.016) & (-0.020,-0.002) & (-0.011,0.005) & (-0.009,0.007) \\
\multicolumn{1}{c}{\multirow{2}{*}{3MTB}} & 0.036& 0.034& 0.038& 0.043& 0.040& 0.038& 0.046& 0.040& 0.031& 0.063& 0.021& 0.019& 0.033& 0.037& -0.001& 0.009& 0.018& 0.059& 0.031\\
& (0.012,0.061) & (0.018,0.065) & (0.034,0.072) & (0.030,0.075) & (0.038,0.073) & (0.018,0.050) & (0.019,0.056) & (0.019,0.066) & (0.022,0.057) & (0.028,0.072) & (0.010,0.046) & (0.025,0.053) & (0.028,0.054) & (0.011,0.054) & (0.002,0.042) & (0.001,0.043) & (0.002,0.040) & (0.046,0.074) & (0.021,0.063) \\
\multicolumn{1}{c}{\multirow{2}{*}{TERMSPR}} & 0.021& 0.011& 0.030& 0.022& 0.035& 0.031& 0.030& 0.035& 0.022& 0.045& 0.021& 0.012& 0.033& 0.039& 0.006& 0.024& 0.030& 0.048& 0.038\\
& (0.002,0.035) & (0.007,0.040) & (0.017,0.045) & (0.008,0.040) & (0.027,0.051) & (0.016,0.038) & (0.012,0.037) & (0.017,0.053) & (0.010,0.042) & (0.028,0.056) & (0.015,0.041) & (0.020,0.044) & (0.028,0.052) & (0.024,0.056) & (0.006,0.034) & (0.016,0.045) & (0.014,0.038) & (0.040,0.062) & (0.032,0.060) \\
\multicolumn{1}{c}{\multirow{2}{*}{CREDSPR}} & -0.002& -0.011& 0.007& 0.066& 0.017& 0.030& 0.036& 0.035& 0.037& 0.009& -0.021& 0.021& 0.020& 0.075& 0.015& -0.012& 0.025& 0.064& 0.008\\
& (-0.013,0.052) & (-0.037,0.030) & (-0.008,0.037) & (0.020,0.084) & (0.005,0.057) & (0.008,0.044) & (0.003,0.050) & (0.030,0.103) & (0.018,0.066) & (-0.010,0.040) & (-0.025,0.021) & (0.009,0.046) & (0.001,0.041) & (0.034,0.093) & (0.000,0.050) & (-0.002,0.060) & (-0.014,0.028) & (0.027,0.073) & (-0.006,0.044) \\
\multicolumn{1}{c}{\multirow{2}{*}{DJUSRE}} & 0.293& 0.319& 0.347& 0.239& 0.375& 0.292& 0.292& 0.366& 0.301& 0.290& 0.326& 0.294& 0.283& 0.286& 0.307& 0.295& 0.268& 0.274& 0.350\\
& (0.246,0.341) & (0.271,0.373) & (0.275,0.358) & (0.222,0.319) & (0.271,0.377) & (0.239,0.308) & (0.269,0.342) & (0.287,0.396) & (0.236,0.310) & (0.249,0.324) & (0.294,0.368) & (0.255,0.314) & (0.278,0.341) & (0.219,0.300) & (0.271,0.353) & (0.260,0.347) & (0.246,0.319) & (0.246,0.318) & (0.323,0.406) \\
\multicolumn{1}{c}{\multirow{2}{*}{LEV}} & 2.630& -6.060& 4.650& 3.058& -3.064& 0.268& -0.064& 0.860& -0.366& 8.045& -22.402& 7.352& -35.741& -1.119& 9.080& 10.716& -0.779& -11.400& 9.890\\
& (2.268,2.707) & (-7.307,-5.267) & (4.248,6.207) & (2.416,3.528) & (-3.571,-2.549) & (0.287,0.856) & (-0.121,0.227) & (0.782,1.070) & (-0.706,1.757) & (-0.980,9.195) & (-24.988,-18.350) & (1.225,10.041) & (-38.175,-34.204) & (-1.154,-1.111) & (8.050,10.669) & (6.473,18.929) & (-1.657,3.050) & (-13.780,-6.341) & (8.222,11.550) \\
\multicolumn{1}{c}{\multirow{2}{*}{MK2BK}} & 0.300& -8.786& -3.066& -1.627& -8.476& 1.076& 25.007& 0.000& 5.322& -0.279& -0.787& -1.459& -6.012& -0.003& -6.857& -4.200& -0.396& -4.331& -0.780\\
& (-1.643,0.835) & (-11.358,-6.699) & (-14.052,-0.820) & (-4.903,3.550) & (-14.151,-7.705) & (-0.043,2.216) & (20.889,34.748) & (-0.000,0.000) & (4.433,6.322) & (-1.372,-0.065) & (-1.913,-0.332) & (-2.508,0.178) & (-6.691,-4.370) & (-0.005,0.000) & (-12.815,-6.164) & (-18.357,-2.606) & (-0.807,0.152) & (-4.284,1.753) & (-1.347,0.237) \\
\multicolumn{1}{c}{\multirow{2}{*}{SIZE}} & 3.978& -4.920& -0.179& 11.516& -4.149& -0.405& -15.947& 26.698& -5.300& 0.296& -1.783& -0.274& -0.263& -1.691& -0.400& 13.690& 7.174& -3.359& 1.080\\
& (4.127,4.899) & (-5.530,-3.787) & (-0.803,0.159) & (11.033,11.849) & (-4.625,-2.734) & (-1.208,-0.310) & (-17.190,-15.678) & (26.549,27.138) & (-6.802,-4.744) & (0.129,2.058) & (-2.431,-1.403) & (-0.793,0.752) & (-0.460,-0.138) & (-1.796,-1.624) & (-0.922,0.768) & (12.511,17.506) & (6.381,7.386) & (-5.721,-2.889) & (0.501,1.572) \\
\multicolumn{1}{c}{\multirow{2}{*}{MM}} & 8.429& -0.878& 0.429& -0.759& -3.787& 4.763& 6.950& -0.906& 2.560& -1.681& 3.131& 1.197& 3.288& 5.928& -1.908& 6.849& -0.743& 3.734& -1.384\\
& (-2.087,9.599) & (-8.888,3.136) & (-5.980,5.878) & (-4.267,7.510) & (-8.054,4.351) & (-5.082,5.185) & (0.364,11.800) & (-3.319,3.090) & (-2.423,10.332) & (-7.046,0.119) & (-0.371,5.001) & (-3.896,4.845) & (-2.685,8.078) & (-7.234,5.686) & (-7.929,2.842) & (1.663,17.154) & (-1.229,0.620) & (-3.442,6.773) & (-6.165,3.729) \\
\multicolumn{1}{c}{\multirow{2}{*}{$\sigma_k$}} & 0.123  & 0.121  & 0.094  & 0.100  & 0.089  & 0.112  & 0.094  & 0.100  & 0.080  & 0.106  & 0.107  & 0.092  & 0.091  & 0.068  & 0.091  & 0.114  & 0.100  & 0.094  & 0.107  \\
& (0.103,0.126) & (0.101,0.129) & (0.078,0.103) & (0.097,0.122) & (0.075,0.096) & (0.094,0.113) & (0.083,0.102) & (0.084,0.106) & (0.071,0.090) & (0.095,0.115) & (0.097,0.117) & (0.081,0.098) & (0.077,0.094) & (0.054,0.070) & (0.084,0.107) & (0.092,0.118) & (0.096,0.117) & (0.088,0.108) & (0.100,0.121) \\
      \bottomrule
\end{tabular}}
\caption{Parameter estimates obtained by fitting the Dynamic CoVaR model to each of the 19 assets vs SP500 and all the exogenous variables, for the confidence levels $\tau=0.05$. For each regressor the first row reports parameter estimates by Maximum a Posteriori, while the second row reports the 95\% High Posterior Density (HPD) credible sets.}
\label{tab:Model_Dynamic_ParEst_005}
\end{small}
\smallskip
\end{sidewaystable}
%
\clearpage

\newpage
%

%

\begin{thebibliography}{99}

%
%
\bibitem{acharya_etal.2012} \textsc{Acharya, V.V,, Engle, R.F. and Richardson, M., (2012).} Capital shortfall: a new approach to ranking and regulating systemic risks. \textit{American Economic Review}, \textbf{102}, pp. 59--64.
%
\bibitem{acharya_etal.2010} \textsc{Acharya, V.V., E., Philippon, T. and Richardson, M., (2010).} Measuring systemic Risk. \textit{New York University Working Paper}.
%
\bibitem{adams_etal.2010} \textsc{Adams, Z., F\"{u}ss, R. and Gropp, R., (2010).} Modeling spillover effects among financial institutions: a state-dependent sensitivity Value-at-Risk (SDSVaR) approach. \textit{Working paper, European Business School}.
%
\bibitem{adrian_brunnermeier.2011} \textsc{Adrian, T. and Brunnermeier, M.K., (2011).} CoVaR. \textit{Unpublished manuscript}.
%
\bibitem{bernardi.2013} \textsc{Bernardi, M., (2013).} Risk measures for skew normal mixtures. \textit{Statistics \& Probability Letters}, \textbf{83}, pp. 1819--1824.
%
\bibitem{bernardi_etal.2012} \textsc{Bernardi, M., Maruotti, A. and Petrella, L., (2012).} Skew mixture models for loss distributions: a Bayesian approach. \textit{Insurance: Mathematics and Economics}, \textbf{51}, pp. 617--623.
%
\bibitem{bernardi_etal.2013} \textsc{Bernardi, M., Maruotti, A. and Petrella, L., (2013).} Multivariate Markov-switching models and tail risk interdependence.. \textit{Working paper}.
%
\bibitem{billio_etal.2012} \textsc{Billio, M., Getmansky, M., Lo, A.W. and Pellizon, L., (2012).} Econometric measures of connectedness and systemic risk in the finance and insurance sectors. \textit{Journal of Financial Economics}, \textbf{104}, pp. 535--559.
%
\bibitem{brownlees_engle.2012} \textsc{Brownlees, C.T. and Engle, R. (2012).} Volatility, correlation and tails for systemic risk measurement. \textit{Working paper}.
%
\bibitem{chao_etal.2012} \textsc{Chao, S.-K., H\"ardle, W.F. and Chang, W. (2012).} Quantile Regression in Risk Calibration. \textit{Handbook for Financial Econometrics and Statistics} in Cheng-Few Lee, ed., Springer Verlag.
%
\bibitem{chernozhukov_du.2008} \textsc{Chernozhukov, V. and Du, S., (2008).} Extremal Quantiles and Value-at-Risk. \textit{The New Palgrave Dictionary of Economics. Second Edition. Eds. Steven N. Durlauf and Lawrence E. Blume.} Palgrave Macmillan.
%
\bibitem{carter_kohn.1994} \textsc{Carter, C.K. and Kohn, R., (1994).} On Gibbs sampling for state space models. \textit{Biometrika}, \textbf{81}, pp. 541--553.
%
\bibitem{carter_kohn.1996} \textsc{Carter, C.K. and Kohn, R., (1996).} Markov chain Monte Carlo in conditionally Gaussian state space models. \textit{Biometrika}, \textbf{83}, pp. 589--601.
%
%
\bibitem{dejong.1991} \textsc{De Jong P., (1991).} The diffuse Kalman filter. \textit{Annals of Statistics}, \textbf{19}, pp. 1073--1083.
%
\bibitem{dejong_shephard.1995} \textsc{De Jong P. and Shephard, N., (1995).} The simulation smoother for time series models. \textit{Biometrika}, \textbf{82}, pp. 339--350.
%
\bibitem{derossi_harvey.2009} \textsc{De Rossi, G. and Harvey, A., (2009).} Quantiles, expectiles and splines. \textit{Journal of Econometrics}, \textbf{152}, pp. 179--185.
%
\bibitem{durbin_koopman.2012} \textsc{Durbin, J. and Koopman, S.J., (2012).} \textit{Time Series Analysis by State Space Methods, 2nd Eds}, Oxford: Oxford University Press.
%
\bibitem{durbin_koopman.2002} \textsc{Durbin, J. and Koopman, S.J., (2002).} A simple and efficient simulation smoother for state space time series analysis. \textit{Biometrika}, \textbf{89}, pp. 1603--616.
%
%
\bibitem{engle_manganelli.2004} \textsc{Engle, R.F. and Manganelli, S., (2004).} CAViaR: Conditional autoregressive value at risk by regression quantiles. \textit{Journal of Business and Economic Statistics}, \textbf{22}, pp. 367--381.
%
\bibitem{fan_etal.2013} \textsc{Fan, Y., H\"ardle, W.K., Wang, W. and Zhu L. (2013).} Composite quantile regression for the single-index model. \textit{SFB 649 Discussion Paper}, 2013--010.
%
\bibitem{fruhwirth-schnatter.1994} \textsc{Fr\"uhwirth-Schnatter, S., (1994).} Data augmentation and dynamic linear models, \textit{Journal of time series analysis}, \textbf{15}, pp. 183--202.
%
\bibitem{gelfand_smith.1990} \textsc{Gelfand, A.E. and Smith, F.M., (1990).} Sampling-based approaches to calculating marginal densities. \textit{Journal of the American Statistical Association}, \textbf{85}, pp. 398--409. 
%
\bibitem{geman_geman.1984} \textsc{Geman, S. and Geman, D., (1984).} Stochastic relaxation, Gibbs distributions, and the Bayesian restoration of images. \textit{IEEE Trans. Pattern Analysis and Machine Intelligence}, \textbf{6}, pp. 721--741. 
%
\bibitem{gerlach_etal.2011} \textsc{Gerlach, R.H., Chen, C.W.S. and Chan, N.Y.C., (2011).} Bayesian time-varying quantile forecasting for Value-at-Risk in financial markets. \textit{Journal of Business \& Economic Statistics}, \textbf{29}, pp. 481--492.
%
\bibitem{geweke_1992} 
\textsc{Geweke, J., (1992).} Evaluating the Accuracy of Sampling-Based
Approaches to the Calculation of Posterior Moments. In J. M.
Bernardo, J. Berger, A. P. Dawid, and A. F. M. Smith, eds.,
\textit{Bayesian Statistics 4}, Oxford University Press, pp. 169--193.
%
\bibitem{geweke_2005}
\textsc{Geweke, J., (2005).} \emph{Contemporary Bayesian Econometrics and Statistics}. Wiley Series in Probability and Statistics. Wiley, Hoboken.
%
\bibitem{girardi_ergun.2013}
\textsc{Girardi, G., Erg\"{u}n, A.T., (2013).} \emph{Systemic risk measurement: Multivariate GARCH estimation of CoVaR}. \textit{Journal of Banking \& Finance}, \url{http://dx.doi.org/10.1016/j.jbankfin.2013.02.027}.
%
\bibitem{gourieroux_jasak.2008} \textsc{Gourieroux, C. and Jasiak. J., (2008).} Dynamic quantile models. \textit{Journal of Econometrics}, \textbf{147}, pp. 198--205.
%
\bibitem{hautsch_etal.2011} \textsc{Hautsch, N., Schaumburg, J. and Schienle, M., (2011).} Quantifying time-varying marginal systemic risk contributions. \textit{Unpublished manuscript}.
%
\bibitem{huang.2012} \textsc{Huang, A.Y., (2012).} Value at risk estimation by quantile regression and kernel estimator. \textit{Review of Quantitative Finance and Accounting}, \url{doi 10.1007/s11156-012-0308-x}
%
\bibitem{johannes_polson.2009} \textsc{Johannes, M. and Polson, N., (2009).} MCMC methods for Financial Econometrics, in \textit{Handbook of Financial Econometrics}, Volume 2, pp. 1--72, edited by Y. Ait-Sahalia and L.P. Hansen, North Holland.
%
\bibitem{jorion.2007} \textsc{Jorion, P., (2007).} \textit{Value-at-Risk: The new benchmark for managing financial risk}. McGraw-Hill, Chicago.
%
%
\bibitem{koenker.2005} \textsc{Koenker, P., (2005).} \textit{Quantile Regression}. Cambridge University Press, Cambridge.
%
\bibitem{koenker_basset.1978} \textsc{Koenker, B. and Basset, G., (1978).} Regression Quantiles. \textit{Econometrica}, \textbf{46}, pp. 33--50.
%
\bibitem{koenker_xiao.2006} \textsc{Koenker, R. and Xiao, Z., (2006).} Quantile autoregression. \textit{Journal of the American Statistical Association}, \textbf{101}, pp. 980--990.
%
%
\bibitem{koopman.1991} \textsc{Koopman, S.J., (1991).} Efficient smoothing algorithms for time series models. \textit{Doctoral Thesis}. 
%
\bibitem{koopman_etal.1998} \textsc{Koopman, S.J., Shephard, N. and Doornik, J.A., (1998).} Statistical algorithms for models in state space using Ssf Pack 2.2. \textit{Econometrics Journal}, \textbf{1}, pp. 1--55.
%
\bibitem{kottas_gelfand.2001} \textsc{Kottas, A. and Gelfand, A.E., (2001).} Bayesian semiparametric median regression modeling. \textit{Journal of the American Statistical Association}, \textbf{96}, pp. 1458-1468.
%
\bibitem{kottas_krnjajic.2009} \textsc{Kottas, A., Krnjajic, M., (2009).} Bayesian semiparametric modelling in quantile regression. \textit{Scandinavian Journal of Statistics}, \textbf{36}, pp.297-319.
%
\bibitem{kotz_etal.2001} \textsc{Kotz, S., Kozubowski, T.J. and Podg\'orski, K.}  (2001). \textit{The Laplace distribution and generalizations. A revisit with Applications to Communications, Economics, Engineering, and Finance}, Birkh\"auser.
%
\bibitem{kozumi_kobayashi.2011} \textsc{Kozumi, H. and Kobayashi, G., (2011).} \textit{Gibbs Sampling Methods for Bayesian Quantile Regression}, \textit{Journal of Statistical
Computation and Simulation}, \textbf{81}, pp. 1565--1578.
%
\bibitem{kuester_etal.2006} \textsc{Kuester, K., Mittnik, S. and
Paolella, M.S., (2011).} Value-at-Risk Prediction: A Comparison of Alternative Strategies, \textit{Journal of Financial Econometrics}, \textbf{4}, pp. 53--89.
%
\bibitem{kurose_omori.2012} \textsc{Kurose, Y. and Omori, Y., (2012).} Bayesian analysis of time-varying quantiles using a smoothing spline, \textit{Journal of the Japan Statistical Society}, \textbf{42}, pp. 23--46.
%
%
%
\bibitem{lee_su.2012} \textsc{Lee, C.F. and Su, J.B., (2012).} Alternative statistical distributions for estimating value-at-risk: theory and evidence, \textit{Review of Quantitative Finance and Accounting}, \textbf{39}, pp. 39:309--331.
%
\bibitem{lin_chang.2012} \textsc{Lin, N. and Chang, C., (2012).} Comment on Article by K. Lum and A.E. Gelfand, ``Spatial Quantile Multiple Regression Using the Asymmetric Laplace Process'', \textit{Bayesian Analysis}, \textbf{7}, pp. 263--270.
%
\bibitem{liu.1994} \textsc{Liu, J.S. (1994).} The Collapsed Gibbs Sampler in Bayesian Computations with Applications to a Gene Regulation
Problem, \textit{Journal of the American Statistical Association}, \textbf{89}, pp. 958--966.
%
\bibitem{lum_gelfand.2012} \textsc{Lum, K. and Gelfand, A.E., (2012).} Spatial Quantile Multiple Regression Using the Asymmetric Laplace Process, \textit{Bayesian Analysis}, \textbf{7}, pp. 235--258.
%
\bibitem{mcneil_etal.2005} \textsc{McNeil, A., Frey, R. and Embrechts, P., (2005).} \textit{Quantitative risk management: Concepts, Techniques, Tools}. Princeton Series in Finance, Princeton.
%
\bibitem{park_casella.2008} \textsc{Park, T. and Casella, G., (2008).} The Bayesian Lasso. \textit{Journal of the American Statistical Association}, \textbf{103}, pp. 681--686.
%
\bibitem{tanner_wong.1987} \textsc{Tanner, M.A. and Wong, W.H., (1987).} The calculation of posterior distributions by data augmentation. \textit{Journal of the American Statistical Society}, \textbf{82}, pp. 528--550.
%
%
\bibitem{schaumburg.2010} \textsc{Schaumburg, J., (2010).} Predicting extreme VaR: nonparametric quantile regression with refinements from extreme value theory, \textit{SFB 649 Discussion Paper}, 2010--009.
%
\bibitem{taylor.2008} \textsc{Taylor, J.W., (2008).} Simultaneous linear quantile regression: a semiparametric Bayesian approach. \textit{Bayesian Analysis}, \textbf{7}, pp. 51--72.
%
\bibitem{tokdar_kadane.2012} \textsc{Tokdar, S.T and Kadane, J.B., (2008).} Using exponentially weighted quantile regression to estimate value at risk and expected shortfall. \textit{Journal of Financial Econometrics}, \textbf{6}, pp. 382--406.
%
\bibitem{van_dyk_park.2008} \textsc{van Dyk, D.A. and Park, T., (2008).} Partially collapsed Gibbs samplers: Theory and Methods. \textit{Journal of the American Statistical Association}, \textbf{110}, pp. 790--796.
%
%
\bibitem{yu_moyeed.2001} \textsc{Yu, K. and Moyeed, R., (2001)}. Bayesian quantile regression. \textit{Statistics and Probability Letters}, \textbf{54}, pp. 437--447.
%
\end{thebibliography}
\end{document}